# Shock wave formation in the thermosphere by an earthgrazing fireball: Empirical evidence for volatile-enhanced hydrodynamic shielding

Elizabeth A. Silber[1,*], Denis Vida[2,3], Miro Ronac Giannone[1], Jamie Shepherd[4], Sarah Albert[1], Daniel C. Bowman[1,¥], Tammy Do[2], Margaret Campbell-Brown[2,3], Peter Jenniskens[5], Reynold E. Silber[6]

[1]Sandia National Laboratories, Albuquerque, NM, 87123, USA; [2]Department of Physics and Astronomy, University of Western Ontario, London, Ontario, N6A 3K7, Canada; [3]Western Institute for Earth and Space Exploration, University of Western Ontario, London, Ontario, N6A 5B7, Canada; [4]Easter Auchraw Cottage, Lochearnhead, FK19 8PU, Scotland, UK; [5]SETI Institute, Mountain View, CA, 94043, USA; [6]Sentinel Scientific Consulting Group LLC, Rio Rancho, NM, 87144, USA

¥now at Urban Sky Theory, Denver, CO, 80216, USA

*Corresponding author: esilber[at]sandia.gov

**Article accepted for publication in Icarus on 21 May 2026**

**Special Issue: Meteoroids 2025 - Recent Advances in Meteor Science**

## Abstract

Hydrodynamic shielding is a theoretically well-established but observationally elusive and experimentally difficult-to-replicate phenomenon with implications that extend far beyond meteor physics. Rare earthgrazing meteoroids with infrasound signatures that penetrate to the ground can be used to probe hydrodynamic shielding that leads to strong shock formation at high altitude. Here, we report the first coordinated optical and multi-station infrasound observations of a centimeter-scale earthgrazing fireball that generated sustained cylindrical line shock at thermospheric altitudes near 92 km. The event was recorded by numerous optical stations and three infrasound arrays, allowing trajectory reconstruction, ablation behavior, acoustic source localization, and shock characteristics. Optical observations indicate early mechanical erosion and ablation/evaporation at exceptionally low dynamic pressure, consistent with a cometary or a porous, volatile-bearing CM chondritic object. Independent infrasound detections localize shock generation to multiple points along a 164 km trajectory segment near perigee. Weak-shock modeling yields a consistent blast radius of ~30 m, implying an acoustic-equivalent source size far exceeding the physical dimensions of the ~45 g nucleus. We demonstrate that classical gas dynamics and ablation-driven hydrodynamic shielding alone cannot account for these observations under ambient thermospheric conditions. We show that volatile release provides the additional flow-field density enhancement required to amplify hydrodynamic shielding, reduce the effective local Knudsen number, and sustain a shock envelope capable of radiating detectable infrasound. These results demonstrate that small, volatile-rich meteoroids can transiently establish continuum-like flow in rarefied environments.

## Highlights

- First optical–infrasound detection of thermospheric shock from an earthgrazer
- Shock formation occurs at ~92 km under exceptionally low dynamic pressure
- Volatile release amplifies hydrodynamic shielding in rarefied flow

# 1. Introduction

## *1.1 Motivation*

Meteoroids entering Earth's atmosphere serve as natural probes of hypersonic flow physics under extreme conditions, enabling investigation of ablation, fragmentation, physico-chemical processes, and acoustic coupling (Ceplecha et al., 1998; Silber et al., 2018). Their passage through the upper atmosphere provides direct evidence of energy transfer mechanisms, thermal disequilibrium, and dynamic response of matter exposed to high-enthalpy environments. Depending on their size, velocity, composition, and internal structure, these bodies undergo initial sputtering, gradual differential ablation, progressive fragmentation, or complete disruption leading to explosive airbursts, with only a small fraction surviving to deliver meteorites to the surface (Ceplecha et al., 1998). Hydrodynamic shielding, defined as the accumulation of ablated and vaporized material that modifies the near-field flow around a hypersonic body, is a well-established theoretical concept in atmospheric entry physics and has been invoked across a wide range of flow regimes to explain enhanced coupling and shock formation (Boyd, 2000; Jenniskens et al., 2000; Josyula and Burt, 2011; Popova et al., 2001; Popova et al., 1998; Silber et al., 2018). It is also a necessary precursor for meteoroid generated shock waves. However, direct observational constraints on the development and effectiveness of hydrodynamic shielding, and by extension shock waves remain limited, particularly for small bodies in rarefied or transitional atmospheric conditions.

Within this broader context, earthgrazing meteoroids constitute a particularly rare but scientifically valuable class that can be used to test how hydrodynamic shielding and related flow-field processes develop under prolonged hypersonic entry in rarefied conditions. Entering the atmosphere at extremely shallow, almost tangential angles, they may traverse distances exceeding 1000 km through the upper atmosphere before either exiting back into space or becoming gravitationally bound to Earth (Shober et al., 2020). The long-duration, near-horizontal trajectories expose earthgrazers to sustained but relatively low dynamic pressures, providing a natural means to examine the potential formation of hydrodynamic shielding and shock waves as a function of the ablation dynamics, strength, porosity, and fragmentation thresholds of fragile, volatile-bearing materials under conditions that are difficult to reproduce in conventional laboratory settings.

Despite their diagnostic potential, earthgrazers remain among the least observed classes of meteoroid entries. In practice, the tangential passage of centimeter-sized earthgrazing objects through the rarefied lower thermosphere (above 90 km altitude) has never been observed simultaneously by both optical and acoustic methods. In principle, such trajectories are expected to produce only limited luminous output and weak ablation (or sputtering), conditions that suppress formation of the dense, high-temperature flow fields required for significant radiative emission. The absence of an ablationally amplified flow field limits deviation of the local Knudsen number from its ambient value, inhibiting the formation of a detached shock front (Bronshten, 1983). Under these circumstances, even high-velocity entries are not expected to generate coherent shocks in the rarefied thermosphere. Here, the local Knudsen number is understood as the ratio between the mean free path and a representative scale length of the object itself within the immediate flow field (Anderson, 2006).

Theory and numerical modeling have long suggested that ablation-driven hydrodynamic shielding can reduce the effective Knudsen number by orders of magnitude, promoting a transition toward continuum-like flow and enabling shock formation even when ambient conditions remain rarefied. A closely related physical analogy exists in the generation of shock waves by rocket vehicles at orbital altitudes near 188 km. Cotten and Donn (1971) demonstrated that infrasound detected at the ground from Apollo launch vehicles could only be explained if the rocket exhaust plume was treated as an enlarged, dense, conical source moving supersonically through a highly rarefied atmosphere. However, a cm-sized natural body traversing the rarefied atmosphere may not generate sufficient flow field density enhancements through typical ablation mechanisms, and such a deficit could only be compensated through release of volatiles. Although fundamentally different in origin, the volatile- and ablation-driven vapor envelope surrounding an earthgrazing meteoroid provides a natural analog to this plume-mediated enlargement of the effective hypersonic source. Earthgrazers traversing the lower thermosphere provide a rare observational test of this problem.

### *1.2 Background*

Historically, only a handful of earthgrazers have contributed to our understanding of such shallow entries, including the 1972 Great Daylight Fireball (Ceplecha, 1979; Ceplecha, 1994), the 1990 Czechoslovakia–Poland event (Borovička and Ceplecha, 1992), the 1992 Peekskill fireball, which produced a recovered meteorite (Beech et al., 1995; Brown et al., 1994; Ceplecha et al., 1995), and later observations over Japan (Abe et al., 2006) and Spain (Moreno et al., 2016). Yet, even these landmark cases produced no simultaneous optical and multi-point acoustic observations from thermospheric altitudes, and thus none provided the direct constraints needed to characterize shock formation and flow-field evolution in rarefied conditions. The deficiency is compounded by a systemic detection bias in current planetary-defense networks. Operational detection architectures, such as satellite-based systems contributing data to National Aeronautics and Space Administration (NASA) Jet Propulsion Laboratory (JPL) Center for Near-Earth Object Studies (CNEOS) fireball catalog, are not designed for dedicated meteoroid observation (e.g., Brown et al., 2002; Silber, 2024; Silber and Sawal, 2025). They primarily report bright, high-energy events associated with small asteroid impacts that may pose immediate impact risks (Peña-Asensio et al., 2022). As a result, shallow, non-terminal trajectories characterized by weak ablation and low luminosity often escape detection, creating a significant observational gap in the study of rarefied hypersonic flow.

One of the most effective diagnostic tools for identifying shock formation is the detection of infrasound. The decay of a ballistic shock produces acoustic waves that propagate efficiently with minimal attenuation over hundreds and sometimes thousands of kilometers (Campus and Christie, 2009; Evans et al., 1972; ReVelle, 1976). Combined optical and infrasound analyses have demonstrated that centimeter-scale meteoroids can generate detectable shocks at altitudes much higher than previously assumed, confirming that shock formation can occur within the slip and early transitional flow regimes (Brown et al., 2007; Moreno-Ibáñez et al., 2018). These findings provided direct observational support for hydrodynamic shielding in the upper atmosphere (Moreno-Ibáñez et al., 2018). These studies, however, did not explicitly consider the contribution of volatile release in the formation of acoustically detected shock waves from high altitude meteoroids.

While progress has been made for more typical meteoroid entries, the acoustic characterization of earthgrazers remains an almost entirely unexplored frontier. Aside from a single earthgrazing event identified in a broad survey but not analyzed in detail (Silber and Brown, 2014) and non-public acoustic records from the 1972 Great Daylight Fireball (ReVelle, 1997), no open-literature study has quantitatively linked an earthgrazer's optical dynamics to a robust infrasound signature detected by multiple arrays. This absence of coordinated optical and acoustic data represents a significant observational and analytical gap, preventing a comprehensive understanding of shock onset and propagation at thermospheric altitudes for this unique flight geometry.

### *1.3 Aim*

The fundamental question we seek to answer here is how small, centimeter-scale earthgrazers can generate strong shock waves, in the absence of strong ablation and in the region near the upper transitional to free-molecular flow boundary. In this work, we analyze a rare centimeter-scale earthgrazing meteoroid that on 22 September 2020 produced simultaneous optical and infrasonic detections near 92 km altitude over northern Europe. This fireball was recorded by 23 dedicated meteor patrol cameras as well as three infrasound stations of the Royal Netherlands Meteorological Institute (KNMI), marking the first documented earthgrazer with multi-point, multi-station infrasound detections, thereby providing the first quantitative linkage between optical ablation dynamics and acoustic signals. Through combined observational analysis and simple theoretical considerations, we show that hydrodynamic shielding, amplified by rapid volatile emission, can sustain shock formation in the lower thermosphere where continuum gas dynamics would otherwise predict no shock development. These findings thereby resolve a key paradox in meteor physics regarding the persistence of shock phenomena at thermospheric altitudes (Boyd et al., 1995; Josyula and Burt, 2011; Silber et al., 2018).

## 2. Data and Methods

Here we describe the observational data collected for the earthgrazing fireball that occurred over Northern Europe on 22 September 2020. The event was detected at 03:53:27 UTC (05:53 CEST / 04:53 BST), with a luminous trajectory spanning from northern Germany to the United Kingdom (**Figure 1**). In the following sections, we outline our methodological approach for reconstructing the meteoroid's atmospheric path, modeling its physical degradation, and characterizing the infrasound wavefield generated during its transit.

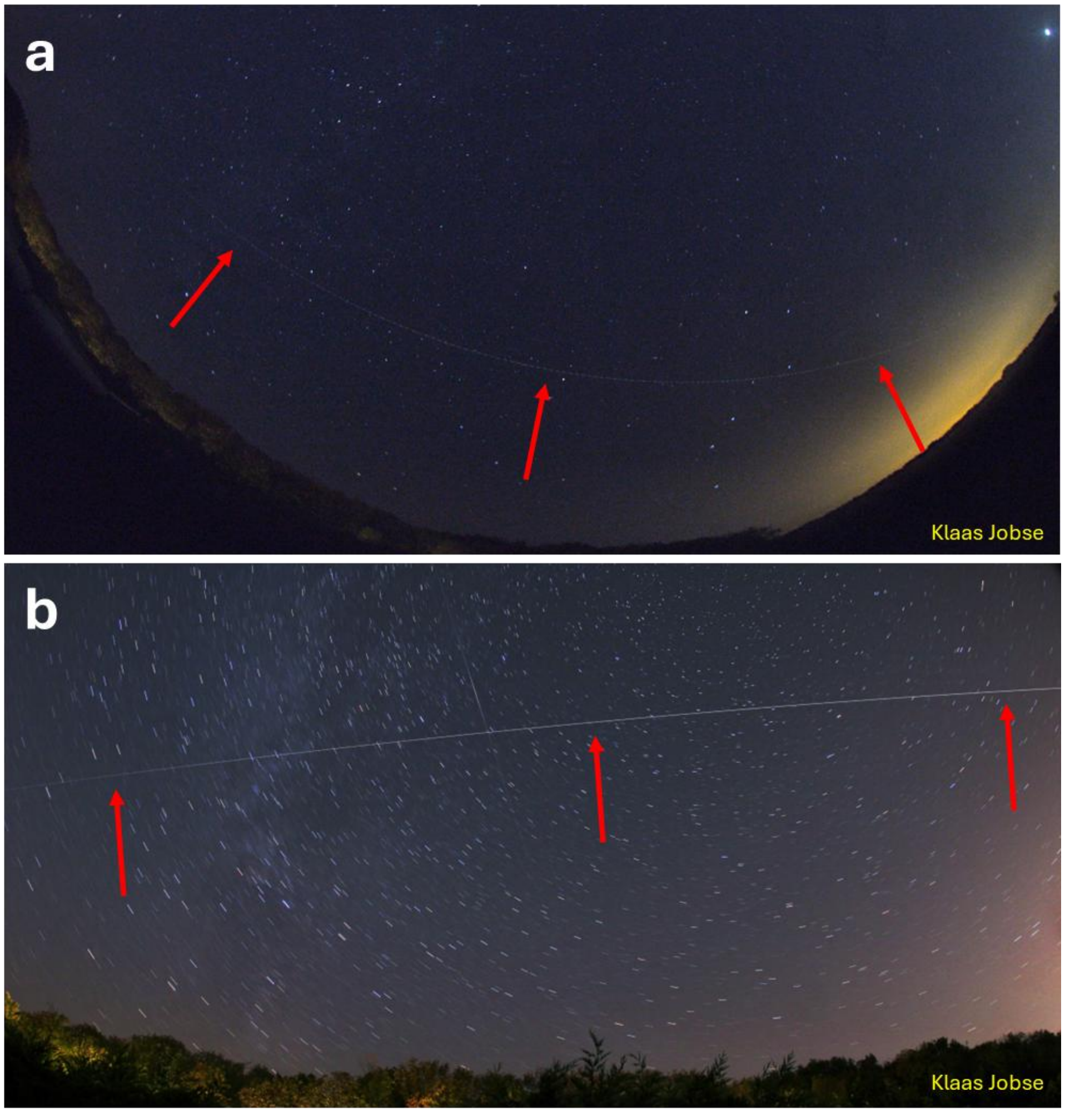

**Figure 1:** Images captured by the cameras of Klaas Jobse in Oostkapelle. (a) All-sky camera image taken from the Observatory. (b) Image taken with a 15 mm lens, showing only the last part of the meteor flight. The fireball motion is from right to left. There is the fading brightness at the end. Images credit: Klaas Jobse. Reproduced with permission of the author[1].

[1] https://sattrackcam.blogspot.com/2020/09/a-very-unusual-fireball-over-nw-europe.html

### *2.1. Optical Trajectory and Orbit Reconstruction*

The luminous trajectory was reconstructed using data from 22 cameras across six meteor networks: European Fireball Network (EN; Borovička et al. (2022)), Fripon (Colas et al. (2020)), Global Meteor Network (GMN; Vida et al. (2021)), International Meteor Organization's Video Meteor Network (IMO VMN; Molau and Barentsen (2014)), Network for Meteor Triangulation and Orbit Determination (NEMETODE; Stewart et al. (2013)), and the United Kingdom Meteor Network (UKMON; Campbell-Burns and Kacerek (2014)). This extensive coverage provided a dense triangulation geometry spanning Germany, the Netherlands, and the United Kingdom (**Figure 2**).

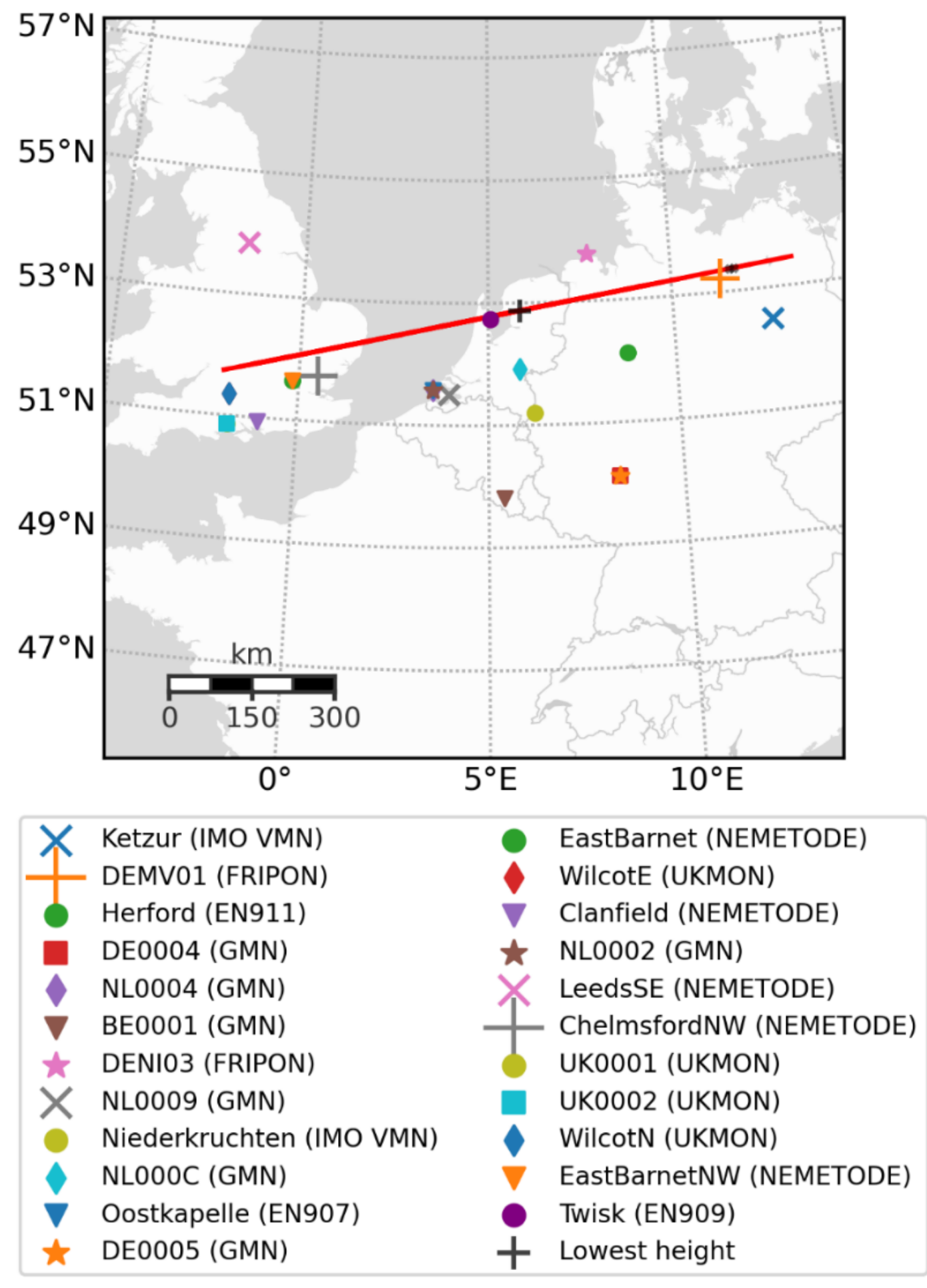

**Figure 2:** Gound map of the fireball showing the extent of its luminous trajectory. Individual stations are marked and ordered in the legend by the order in which they first observed the fireball. Co-located cameras are: DE0004 & DE0005; EastBarnet & EastBarnetNW; UK0001 & UK0002; EN907, NL0002 & NL0004. Camera-network attribution and station-source information are provided in the Data Availability section. The point at which the lowest height of 91.3 km was reached is marked.

Due to the large number of available data points, the trajectory solution was essentially over-constrained. Consequently, for the final orbit determination, we selected only those stations with the optimal intersection geometry, no pixel saturation, and available raw data. Trajectory solutions were obtained using the Monte Carlo astrometric intersection procedure described by Vida et al. (2020), with specific adaptations required for the event's unique duration and geometry. The long duration of the fireball presented specific challenges for deceleration measurements. Standard solvers attempt to optimize solutions by correcting observation timings for clock drifts, which can obscure the true deceleration signal due to atmospheric drag. For this event, drag caused a positional lag of ~6 km over the trajectory. To prevent the solver from interpreting this physical lag as timing errors, we aggregated long-duration segments from high-quality cameras to establish a clean deceleration curve. This approach necessitated sacrificing data from the first 5 seconds of flight but ensured a valid solution for the start and end points of the complete luminous path.

The final pre-encounter and post-encounter orbits were computed using the REBOUND N-body code (Rein and Liu, 2012), accounting for planetary perturbations. Because of the exceptionally long duration of the event (31.6 s) and shallow earthgrazing geometry, the trajectory was measurably curved rather than linear, requiring the gravity-correction procedure described in **Appendix A**.

### *2.2. Photometry and Ablation Modeling*

Photometric data were extracted from unsaturated stations, where camera Twisk EN909 of the European Fireball Network captured the peak brightness without saturation and served as the reference standard. This camera provided photometric data only. Observations from other GMN cameras were cross-calibrated to this reference to normalize for differences in spectral sensitivity. The calibrated light curve was modeled using the erosion model of Borovička et al. (2007), which treats the meteoroid not as a monolithic object but as an aggregate of grains released under aerodynamic loading. To perform calculations, we used the implementation of Vida et al. (2023). Once released, each grain is assumed to ablate independently following classical single-body ablation theory (Bronshten, 1983; Ceplecha et al., 1998). Particular emphasis was placed on the dynamic pressure, $q = (1/2)\rho v^2$, to constrain the aerodynamic stress at which the body might fragment. This metric provided an independent estimate of the bulk strength, enabling comparisons between observed fragmentation thresholds and the conditions required for shock formation.

Input parameters for the light curve model included the luminous efficiency of Borovička and Spurný (2020), a bulk density spanning 300–1000 kg/m$^3$ to represent cometary material (Kikwaya et al., 2011), and an intrinsic ablation coefficient $\sigma$ = 0.005 kg/MJ. We also incorporated variable ablation coefficients and a log-normal grain mass distribution. Given the orbital classification (see **Section 3.1**), and high beginning altitude, we assumed a bulk density of 1000 kg/m$^3$ in our optical analysis (Buccongello et al., 2024) because lower densities produced a fireball that was too bright at such heights. This choice reflects the expected porous, volatile-bearing nature of the body, and does not by itself uniquely constrain composition (e.g., Borovička, 2005). We utilized a grain mass distribution range of $10^{-10}$ to $10^{-8}$ kg with a mass index of $s$ = 2.0, and a grain density of 2000 kg/m$^3$, consistent with parameters for similar fireballs, such as Taurids (Borovička and Spurný, 2020). The power of a zero-magnitude meteor of 1500 W was used, as appropriate for the European Fireball Network sensors (Borovička et al., 2022).

To distinguish between thermally driven ablation and mechanically driven fragmentation, we calculated the theoretical power received by the meteoroid assuming conservation of energy. The received power $P$ was derived using relation $P = \frac{1}{2}\rho_{atm} S v^3 \left(\frac{\Gamma}{2-\Gamma}\right)$, where $\rho_{atm}$ is atmospheric density, $S$ is cross-sectional area, $v$ is velocity, and $\Gamma$ is the drag coefficient. We compared the evolution of this theoretical power curve against the observed light curve and the calculated dynamic pressure to identify the primary driver of mass loss.

### *2.3 Infrasound Detection and Signal Processing*

Infrasound data were acquired from three permanent arrays of the Royal Netherlands Meteorological Institute (KNMI): EXL, DBN, and CIA (e.g., Smink et al., 2019). Their locations relative to the fireball trajectory and modeled acoustic source region are shown in **Section 3.4**. Each array consists of multiple microbarometers designed to detect low-frequency acoustic waves. Continuous recordings from the time of the event were processed to identify signals associated with the fireball. Power spectral densities (PSDs) for signal and pre-event noise segments were computed for each station to determine the optimal frequency cutoffs for bandpass filtering (Ens et al., 2012). The frequency range that maximized signal amplitude was selected for each array.

The analysis was performed using the InfraPy software package (Blom et al., 2016). Signal detection and characterization were then performed using a Bartlett beamformer in sliding 4-second windows with a 95% overlap. The Fisher-statistic (F-stat) was employed as a coherence metric to identify statistically significant detections above the noise, corresponding to coherent wavefronts crossing the arrays. For each confirmed detection, the back azimuth and trace velocity of the incoming wave were measured. Finally, the beamformed signal (the "best beam") corresponding to the peak F-stat value was extracted for waveform analysis, and the dominant signal period was determined using the zero-crossings method (Ens et al., 2012; ReVelle, 1997).

A complementary, multi-frequency analysis was performed using the Cardinal software package to refine the signal time-frequency characteristics (Ronac Giannone et al., 2025). Cardinal applies frequency-wavenumber analysis to time-frequency space that has been pixelated by pre-defined time windows and frequency bands. Statistically significant pixels were then grouped into detection families using an aggregation function. The measured slowness vector from the resulting detection family at each array was used to produce a final, optimized beamformed signal.

### *2.4 Acoustic Propagation and Source Localization*

To localize the acoustic sources along the earthgrazer's path, we performed propagation modeling using the InfraGA ray-tracing software, which calculates acoustic paths, or eigenrays, in a realistic 3D atmosphere (Blom, 2014; Blom and Waxler, 2017). Atmospheric specifications for the time and locations of the three KNMI infrasound stations were derived using the Ground-to-Space (G2S) model (Drob et al., 2003), hosted by the National Center for Physical Acoustics (NCPA) (Hetzer, 2024). G2S combines real-time observational data with climatological models to generate profiles of temperature, pressure, and wind from the surface to the lower thermosphere. The acoustic waves were assumed to emanate along the previously-calculated optical trajectory, creating a series of emission points (Silber and Brown, 2014). For each infrasound station, we performed an eigenray search to find the most probable source point for the received signal (Silber and Bowman, 2025). The search parameter space included rays launched at a wide range of inclination angles ($\theta$) relative to the horizontal (-89.9° $<\theta<$89.9°), in 0.1° increments. The search followed the along-trajectory source-search approach described by Silber and Brown (2014), with rays launched over the full inclination-angle fan at each candidate source point. We then systematically compared modeled parameters for all possible eigenrays with the measurements at each station. The best-fit source location was identified by finding the eigenrays that simultaneously minimized the misfit between the modeled and observed travel times and back azimuths (Silber and Bowman, 2025), as demonstrated in **Figure 3**. Given the event's long duration, this analysis accounted for the fireball's position and the signal's origin at different points along the trajectory, ensuring travel times were calculated from the correct point in both space and time. This procedure allowed for the robust correlation of each infrasound detection to a specific source altitude and location along the luminous path.

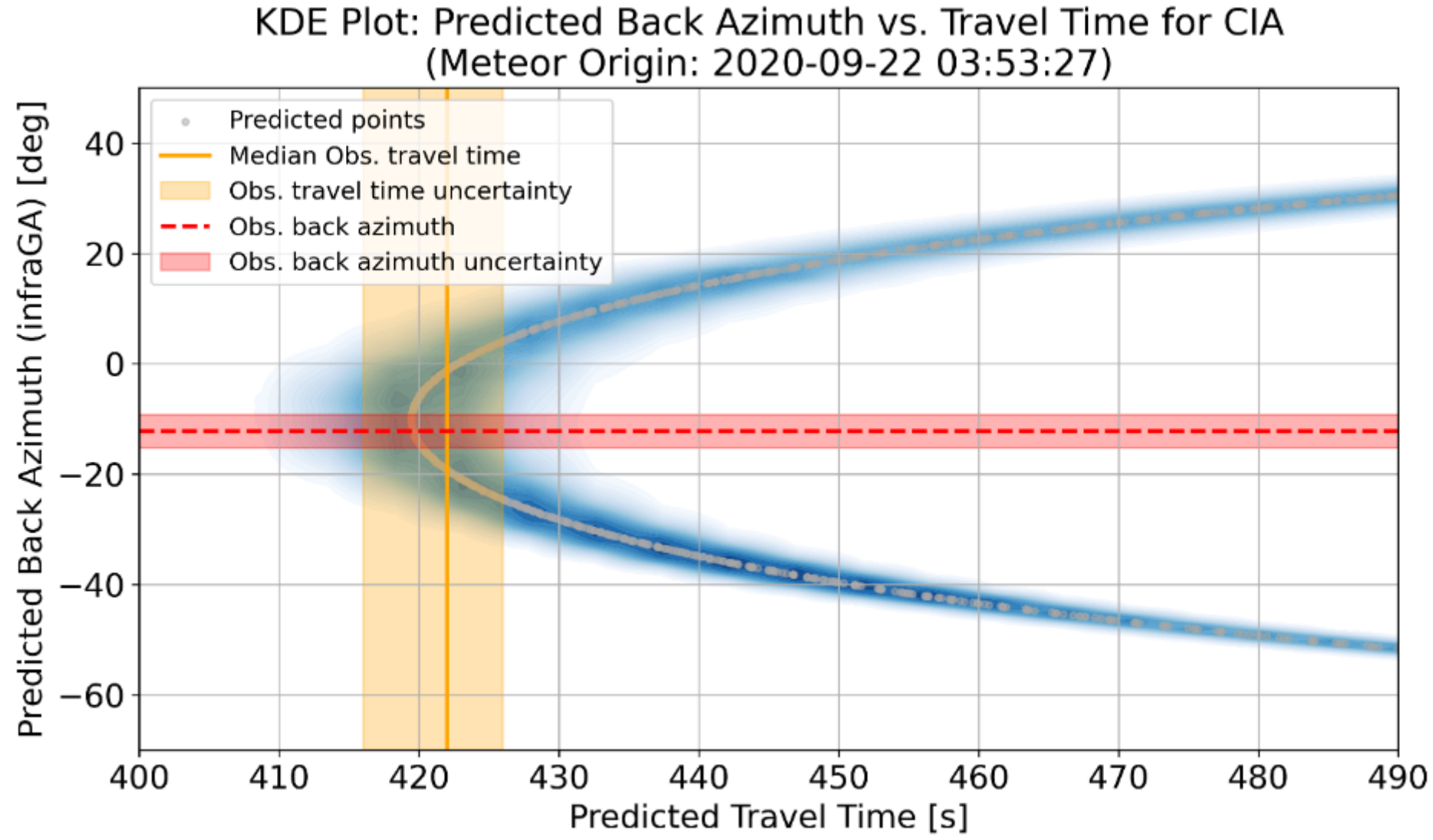

**Figure 3:** Kernel Density Estimate (KDE) plot illustrating the source localization procedure for station CIA. The plot shows the modeled back azimuth and travel time for all potential source points along the trajectory. The intersection of the modeled values (blue density) with the observed travel time (vertical orange band) and observed back azimuth (horizontal red band) uniquely constrains the best-fit solution.

### *2.5 Source Characterization and Energy Estimation*

To determine the source characteristics of the earthgrazer, we applied the refined ReVelle (1974) weak shock model, following the inversion procedure described in Silber et al. (2015). The input source for the weak shock modeling was the acoustic point determined in the previous step (**Section 2.4**). This approach uses the model iteratively to find the source blast radius ($R_0$) required to reproduce the observed infrasound signal period ($\tau$) at each ground station. The blast radius provides a direct physical measure of the energy deposited per unit path length by the meteoroid and thus characterizes the strength of the acoustic source. The model incorporated the G2S atmospheric profiles derived during the source localization step to account for propagation effects between the high-altitude source and the ground-based receiver.

Once the best-fit blast radius was determined for each source-station pair, it was used to estimate the effective acoustic source parameters, including an acoustic-equivalent diameter ($d_{ws}$) derived using the relationship:

$$d_{ws} = R_0/M, \qquad (1)$$

where $M$ is the meteoroid Mach number at the source altitude (ReVelle 1976). The Mach-diameter approximation is appropriate for ablating meteoroids and has been used for other events to successfully interpret the source function (Silber, 2026; Silber et al., 2015). The Mach number is defined as the ratio of the flow (or object) speed to the local speed of sound, $M = v/v_a$, and quantifies the degree to which compressibility and shock formation dominate the flow (Anderson, 2006; Zel'dovich and Raizer, 2002).

For completeness and comparison with prior work, we also estimated the total source energy using empirical period–yield relations (Ens et al., 2012; ReVelle, 1997; Silber et al., 2025b). These relations take the form $\log(E) = A \log(\tau) + B$, where A and B are empirically derived coefficients, $E$ is the equivalent explosive yield in kilotons of trinitrotoluene (TNT) equivalent (1 kt TNT = $4.184 \cdot 10^{12}$ J), and $\tau$ is the dominant signal period in seconds. The dominant signal periods from the three detecting arrays were averaged to produce a single value $\tau_{avg}$, and the global fit coefficients for averaged-period detections (A = 3.71, B = –2.07) from Silber et al. (2025b) were applied to derive an independent energy estimate. It is important to emphasize that these period–yield relations were calibrated primarily using stratospheric and mesospheric bolide events and are not optimized for thermospheric sources such as the present case. At thermospheric altitudes, enhanced dispersion, reduced atmospheric density, and complex propagation effects can modulate apparent signal periods (Chunchuzov et al., 2025), leading to systematic overestimation of source energy when standard period–yield relations are applied. Consequently, the energy, mass, and diameter values derived from this approach should be regarded as approximate upper bounds. In this study, period–yield estimates are included for methodological completeness, while the physically constrained weak-shock inversion provides the primary basis for interpreting the acoustic source characteristics of the earthgrazing meteoroid.

## 3. Results

### *3.1 Optical Trajectory and Orbit*

The 22 September 2020 earthgrazer produced a luminous trajectory observed for 31.6 seconds, covering a ground distance of 1060.7 km (**Figure 2**). The fireball was first detected at an altitude of 109.7 km approximately 100 km north of Berlin, Germany. It traveled westward at an azimuth of 265.5°, descending to a perigee (minimum height) of 91.3 km between Amsterdam and Groningen, Netherlands. Following perigee, the object ascended, ceasing luminous flight at an altitude of 113.6 km over England. The trajectory solution yielded fit residuals generally within ±100 m across the track length (**Figure 4a**). The meteoroid entered the atmosphere with an initial velocity of 34.17±0.01 km/s and exited at 33.59±0.01 km/s, corresponding to a total deceleration of approximately 580 m/s.

Orbital integration indicates that the meteoroid survived the atmospheric passage, though its heliocentric orbit was modified by the encounter. The semi-major axis decreased from 2.54 AU (pre-encounter) to 1.69 AU (post-encounter) (**Figure 4b**). Over the observed luminous flight, the geocentric radiant underwent an angular radiant shift of 6.8° between the pre- and post-encounter solutions (also see **Figures A2** and **A3**, **Appendix A**). The full trajectory and orbital parameters are detailed in **Table 1** and **Table 2**, respectively. The calculated Tisserand parameter with respect to Jupiter is $T_J$=2.685±0.002, placing the object in the Jupiter-family comet–like dynamical regime. Nevertheless, dynamical class does not uniquely determine composition; meteorite-delivering orbits, including those of carbonaceous chondrites such as CM, overlap this region of orbital-element space (Jenniskens and Devillepoix, 2025). This motivates considering multiple volatile-rich end members in the discussion (**Section 4.2**).

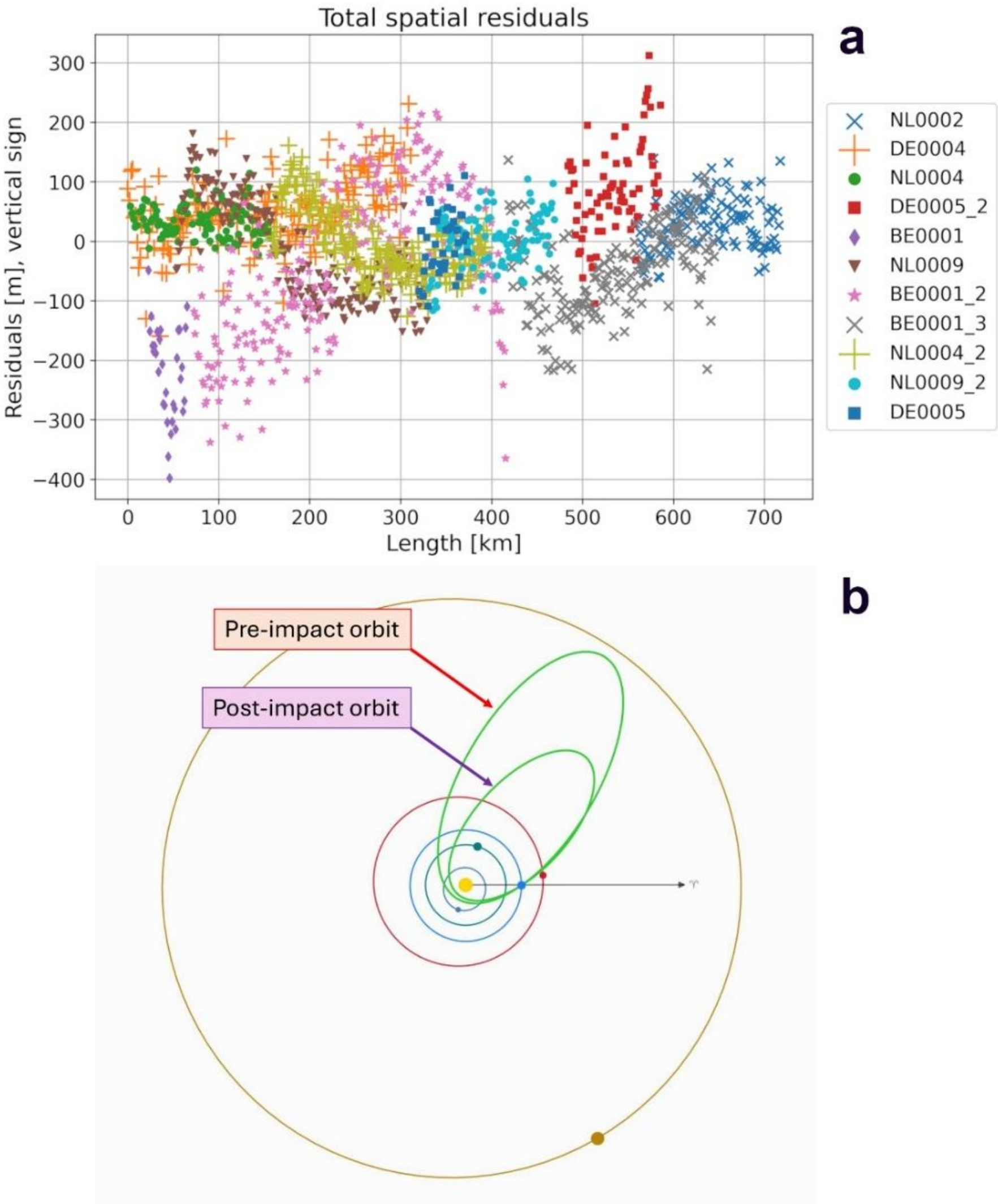

**Figure 4:** (a) Trajectory fit residuals from all cameras used in the reference solution. Where fireball records from individual cameras were divided into multiple video segments, the subsequent parts are labelled _2 and _3. (b) Pre- and post-encounter orbits. Although the fireball only slowed by 0.58 km/s in the atmosphere, a reverse slingshot effect robbed it of nearly 3 km/s in heliocentric velocity, explaining the reduced semi-major axis.

**Table 1:** Trajectory parameters of the fireball (apparent ground-fixed, epoch of date). Heights are given above the WGS84 reference ellipsoid. The total duration of the fireball was 31.51 s.

| | **Begin point** | |
|---|---|---|
| Time (UTC) | 2020-09-22 03:53:26.85 | ± 0.05 s |
| Latitude | 53.455180° | ± 65 m |
| Longitude | 13.315790° | ± 41 m |
| Height | 109.75 km | ± 38 m |
| | **Endpoint** | |
| Time (UTC) | 2020-09-22 03:53:58.40 | |
| Latitude | 51.729796° | ± 21 m |
| Longitude | -1.975294° | ± 155 m |
| Height | 113.59 km | ± 51 m |
| | **Lowest point** | |
| Latitude | 52.874563° | ± 237 m |
| Longitude | 5.847300° | ± 1280 m |
| Height | 91.3 km | ± 17 m |
| | **Radiant and speed** | |
| Azimuth (at begin point) | 85.554° | ± 0.008° |
| Elevation (at begin point) | 4.179° | ± 0.006° |
| Initial velocity | 34.17 km/s | ± 0.01 km/s |
| Final velocity | 33.59 km/s | ± 0.01 km/s |

**Table 2:** Geocentric radiant and orbit of the fireball (epoch J2000).

| | **Pre-encounter** | **Post-encounter** |
|---|---|---|
| **Geocentric radiant** | | |
| R.A. (°) | 165.83574 ± 0.00303 | 161.65 ± 0.02 |
| Declination (°) | +3.83590 ± 0.00901 | 9.21 ± 0.02 |
| Velocity (km/s) | 32.020 ± 0.006 | 31.4376 ± 0.0051 |
| **Orbit** | | |
| Perihelion, $q$ (AU) | 0.29320 ± 0.00035 | 0.22018 ± 0.00022 |
| Semimajor axis, $a$ (AU) | 2.54301 ± 0.00474 | 1.68568 ± 0.00247 |
| Eccentricity, $e$ | 0.88470 ± 0.00035 | 0.86938 ± 0.00032 |
| Inclination, $i$ (°) | 3.00161 ± 0.03924 | 2.12268 ± 0.04503 |
| Argument of perihelion, $\omega$ (°) | 239.48970 ± 0.02851 | 47.23821 ± 0.01101 |
| Ascending node, $\Omega$ | 359.31050 ± 0.00029 | 179.36128 ± 0.00066 |

### *3.2 Light Curve and Ablation Behavior*

The best-fit erosion model (***Figure 5***) indicates an initial meteoroid mass of approximately 45 g, with ~50% of this mass lost during the luminous flight. The modeling results suggest that fragmentation (erosion) may have been initiated at a dynamic pressure of ~0.35 kPa during the descent (inbound) and ceased at nearly the same pressure during the ascent phase. The erosion coefficient required to fit the observations was not constant. After the initial onset of fragmentation, the coefficient dropped by a factor of three before rising again when the dynamic pressure reached ~2.5 kPa. Furthermore, during the ascent (outbound) phase, the erosion coefficient required to explain the luminosity was approximately three times higher than that during the descent, despite similar air density and velocity conditions. This behavior, however, could be explained by the release of volatiles contained in the object (see **Section 4.1**).

A comparison of the observed light curve with the theoretical energy deposition reveals a distinct temporal offset (**Figure 6**). The light curve peak aligns with the peak in dynamic pressure rather than the peak in theoretical thermal power. The theoretical power peaks while the object is still optically faint, ~1 second before the maximum brightness, whereas the observed luminosity remains high even as the theoretical power drops. The modeling indicates that in the second half of the flight, the primary source of luminosity was the ablation of released fragments rather than the main body.

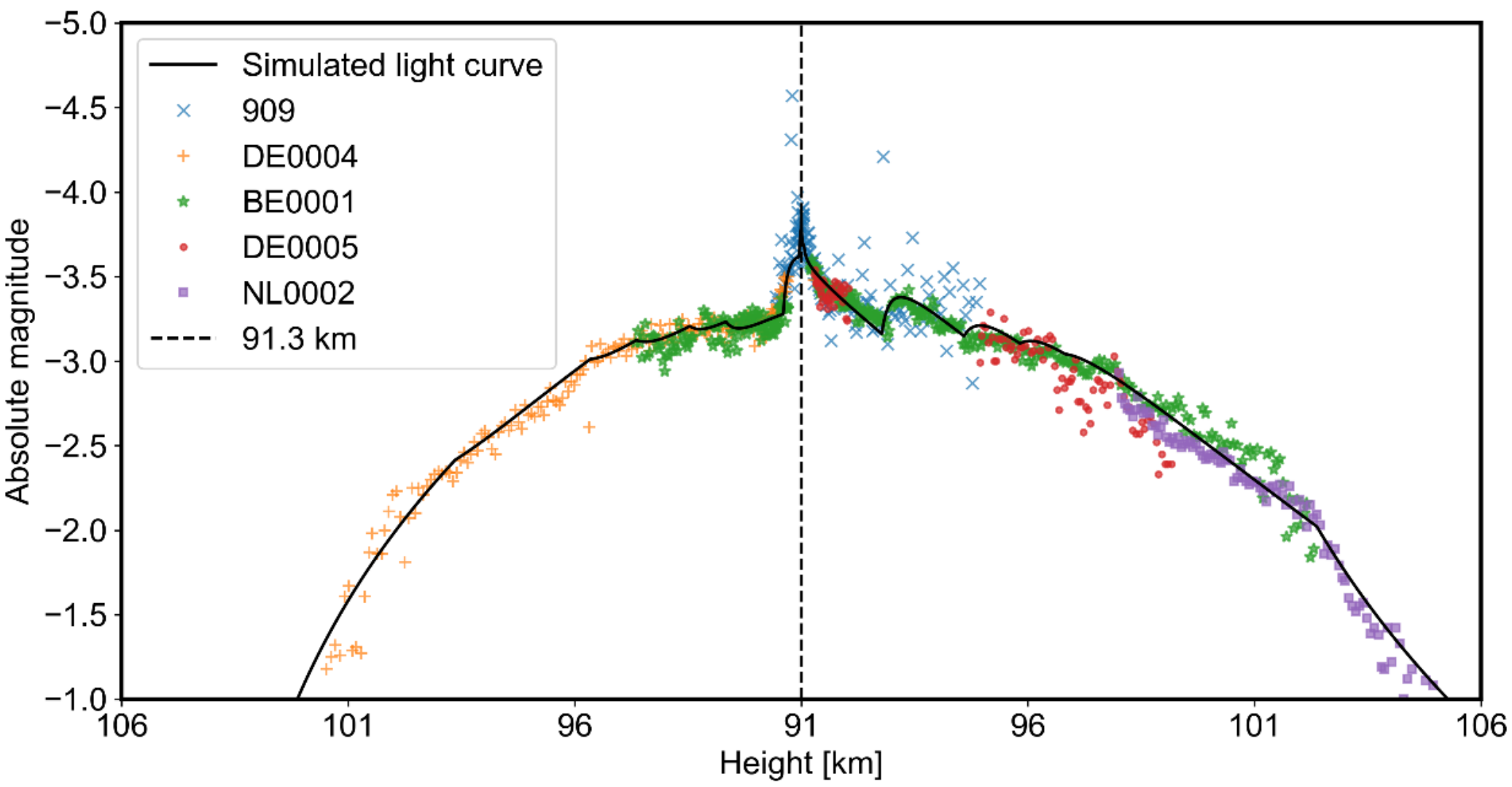

**Figure 5:** Light curve modelled with the Borovička et al. (2007) erosion model. The plot is symmetrical around the bottom height.

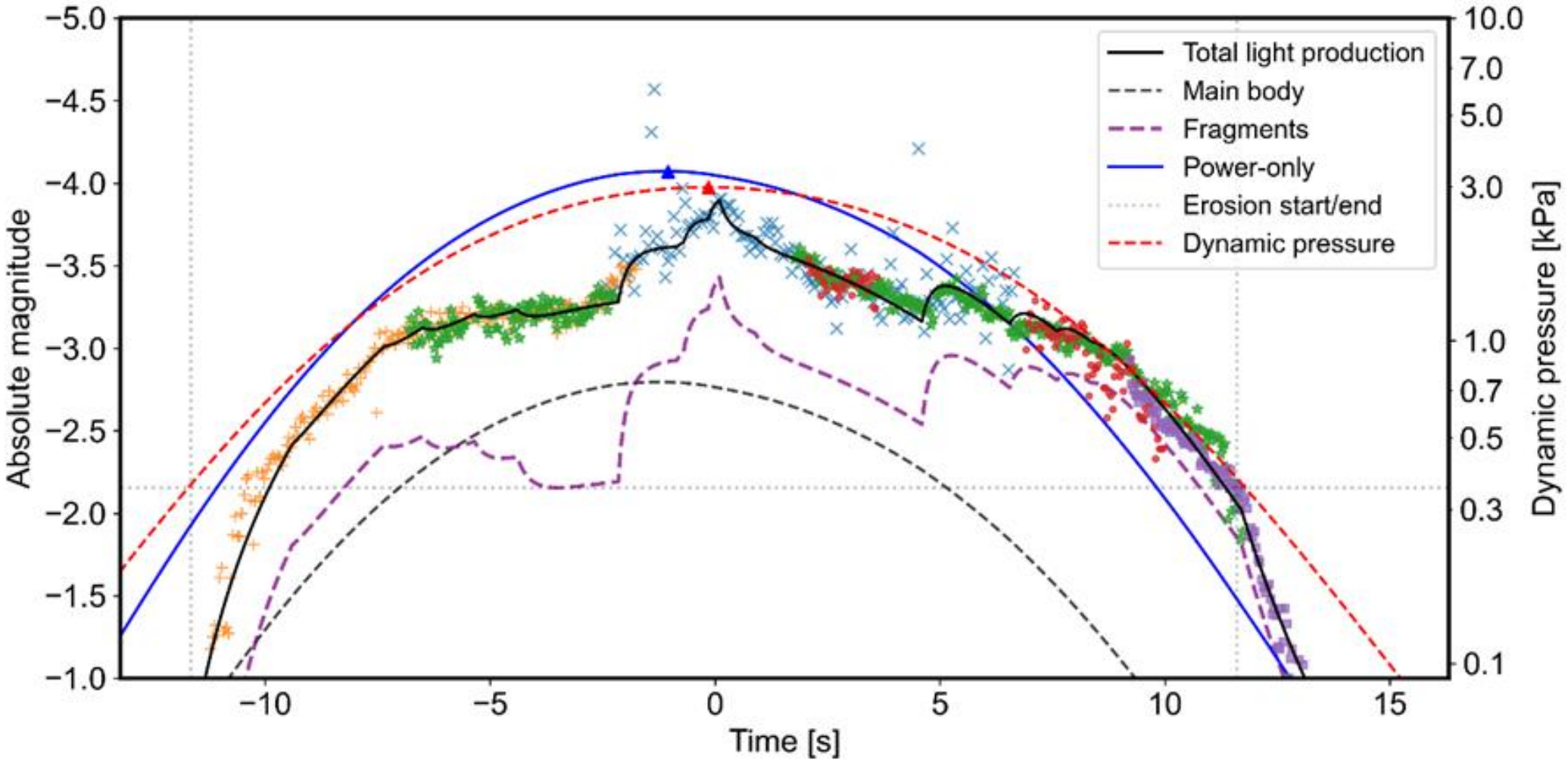

**Figure 6:** Comparison between the observed and the modelled light curve together with a theoretical light curve solely based on the received power. The dynamic pressure is plotted on the right-hand axis and scaled to align with the light curve showing the total light production. Triangles mark the times of peak power and dynamic pressure. Modelled brightnesses of the main body and the fragments are also shown. The time is aligned such that the moment the fireball reaches the bottom height is at $t = 0$.

### *3.3 Infrasound Signal Characteristics*

The earthgrazer produced unambiguous infrasound signals that were detected at all three KNMI arrays: EXL, DBN, and CIA. The measured signal parameters for each of the three detecting stations are summarized in **Table B1(A)** (see **Appendix B**). The arrival sequence of the signals, with the first detection at EXL (03:58:45 UTC) followed by DBN (04:00:20 UTC) and CIA (04:00:46 UTC), is consistent with the fireball's westward motion and the geometry of the network. All three detections exhibit short duration (2.23±0.15 s) signals with the characteristic N-wave shape (**Figure 7**, left), consistent with ballistic arrivals. The signal energy is concentrated at low frequencies, with peak frequencies between 0.7 Hz and 1.1 Hz (**Figure B1**). The dominant period, as measured by the zero-crossings method, averages 1.24±0.16 seconds across the three arrays (1.23±0.38 s at EXL, 1.10±0.07 s at DBN, and 1.38±0.29 s at CIA). A cross-correlation analysis shows high inter-element correlation for the signals detected at arrays EXL (**Figure 7**) and CIA. The correlation between elements at array DBN is lower. The measured signal parameters also varied significantly between the stations. The EXL array recorded the highest peak-to-peak amplitude (0.788 Pa) and the highest trace velocity (1702 m/s). The DBN array recorded the lowest amplitude (0.347 Pa) and a trace velocity of 537 m/s. The CIA array recorded an amplitude of 0.608 Pa and the lowest trace velocity at 445 m/s.

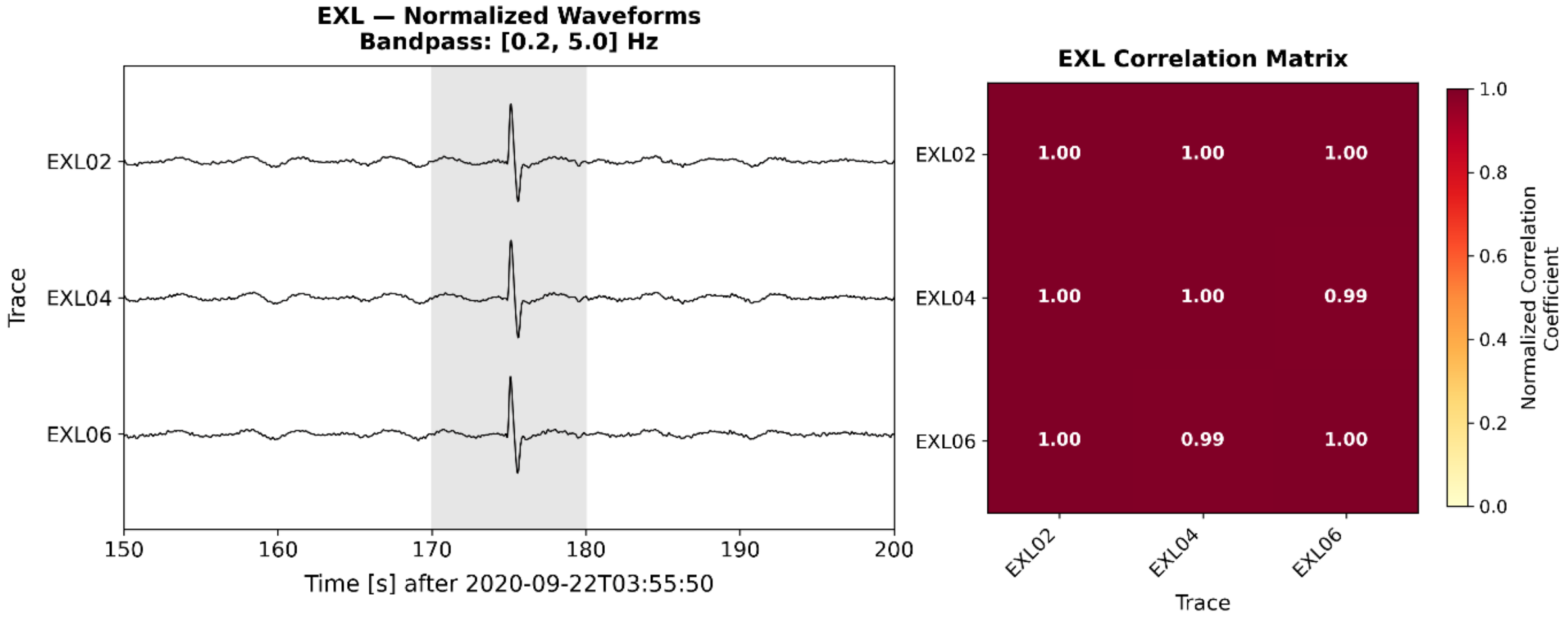

**Figure 7:** (left) Normalized waveforms recorded at each element of the EXL array, showing the N-wave signal. (right) Corresponding cross-correlation matrices for each array.

### *3.4 Acoustic Source Localization*

Acoustic propagation modeling localized the best-fit source for the three array detections to three distinct points along the trajectory corresponding to altitudes between 91.6 km and 91.8 km (**Table B1(B)**). These source points are situated near the fireball's point of lowest altitude over the Netherlands (**Figure 8**). The calculated lateral separations between these distinct shock origins along the trajectory are: 131.5 km (EXL to DBN) and 32.3 km (DBN to CIA), resulting in a total lateral distance of 163.8 km between the first and last sources (EXL to CIA). The smallest separation (32.3 km) is over an order of magnitude larger than the maximum ground-projected point location uncertainty (~2.1 km at DBN), confirming that these three sources represent truly distinct physical shock-production events rather than artifacts of modeling uncertainty. This spatial distribution is consistent with the ray-path geometry predicted by the propagation analysis. For instance, the signal detected at EXL corresponds to the modeled ray inclination of −84.8°, indicating an almost vertical arrival consistent with the array's proximity (8 km) to the ground track. The signal detected at CIA has a modeled ray inclination of −50.1°, consistent with its much larger lateral offset of 88.5 km. The modeled ray inclinations are further corroborated by the observed trace velocities at each array, supporting the reliability of the acoustic source localization. During the flight segment associated with infrasound detections, the object was travelling at a velocity of 33.5±0.3 km/s.

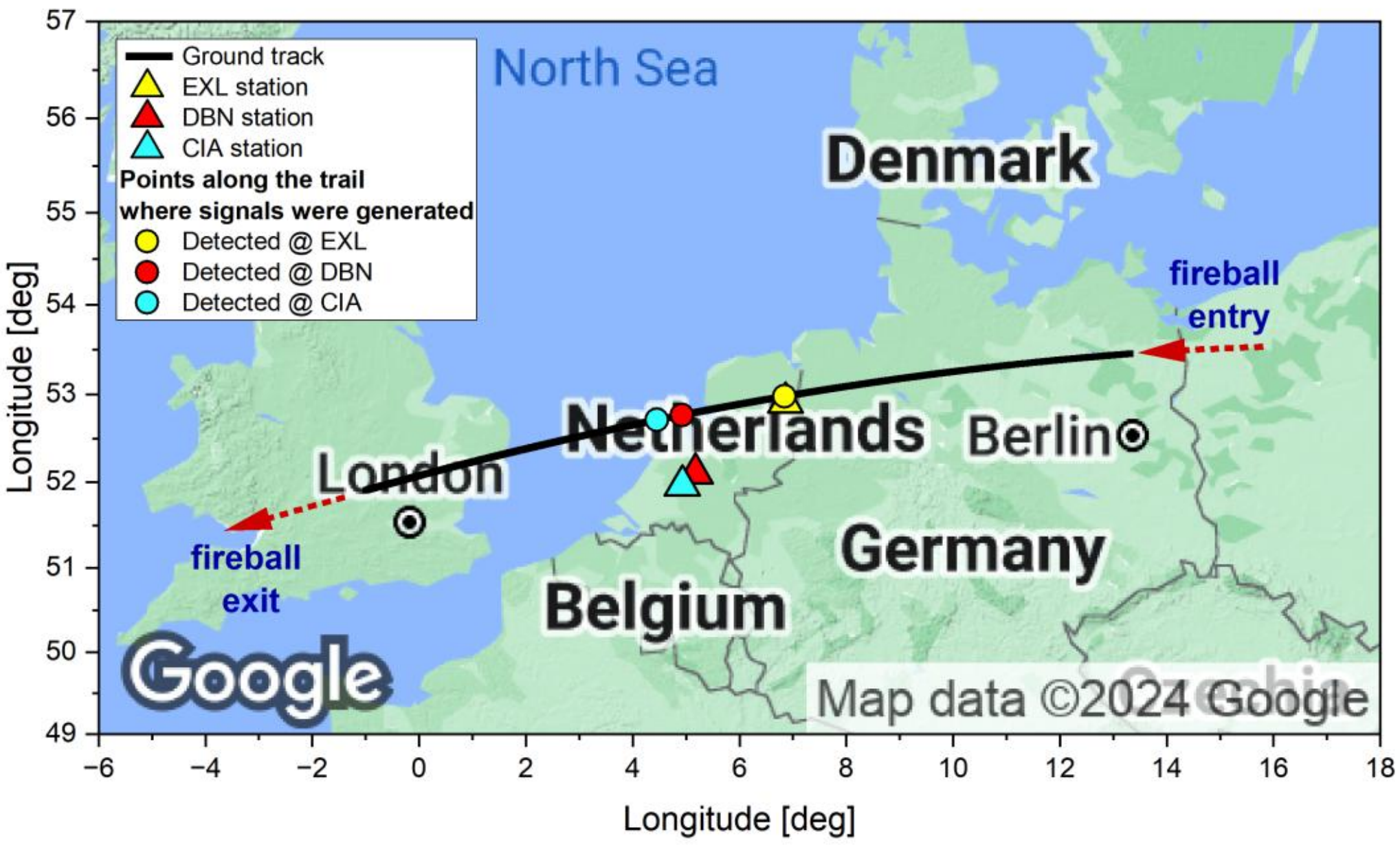

**Figure 8:** Map showing the reconstructed ground track of the fireball (black line) and the geolocated infrasound source points for each detecting station: EXL (yellow circle), DBN (red circle), and CIA (cyan circle). The sources are clustered near the fireball's point of lowest altitude over the Netherlands.

### *3.5 Weak Shock Modeling: Blast Radius and Source Size*

The weak shock modeling analysis yielded a remarkably consistent best-fit blast radius of 30.7±0.1 m across the three distinct source-receiver paths. The model-predicted dominant periods in the weak shock regime (ranging from 1.27 s to 1.37 s) corresponding to this blast radius are in close agreement with the observed infrasound periods (1.24±0.16 s). For this solution, the model calculates a consistent fundamental period at the source of 0.31 s for all three detections. The results of the weak shock modeling analysis are summarized in **Table B1(C)**. **Figure B2** provides a graphical representation of the model's output for the EXL station, showing the predicted evolution of the signal period as the wave propagates from the source altitude. The blast radius, which represents the energy deposited per unit path length, was then used to estimate the effective acoustic properties of the source. Applying Eq. (1) and using the Mach number at the source altitude ($M$ = ~122), the effective acoustic diameter of the source $d_{ws}$ was calculated to be 25.2 cm. For this event, the inferred $R_0$ corresponds to an acoustic-equivalent energy deposition on the order of $10^2$ J/m.

### *3.6 Infrasound Energy Estimates from Signal Period*

Applying the bolide-adapted period–yield relations to the mean observed signal period of 1.24 s yields an estimated total energy release of ~19 tons of TNT equivalent. As noted in **Section 2.5**, period–yield relations are not calibrated for thermospheric sources and may overestimate energy because high-altitude propagation, dispersion, and frequency-dependent attenuation can increase the observed dominant period. Recent analysis of regional high-altitude meteor infrasound further suggests that period-based energy estimates can be biased high for sources above ~80–90 km because the atmosphere acts as a frequency-dependent low-pass filter (Silber et al., 2026). Accordingly, the energy value reported here is treated solely as an upper-bound comparison and is not used to infer a physical meteoroid mass or to interpret this event. The meteoroid mass used in this study is instead constrained by the optical trajectory and light curve modeling. Physically grounded constraints on acoustic source dimensions and energy coupling are derived from the trajectory-constrained weak-shock modeling presented in **Section 3.5**.

# 4. Discussion

## *4.1 Empirical and Observational Constraints*

### *4.1.1 Optical Evidence for a Mechanically Weak Meteoroid*

The observed optical light curve does not scale monotonically with the theoretical thermal power received by the meteoroid, as would be expected for a single, composite body undergoing smooth ablation. Instead, peak optical emission coincides with the maximum dynamic pressure encountered along the trajectory, while the calculated received power reaches its maximum approximately one second earlier. This phase offset persists across multiple unsaturated observing stations and remains after correction for geometric range effects and timing aggregation, indicating that it is not an artifact of individual camera timing, saturation behavior, or observational geometry. We note that the received-power curve is an idealized proxy (e.g., it assumes a representative cross section and drag coefficient and does not explicitly capture rapid changes in shape or emitting surface area); however, these simplifications would not naturally produce the consistent ~1 s offset observed between the received-power peak and the luminosity peak, nor would they explain why the luminosity maximum tracks the dynamic-pressure maximum more closely than the received-power proxy.

Dynamic pressure provides a direct measure of mechanical loading on the meteoroid. The observed alignment between peak luminosity and peak dynamic pressure therefore indicates that mechanically driven erosion likely dominated the luminous behavior during this phase of trajectory, potentially aided by rapid release of volatile-bearing material. In this regime, optical emission is governed not solely by instantaneous aerodynamic heating of a coherent body, but by the rapid heating and ablation of newly exposed material following any potential structural disruption. Because volatile-bearing phases can accelerate this transition by weakening the matrix and contributing additional gas-phase mass loading, the light-curve morphology is also consistent with sequential or rapid devolatilization as interior material is exposed and thermally processed.

This interpretation is supported by the erosion modeling results. The inferred erosion coefficient exhibits pronounced hysteresis between the inbound and outbound legs of the trajectory, increasing by approximately a factor of three on the outbound leg at comparable dynamic pressures. Such behavior is difficult to reconcile with a homogeneous object. Instead, it is consistent with a heterogeneous or stratified structure in which an initially stronger exterior partially shields an interior that is mechanically weaker, porous, and potentially volatile-rich. Once outer layers are compromised through combined mechanical and thermal erosion, subsequent mass loss can proceed more efficiently even under similar aerodynamic forcing.

The optical evidence on its own therefore favors an erosion-controlled luminous regime rather than purely thermal ablation, which on its own struggles to explain the observed behavior in the rarefied upper-atmospheric flow regime. While uncertainties in luminous efficiency, grain-size distribution, and absolute mass loss introduce non-uniqueness in quantitative estimates, the relative timing of luminosity maxima, the inbound–outbound asymmetry in erosion behavior, and the persistence of these features across independent observations indicate at least some degree of progressive mechanical disruption of a weak body. These same observables are compatible with a scenario in which mechanical disruption and devolatilization act together: disruption exposes fresh, volatile-bearing material, and volatile release can further promote erosion and fragmentation.

Based on the optical trajectory reconstruction and erosion modeling (**Section 3**), the meteoroid is inferred to have had an initial characteristic diameter of approximately 4.4 cm, assuming a bulk density of 1000 kg/m$^3$. Importantly, the light-curve modeling could only be reconciled with a low bulk density of this order, implying a highly porous, low-cohesion body; such a density is consistent with cometary aggregates and other weak, porous materials, but it does not uniquely determine ice fraction. Owing to uncertainties in luminous efficiency, fragmentation behavior, and grain-size distributions, the light-curve/erosion model should be treated as a physically plausible solution rather than a definitive compositional diagnosis.

Additionally, orbital parameters alone do not uniquely determine meteoroid composition. Although the object's Tisserand parameter places it in the Jupiter-family comet–like dynamical regime (**Section 3.1**), this region of orbital-element space overlaps dynamical source regions associated with delivery of both carbonaceous (including CM-like) and ordinary (H) chondritic material to Earth (Jenniskens and Devillepoix, 2025). Compositionally, CM material is comparatively volatile-rich relative to H chondrites, but dynamics alone cannot distinguish between these possibilities. Independently of dynamical class, the long-duration Earth-grazing passage (luminous path >1000 km) indicates that the object retained a mechanically resilient component throughout the encounter. Under classical single-body ablation treatments (Bronshten, 1983), a centimeter-scale body composed almost entirely of water ice would be expected to erode efficiently and rapidly during its extended passage through the atmosphere, whereas a composite silicate matrix consisting of a porous aggregates containing substantial silicate grains can retain mass and continue to ablate over an extended path. Accordingly, the inferred low bulk density is best interpreted as an evidence-based constraint pointing to a highly porous, volatile-bearing aggregate rather than a unique taxonomic identification.

In summary, the optical constraints favor a low-density, mechanically weak, porous aggregate, whereas the long-duration earthgrazing survivability implies that a non-negligible phyllosilicate component with some residual cohesion persisted through the encounter. Because these observations can be satisfied by different mixtures of ice/volatile phases and phyllosilicate grains, we bound the volatile contribution using end-member compositions rather than asserting a unique taxonomic assignment.

We do not attempt to determine the exact composition of the object. The goal of this work is to test whether volatile release can plausibly enhance flow field mass loading enough to support strong shock formation under rarefied thermospheric conditions. Accordingly, we bracket volatile release using two plausible volatile-rich end members that capture (i) an upper-bound volatile supply and (ii) a conservative lower-bound volatile supply consistent with a porous, weak, grain-rich meteoroid.

The first end member is a cometary aggregate, which is expected to be ice-rich and composed of low-density porous material containing silicate grains and other dust components. *In situ* and remote observations of cometary dust show that such grains are typically silicate-rich and can have densities on the order of ~2000–3000 kg/m$^3$ (Krueger and Kissel, 1987; McDonnell et al., 1991), consistent with the grain-density assumptions adopted in the erosion model (e.g., Borovička et al., 2007; Ceplecha et al., 1998). Cometary bodies are inferred to be high-porosity, low-cohesion aggregates of grains with bulk densities of order 100–1000 kg/m$^3$ (e.g., Peale, 1989). In this context, grains may be weakly bound and can be released under moderate heating. Fragile cometary meteoroids may also contain small, more resistant carbonaceous-chondrite-like components (Borovička, 2005). This framing is also consistent with rarefied-flow dustball simulations that show that cometary meteoroids can behave as granular aggregates under aerodynamic loading (Hulfeld et al., 2021 and references therein). As an end member, an ice-rich aggregate provides an upper bound on volatile-driven mass loading and flow field density enhancement.

The second end member is a CM-like, porous and volatile-rich carbonaceous chondrite proxy used as a conservative lower bound on volatile-driven mass loading, not as an assertion of a specific CM meteorite analogue (see **Appendix C**). The event's orbital context is compatible with delivery of carbonaceous material to Earth (Jenniskens and Devillepoix, 2025). Moreover, cometary meteoroids are composed of grains whose ablation behavior and bulk chemistry are broadly chondritic, and fragile cometary meteoroids may contain more resistant carbonaceous-chondrite–like components (Borovička, 2005), motivating the use of a CM-like proxy as a physically grounded representation of the phyllosilicate volatile-bearing granular component. CM chondrites, characterized by mechanically weak matrices, high porosity, and abundant hydrated phases, provide a plausible physical analogue (e.g., Bischoff, 1998; Jenniskens and Devillepoix, 2025; Rubin et al., 2007; Trigo-Rodríguez et al., 2019). These properties naturally promote early and progressive mechanical erosion during atmospheric entry and form an essential physical basis for the volatile-driven processes examined in subsequent sections. Using porous CM-like material as a proxy therefore provides a conservative benchmark for estimating the volatile contribution required for shock-supporting hydrodynamic shielding in the thermosphere.

Under either end member, the key inference from the optical data remains the same: the combination of low bulk density, early onset of erosion at exceptionally low dynamic pressure, and pronounced inbound–outbound hysteresis favors a porous, mechanically weak body whose luminous evolution is controlled by progressive disruption and exposure of volatile-bearing material. This physical interpretation forms the basis for the volatile-enhanced shielding mechanism developed in the following sections and in **Appendix C**. Since a cometary end member provides a straightforward upper bound on volatile-driven mass loading, it is not necessary to explicitly examine it further. Here we test whether the more conservative CM-like proxy representing a small, phyllosilicate-frame body can still supply the near-field density enhancement required to sustain a shock-coupled source region.

#### *4.1.2 Infrasound Detections and Acoustic Source Localization*

The infrasound signals were produced by a centimeter-sized meteoroid traveling on a shallow, earthgrazing trajectory. By the time the meteoroid was detected by infrasound nearly half-way through its trajectory and covering almost 500 km in the thermosphere, it had lost some of its initial mass, and therefore, we estimate its diameter at ~4.0(±0.2) cm. Comparative analysis of the acoustic source geometry and the optical light curve indicates that shock generation occurred along the portion of the trajectory corresponding to radiative emission maxima and the region of highest atmospheric density encountered by the meteoroid along its ~1000 km path. The optical light-curve morphology does not exhibit significant discrete flares, demonstrating that the infrasonic signal originated from a sustained cylindrical shock wave generated by the hypersonic passage of the meteoroid and the associated development of a large and dense flow field, rather than from impulsive fragmentation-related events (Silber, 2024; Wilson et al., 2025).

All three KNMI arrays recorded short-duration N-wave signals, the classic waveform signature of a ballistic shock that has decayed into an infrasonic wave. This morphology indicates that the meteoroid generated strong near-field shock waves and acted as a cylindrical line source during its tangential passage through the lower thermosphere (ReVelle, 1976; Silber and Brown, 2014). Ballistic infrasound arrivals of this type have been documented previously for steeper-angle bright fireballs and artificial re-entries (e.g., Brown et al., 2007; Brown et al., 2011; ReVelle and Edwards, 2006; Scamfer et al., 2026; Silber and Brown, 2014; Wilson et al., 2025). The present event is distinct because the source was an earthgrazing meteoroid at thermospheric altitude, extending the known parameter space for ballistic shock generation and infrasonic detectability to extremely shallow, high-altitude trajectories.

The geometry of this event was particularly favorable for acoustic detection. The near-perigee segment, where dynamic pressure and energy deposition were maximized, passed almost directly above the KNMI acoustic network (***Figure 8***). As a result, the arrays recorded both near-vertical and oblique direct arrivals from the same broad segment of the trajectory, enabling spatially resolved localization of the acoustic source regions. The modeled source points from the three arrays converge within a narrow altitude band near the lowest-altitude portion of the trajectory, consistent with the nearly constant-altitude geometry of the earthgrazer and the concentration of acoustic radiation near perigee. This localization demonstrates that the infrasound originated from separate, spatially resolved source regions along the same earthgrazing trajectory, effectively tracing shock production over the near-perigee segment.

Variations in arrival time, amplitude, coherence, and trace velocity among the arrays are consistent with source–receiver geometry and atmospheric propagation effects (Pilger et al., 2020; Scamfer et al., 2026; Wilson et al., 2025). The high trace velocity measured at EXL (1702 m/s) indicates a near-vertical arrival, consistent with the station's location almost directly beneath the perigee segment (Silber and Bowman, 2025). The strong signal coherence and high signal-to-noise ratio at EXL further support this geometry, because near-vertical arrivals are less affected by path-integrated dispersion and wind advection. The lower trace velocities at DBN (537 m/s) and CIA (445 m/s) reflect increasingly oblique propagation paths, consistent with their larger horizontal offsets from the line-source segment. The weaker and less coherent signal at DBN, despite its comparable range to CIA, likely reflects path-dependent attenuation, wind structure, or scattering along the propagation path. Such differences are expected in the middle and upper atmosphere, where wind and temperature gradients strongly influence acoustic ducting efficiency (e.g., Chunchuzov et al., 2025; Chunchuzov et al., 2026).

Multiple infrasound detections from a single meteoroid event at regional distances are not unusual by themselves. Previous studies have documented multiple arrivals at individual stations and multi-station detections from regional fireballs or controlled re-entries, usually from lower-altitude sources in continuum-flow conditions (e.g., Brown et al., 2011; McFadden et al., 2021; Pilger et al., 2020; Scamfer et al., 2026; Silber and Brown, 2014; Wilson et al., 2025). Long-range studies have also reported signals apparently originating from different portions of extended fireball trajectories, though at large distances geometric uncertainties typically preclude the resolution of distinct acoustic source regions with high confidence (e.g., Pilger et al., 2015; Silber et al., 2011).

The present event differs because multiple infrasound arrays recorded direct arrivals from separate, resolvable source regions along a single earthgrazing trajectory at ~92 km altitude. This was possible because the near-perigee segment passed close to the KNMI arrays, allowing both near-vertical and oblique arrivals to be associated with distinct source points along the trajectory. The resulting localization effectively traces shock production along the thermospheric flight segment and provides direct observational evidence that coherent shock radiation occurred in the lower thermosphere, where classical continuum-flow theory would not predict shock formation around a centimeter-scale solid body alone.

### *4.1.3 Context from Prior Earthgrazers and Artificial Entries*

The present event occupies an unusual observational and physical regime among earthgrazers and shallow-entry fireballs. Having established the infrasound detections and source localization, the main distinction lies in the source geometry and altitude: the acoustic arrivals were resolved along a shallow earthgrazing trajectory in the lower thermosphere. The three acoustic source regions identified here occur at or very near perigee and demonstrate that infrasound can trace the motion of such objects through the atmosphere with spatial resolution along the flight path (Scamfer et al., 2026; Silber and Bowman, 2025). This provides an empirical link between optical trajectory reconstruction and distributed acoustic source generation at thermospheric altitudes.

The altitude and dynamic-pressure regime further distinguish this event from previously studied earthgrazers and other extremely shallow-entry fireballs. Modeling indicates that erosion was initiated high in the thermosphere at a dynamic pressure of only ~0.35 kPa, consistent with the low bulk strength of a weak, porous body. Near the perigee altitude of 91.7 km, the atmospheric dynamic pressure increased to approximately 1.37 kPa (calculated using G2S atmospheric density $\rho_a \sim 2.44 \cdot 10^{-6}$ kg/m$^3$ and $v$ = 33.5 km/s). For comparison, the 1972 Great Daylight Fireball reached a minimum altitude of ~57 km at dynamic pressures of approximately 20–30 kPa (Ceplecha, 1979), while the 1990 Czechoslovakia–Poland earthgrazer reached minimum altitudes of ~80 km, with estimated dynamic pressures of 3–5 kPa (Borovička and Ceplecha, 1992). Peekskill, although sometimes discussed in the context of shallow-entry fireballs, was not a typical earthgrazer because it did not survive atmospheric passage; its minimum observed altitude was ~32 km before terminal disruption (Ceplecha et al., 1995). The present event therefore represents coherent shock-related infrasound generation at substantially lower dynamic pressure and higher altitude than these comparison cases.

This event also provides useful context for artificial atmospheric-entry and high-altitude moving-source observations. Cotten and Donn (1971) reported infrasonic signatures from rocket vehicles at orbital altitudes near 188 km, a physically distinct case in which a continuously supplied exhaust plume acted as an enlarged, dense moving acoustic source in a rarefied atmosphere. Controlled re-entries, including NASA's Orion space craft (Timmons et al., 2018) and the Origins, Spectral Interpretation, Resource Identification, and Security-Regolith Explorer (OSIRIS-REx) Sample Return Capsule (SRC) (Lauretta et al., 2017) provide a different comparison case: coherent, effectively non-ablating re-entry bodies whose acoustic signatures are dominated by ballistic-shock generation as they descend into denser atmospheric layers where continuum or near-continuum flow conditions are progressively established (Silber et al., 2024). For additional context, a non-ablating body with dimensions comparable to the OSIRIS-REx SRC is expected to begin generating shock waves at altitudes near ~85 km, where continuum flow conditions are first established. Space-debris re-entries provide a more complex comparison because debris objects can fragment, tumble, and ablate during descent, producing distributed and time-variable acoustic sources (Hatty et al., 2026). Conversely, an ablating centimeter-scale object, such as the earthgrazer studied here, generates shock waves at significantly higher altitudes (e.g., Jenniskens et al., 2000; Zinn et al., 2004). This stresses the critical role of small ablating bodies to mimic the continuum-like behavior of substantially larger artificial objects (Silber et al., 2025a). This comparison frames the present event as a natural thermospheric case for understanding how small bodies can transiently acquire effective source dimensions and collisional flow conditions far exceeding those expected from the physical nucleus alone.

### *4.2 Physical Interpretation: Hydrodynamic Shielding and Shock Formation*

The empirical constraints established above require a physical mechanism capable of converting a centimeter-scale body in transitional thermospheric flow into an effective acoustic source much larger than the solid nucleus. For the present event, however, the infrasound-derived acoustic source dimensions and coupling strength imply a shock-bound flow-field envelope that is difficult to sustain through meteoroid surface ablation alone at ~92 km altitude.

We therefore first examine the rarefied-flow constraint imposed by the acoustic source scale and then discuss how hydrodynamic shielding and volatile mass loading can satisfy that constraint. We posit that rapid volatile release provides this missing contribution by injecting additional material into the shielding layer, amplifying hydrodynamic shielding and reducing the effective local Knudsen number. Accordingly, we treat a volatile-rich cometary aggregate as an upper bound (sufficient to address that flow field density deficit) and a phyllosilicate-dominated proxy as a conservative lower-bound stress test, in order to determine whether the required flow field mass loading and shock strength can be achieved by considering a release of volatiles from porous volatile-rich CM-like nucleus. The quantitative basis is developed in **Appendix C**, where we calculate the atmospheric impingement rate, sputtering-only yield, total non-volatile ablation yield, thermal evaporation contribution, grain-evaporation length scale, and remaining gas-phase line-density deficit. The main text summarizes the key quantities needed to connect those calculations to the observed infrasound source scale and shock-generation requirement.

### *4.2.1 Rarefied-Flow Constraint and Acoustic Source Scale*

Under classical hypersonic flow theory, the formation of a shock in a fluid-like medium requires that the mean free path of particles be much smaller than the characteristic dimensions of the object. The physical interaction between a hypersonic body and the atmosphere is governed by the Knudsen number, $Kn = \lambda/L$, defined as the ratio of the atmospheric mean free path $\lambda$ to the characteristic length scale $L$ of the object or flow field (Anderson, 2006; Probstein, 1961) . Principally, there are four flow regimes: continuum flow, $Kn < 0.01$; slip flow, $0.01 \leq Kn \leq 0.1$; transitional flow, $0.1 \leq Kn < 10$; and free-molecular flow, $Kn \geq 10$. For a centimeter-scale meteoroid, such as this earthgrazer, traveling at altitudes of 90–110 km, the ambient conditions are sufficiently rarefied that the flow regime is transitional or free molecular. Classical gas dynamics therefore predicts that a stable, detached shock envelope should not form around the solid body alone because the large mean free path does not support a coherent, shock-front-bound flow envelope (Anderson, 2006; Cercignani, 2000; Probstein, 1961; Shen, 2006).

For the present event, the meteoroid had an inferred diameter of $4.0 \pm 0.2$ cm during the infrasound-generating segment and traversed the atmosphere near an altitude of ~92 km. The G2S-derived mean free path near 91.7 km is $\lambda \sim 0.032$ m, giving $Kn \sim 0.8$ for the solid body. In the absence of substantial modification of the surrounding flow, the scarcity of intermolecular collisions limits the establishment of the sharp pressure, density, and temperature discontinuities required to satisfy the Rankine–Hugoniot conditions for shock formation (Hayes and Probstein, 1959; Probstein, 1961). This establishes the rarefied-flow constraint that must be reconciled with the observed ballistic infrasound.

Modern theoretical work provides a physically consistent mechanism for resolving this apparent contradiction. Specifically, ablation-amplified hydrodynamic shielding can form when frictional heating, direct molecular impacts, sputtering, and evaporation produce an extended vapor-particle envelope around the meteoroid (e.g., Boyd, 2000; Bronshten, 1983; Campbell-Brown and Koschny, 2004; Öpik, 1958; Popova et al., 2000). In this interpretation, frictional and evaporative mass loss from a meteoroid subjected to aerodynamic stress actively contributes to the formation of an extended vapor cap that shields the body from direct impact of the ambient atmospheric flow while simultaneously enlarging the surrounding flow field. This expanded, high-temperature flow field exhibits a substantially higher particle number density than the ambient atmosphere and encompasses the viscous layer surrounding the body (Silber et al., 2018; Silber et al., 2017). In turn, elevated temperatures and enhanced collisionality within this region promote additional evaporation-dominated mass loss from the meteoroid surface (Bronshten, 1983; Silber et al., 2018). For the present event, however, the ambient density and limited non-volatile yield raise the question of whether ordinary ablation alone can generate a sufficiently dense vapor cap and subsequent shock-bound flow. This density-budget problem is evaluated in **Appendix C**.

The infrasound observations impose an additional empirical constraint on the effective source scale. Consistent with the weak-shock modeling results in **Section 3.5**, the observed infrasound periods imply an acoustic-equivalent source diameter of $\mathrm{d_{ws}} \sim 0.25$ m (Eq. (1)). This is not the physical diameter of the meteoroid; it is the characteristic diameter of the energy-coupled near-field flow required to reproduce the observed infrasound periods. For a solid nucleus of mass ∼45 g, such dimensions exceed expectations based on classical size-based scaling arguments at comparable altitudes. The implied flow field must therefore have been substantially larger than the physical body itself, despite the rarefied thermospheric environment.

An independent scale estimate is obtained from the classical disturbed-flow scaling of Lin (1954), in which the radius of the disturbed region behind a rigid hypersonic body depends on drag, ambient density, downstream distance, and velocity. Using the optical meteoroid cross-section, $C_D \approx 0.8$, $x = 1$ m, $v = 33.5$ km s$^{-1}$, and the G2S density at 91.7 km, this scaling gives a characteristic disturbed-flow radius of $R \approx 0.15$ m. This value is remarkably close to the acoustic-equivalent radius $d_{ws}/2 \approx 0.125$ m inferred independently from weak-shock modeling. Although approximate, the agreement supports interpreting $d_{ws}$ as a physically meaningful near-field flow scale rather than as the solid-body diameter.

The blast radius $R_0$ provides a diagnostic measure of the energy deposited per unit path length into the surrounding atmosphere (e.g., ReVelle, 1976; Tsikulin, 1970) . For this event, the inferred $R_0$ corresponds to an acoustic-equivalent energy deposition on the order of $10^2$ J m$^{-1}$, while the kinetic energy loss inferred from optical observations is smaller. This relationship is consistent with established energy-partitioning behavior (Romig, 1965) and indicates efficient transfer of mechanical energy into acoustic excitation through a continuous, high-temperature, quasi-neutral vapor–plasma mixture. Importantly, $R_0$ and $d_{ws}$ reflect the dimensions and density of the energy-coupled flow field rather than the physical size or mass of the solid nucleus, indicating that shock generation is controlled by the properties of the surrounding vapor-particle envelope.

These constraints indicate that the observed infrasound cannot be explained by the solid body alone. A dense, enlarged near-field flow must have formed, but **Appendix C** shows that non-volatile ablation and thermal evaporation alone cannot supply the required particle line density for a centimeter-scale earthgrazer at ∼92 km. An additional gas-phase mass-loading process is therefore required to increase local number density, reduce the effective mean free path, expand the effective flow-field dimensions, and permit transition to shock-bound flow.

### *4.2.2 Initial Collisional Heating and Early Vaporization*

When a centimeter-scale meteoroid enters the rarefied upper atmosphere at hypersonic velocity, it initially encounters a flow regime in which the mean free path of ambient gas molecules is comparable to, or greater than, the characteristic dimensions of the body (e.g., Campbell-Brown and Koschny, 2004). At altitudes of approximately 90–95 km, the ambient atmosphere is composed predominantly of molecular nitrogen ($N_2$) and molecular oxygen ($O_2$), with minor contributions from atomic oxygen and other trace species. Under these conditions, individual atmospheric molecules can directly impinge upon the meteoroid surface rather than interacting through a fully collective fluid response.

The momentum and kinetic energy of $N_2$ and $O_2$ molecules are transferred locally to the meteoroid surface during direct collisions, producing intense localized heating, sputtering, and potentially the cascading release of surface atoms and molecules governed by material properties. This interaction is described by classical meteoroid ablation theory, in which the translational energy of incident atmospheric molecules is converted primarily into thermal energy at the surface (Bronshten, 1983; Silber et al., 2018). The effectiveness of this process depends on both the kinetic energy of the impinging molecules and the binding energies of the meteoroid's constituent materials. Under the rarefied atmospheric conditions at 91.7 km altitude, these processes would need to produce a very high collisionally and thermally induced evaporative or sputtering yield to contribute to the formation of sufficiently dense hydrodynamic shielding (**Figure 9**), precursor to a strong shock in the near field (**Figure 10**).

The partitioning of kinetic energy during these collisions is dominated by heating of the meteoroid surface, leading to the initiation of evaporation or sputtering depending on material composition, surface binding energy, and evolving surface temperature (Romig, 1965) (also see **Appendix C**). Laboratory experiments, numerical simulations, and observational studies have shown that a single atmospheric molecule can eject multiple atoms or molecules from the meteoroid surface under favorable conditions, producing a cloud of vaporized meteoric material in the immediate vicinity of the body (e.g., Boyd, 2000; Jenniskens and Stenbaek-Nielsen, 2004; Popova et al., 2000; Zinn et al., 2004). This cascading vaporization, together with secondary and tertiary collisions, produces a highly superheated nonequilibrium gas cloud immediately ahead of and surrounding the meteoroid (**Figure *9***). This cloud consists of vaporized meteoric material mixed with collisionally excited and partially dissociated atmospheric species, forming the initial luminous flow field that envelops the body. Early numerical studies have described the formation and evolution of this vapor cloud and its interaction with the impinging ambient atmospheric flow, demonstrating how it modifies local density, temperature, and collisional coupling in the near field (Popova et al., 2000).

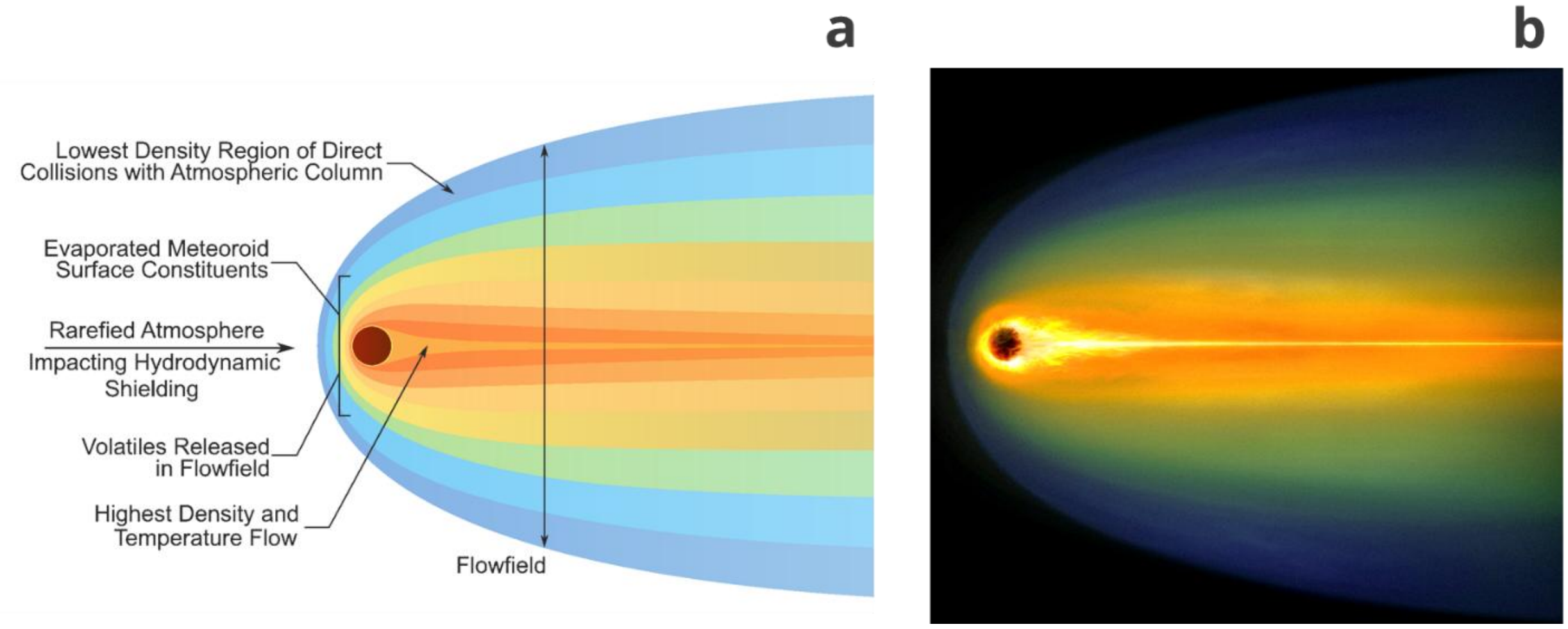

**Figure 9:** Schematic and artistic visualization of flow-field evolution and volatile-enhanced hydrodynamic shielding. (a) Schematic representation of the density and temperature gradients surrounding an entering body in the rarefied thermosphere. (b) Artistic visualization of (a). Note that these figures are not to scale and serve as simplistic conceptual representations of complex, non-equilibrium hypersonic flow phenomena. The vertical black line schematically marks the effective flow-field diameter; its placement and length are illustrative rather than exact. Also see **Appendix C**.

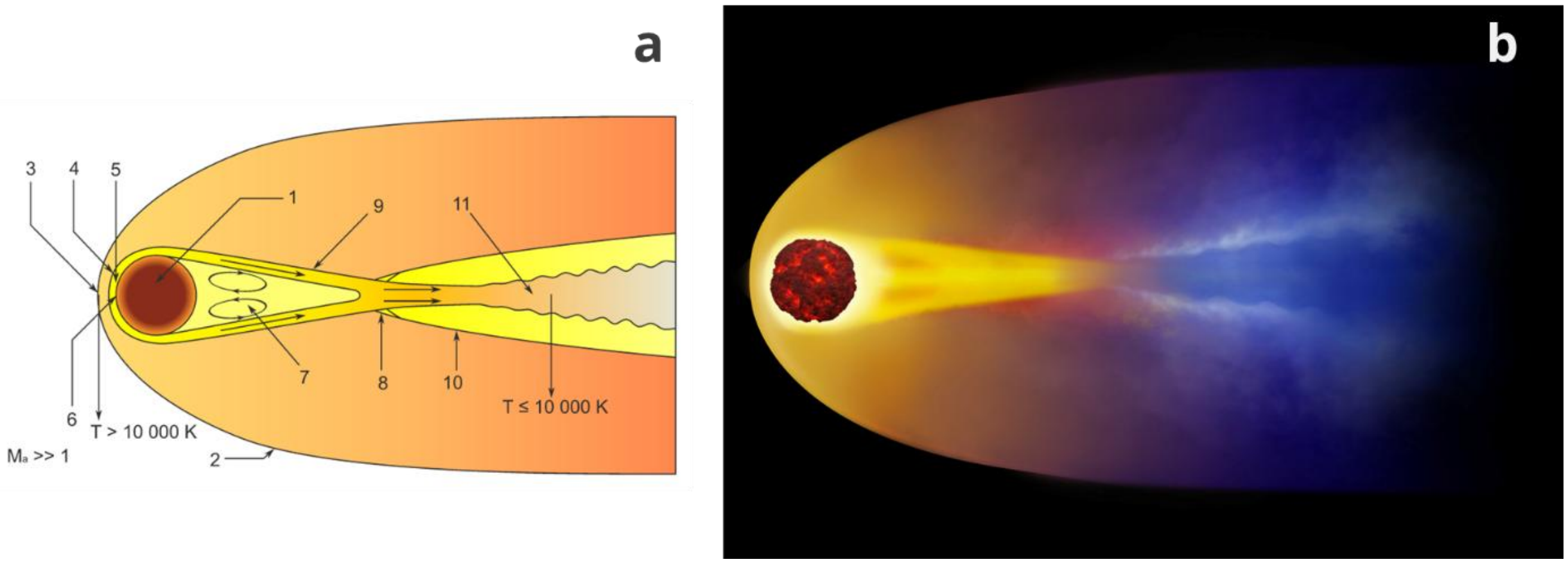

**Figure 10:** Schematic and artistic visualization of thermospheric shock wave formation. (a) Structural components of the hypersonic flow field, as originally illustrated by Silber et al. (2018): (1) meteoroid nucleus; (2) detached bow shock front; (3) shock-front interface; (4) compressed stagnation region; (5) viscous layer; (6) high-temperature near-field source region; (7) recirculating wake; (8) wake closure; (9) boundary layer transition; (10) trailing shock/expansion; and (11) attenuated wake. (b) Visualization of the resulting shock envelope. Note that these figures are not to scale and serve as simplistic conceptual representations of complex, non-equilibrium hypersonic flow phenomena. Also see **Appendix C**.

### *4.2.3 Development of the Vapor Cloud and Hydrodynamic Shielding*

As the density and collisional frequency of vaporized meteoric material surrounding the body increase, the initially sparse vapor cloud evolves into an effective hydrodynamic shielding layer. Previous theoretical and numerical studies have shown that this process is driven by the rapid accumulation of evaporated meteoric species and collisionally heated atmospheric constituents in the near-field region surrounding the meteoroid (Boyd, 2000; Popova et al., 2001; Popova et al., 1998; Silber et al., 2018). The resulting vapor–particle envelope contains entrained dissociated atmospheric atoms that have undergone collisional excitation, dissociation, and partial ionization, together with meteoric atoms and clusters released from the surface. Collectively, this mixture forms a dense, high-temperature, quasi-neutral flow field that co-moves with the meteoroid.

Within this expanding envelope, incoming atmospheric molecules increasingly interact with the vapor cloud rather than directly with the solid surface. These secondary and tertiary collisions transfer momentum and energy within the flow field, redistributing the incoming atmospheric flux over a larger interaction volume. This is further enhanced by continued heating of the meteoroid surface and subsequent high-temperature evaporation of material. As a result, the number density and collisional frequency within the shielding region increase relative to the ambient atmosphere, promoting the formation of a viscous layer enveloping the meteoroid and substantially altering the local flow structure (Bronshten, 1983; Popova et al., 1998; Silber et al., 2018).

The increased vapor pressure within the hydrodynamic shielding envelope can approach or exceed the dynamic pressure of the external flow near the leading surface of the object. When this occurs, the local flow regime within the shielding region transitions away from ambient transitional or free-molecular behavior toward slip-flow or continuum-like conditions, even though the surrounding atmosphere remains rarefied. In effect, the local Knudsen number within the shielding layer is reduced by orders of magnitude compared with its ambient value (Bronshten, 1983; Popova, 2004). This transition is driven by sustained production of vaporized material at the meteoroid surface and continuous entrainment of atmospheric species into the flow field.

The extent and persistence of hydrodynamic shielding depend on the balance between vapor production and advective loss. For objects that penetrate deeper into the atmosphere, increasing ambient density further enhances shielding growth by supplying additional atmospheric material to the flow field. Even for shallow, tangential trajectories such as the earthgrazer examined here, rapid generation of vaporized material can in principle maintain a dense shielding envelope over extended path lengths. Silber et al. (2018) emphasized that this progressive increase in shielding density and volume represents a critical transitional stage in the evolution of meteoroid flow fields, governing whether the system can progress toward shock formation under rarefied conditions.

However, the lower frequency of intermolecular collisions in the surrounding upper transitional regime, together with material properties that limit non-volatile sputtering and evaporation yields, imposes a fundamental limitation. Specifically, that set of limitations is further constrained by the thermally or collisonally driven sputtering yield (Bronshten, 1983; Popova et al., 2007; Silber et al., 2018) that is strongly governed by meteoroid surface composition. While hydrodynamic shielding (**Figure 9**) substantially modifies the local flow field, it does not automatically guarantee the formation of a detached shock envelope capable of efficient coupling to the ambient atmosphere. In particular, when the meteoroid remains confined to a narrow altitude range and does not encounter substantially denser atmospheric layers, the ablationally driven shielding density enhancement alone may be insufficient to transition into and sustain a shock that remains coherent and detectable at the ground. This limitation is central to the present case. The quantitative density-budget analysis in **Appendix C** evaluates this limitation directly and shows that an additional gas-phase mass-loading mechanism is required to amplify near-field density and collisional coupling under the observed thermospheric conditions.

### *4.2.4 Transition to Shock Formation*

As the hydrodynamic shielding envelope grows in size and density, progressively larger pressure, temperature, density, and velocity gradients develop between the dense near-field flow and the surrounding ambient atmosphere. When these gradients become sufficiently steep, the shielding layer can no longer adjust smoothly to the external flow, and a steep $P$, $T$, and $\rho$ discontinuity develops relative to the local atmosphere.

A detached shock envelope then forms at the interface between the dense, high-temperature flow field and the ambient gas. This is the initial detached shock enveloping the flow field, typically termed a bow shock, and it is distinct from the recompressional cylindrical shock, which is comparatively stronger in the meteor-acoustic context (Silber et al., 2018). This transition requires that the local conditions within the shielding envelope become sufficiently collisional for a shock discontinuity consistent with the Rankine–Hugoniot relations to be supported (Anderson, 2006; Ben-Dor et al., 2000; Hayes and Probstein, 1959; Lin, 1954; Probstein, 1961; Zel'dovich and Raizer, 2002).

The key requirement for shock formation in this regime is not that the entire surrounding atmosphere behaves as a continuum, but that the shielding envelope itself attains sufficiently high density and collisional frequency to support shock-bound flow. Once established, the shock envelope encloses a flow field whose effective dimensions may exceed those of the solid nucleus by more than one order of magnitude. The meteoroid–flow system then behaves aerodynamically as a much larger hypersonic body, with shock geometry and strength governed primarily by the properties of the energy-coupled vapor–particle envelope rather than by the physical size of the meteoroid alone (Cotten and Donn, 1971; ReVelle, 2009; Silber et al., 2018; Silber et al., 2015; Silber et al., 2017).

Following its formation, the shock envelope expands radially away from the dense near-field region. In the immediate vicinity of the source, this expansion can be approximated as cylindrical, driven by the rapid release of thermal and mechanical energy from the compressed flow (**Figure 10**) into the surrounding atmosphere (Kornegay, 1965; Masoud et al., 1969). In first approximation, this radial expansion of the high-temperature flow field behind the shock envelope can be treated as hydrodynamic expansion into a very low-pressure, rarefied ambient environment (e.g., Cercignani, 2000; Kornegay, 1965; Kustova et al., 2011; Masoud et al., 1969). As the shock front propagates outward and interacts with the ambient atmosphere, it weakens and transitions into a ballistic acoustic wave at an energy-dependent distance away from the meteoroid trajectory axis, producing the infrasonic signatures detected at the ground (ReVelle, 1976). The weak-shock modeling presented in **Section 3.5** demonstrates that the inferred blast radii are consistent with such cylindrical expansion originating from an extended, dense source region.

Although numerical simulations and analytical models have shown that ablation-driven hydrodynamic shielding, controlled by material properties and atmospheric density, can support shock formation at smaller size scales and higher altitudes than would otherwise be possible in rarefied flow (Boyd, 2000; Popova et al., 2001; Popova et al., 1998; Silber et al., 2018), the present event highlights an important limitation. For an earthgrazing trajectory confined to a narrow altitude range near the upper transitional regime, surface ablation alone may not generate a sufficiently dense shielding envelope to initiate and sustain a shock with the strength and spatial extent implied by the infrasound observations. Constraints imposed by meteoroid velocity, material binding energy, surface composition, and the limited availability of atmospheric collisions restrict the maximum density enhancement achievable through non-volatile ablation alone, particularly for chondritic or phyllosilicate-rich materials. This limitation is quantified in **Appendix C**.

This leads to the central density-budget question addressed here: what additional process can amplify the near-field particle density and energy coupling enough to sustain the shock-bound flow inferred from the infrasound observations?

#### *4.2.5 Non-Volatile Density Deficit*

At the time and location of the infrasound-generating segment, the G2S atmospheric specifications at 91.7 km yield $T = 194$ K, $P = 1.36 \times 10^{-1}$ Pa, and background density $\rho_a = 2.44 \times 10^{-6}$ kg m$^{-3}$. These conditions correspond to a total number density of $5.09 \times 10^{19}$ m$^{-3}$ and a mean free path $\lambda_a \approx 3.2 \times 10^{-2}$ m, using an effective air collision cross-section of $\sim$ 3.65–3.7 Å. At perigee, the meteoroid diameter is estimated to be $\sim$ 4 cm, giving an ambient Knudsen number $Kn = \lambda_a / L \approx 0.8$. The conditions required to sustain a strong shock in the near field imply a substantially higher particle density than the ambient thermosphere provides. Using the Rankine–Hugoniot relations evaluated under these conditions, the required number density for initial shock formation is $n_{\text{shock,req}} \sim$ $3.05 \times 10^{20}$ m$^{-3}$. Expressed as a line density over the acoustically inferred source cross-section, this corresponds to $N_{\text{shock,req}} \sim 1.5 \times 10^{19}$ m$^{-1}$ (**Appendix C**).

As discussed in **Section 3.5** and **Section 4.2.1**, weak-shock inversion yields a blast radius $R_0 \sim 30$ m and an acoustic-equivalent source diameter $d_{ws} \sim 0.25$ m. For a characteristic length $L = 0.25$ m, the ambient Knudsen number becomes $Kn = \lambda_a/L \approx 0.13$, still within the transitional regime. This shows that increasing the effective characteristic length alone is not sufficient. The local mean free path must also be reduced by increasing the number density within the hydrodynamically shielded flow-field envelope. This requirement can, in principle, be met in two non-exclusive ways: (i) increasing the effective characteristic length (leading to expanding the shielding envelope), or (ii) reducing the local mean free path within the flow-field envelope by increasing local number density. Both of these points depend on reducing local $Kn$. The infrasound observations already indicate the first effect through the acoustic-equivalent diameter. The key point here is that collision frequency in the initial shielding region must be sufficiently high to reduce the effective mean free path and preserve collisional coupling leading to the formation of $P$, $T$ and $\rho$ discontinuities, relative to ambient conditions.

To place these constraints in a quantitative flow-regime context, **Figure 11** presents the Knudsen number as a function of altitude and characteristic length scale for both ambient conditions (panel a) and an ablation-enhanced near field (panel b) (also see Campbell-Brown and Koschny, 2004). **Figure 11c** shows that a centimeter-scale body at $\sim 92$ km resides near $Kn \sim 1$, within the transitional regime. An enlarged effective source scale comparable to that inferred from the infrasound observations shifts the local flow toward slip-flow conditions, but this shift still requires corresponding near-field density enhancement to approach shock-supporting collisional behavior. Thus, the figure illustrates why both source enlargement and additional mass loading are required.

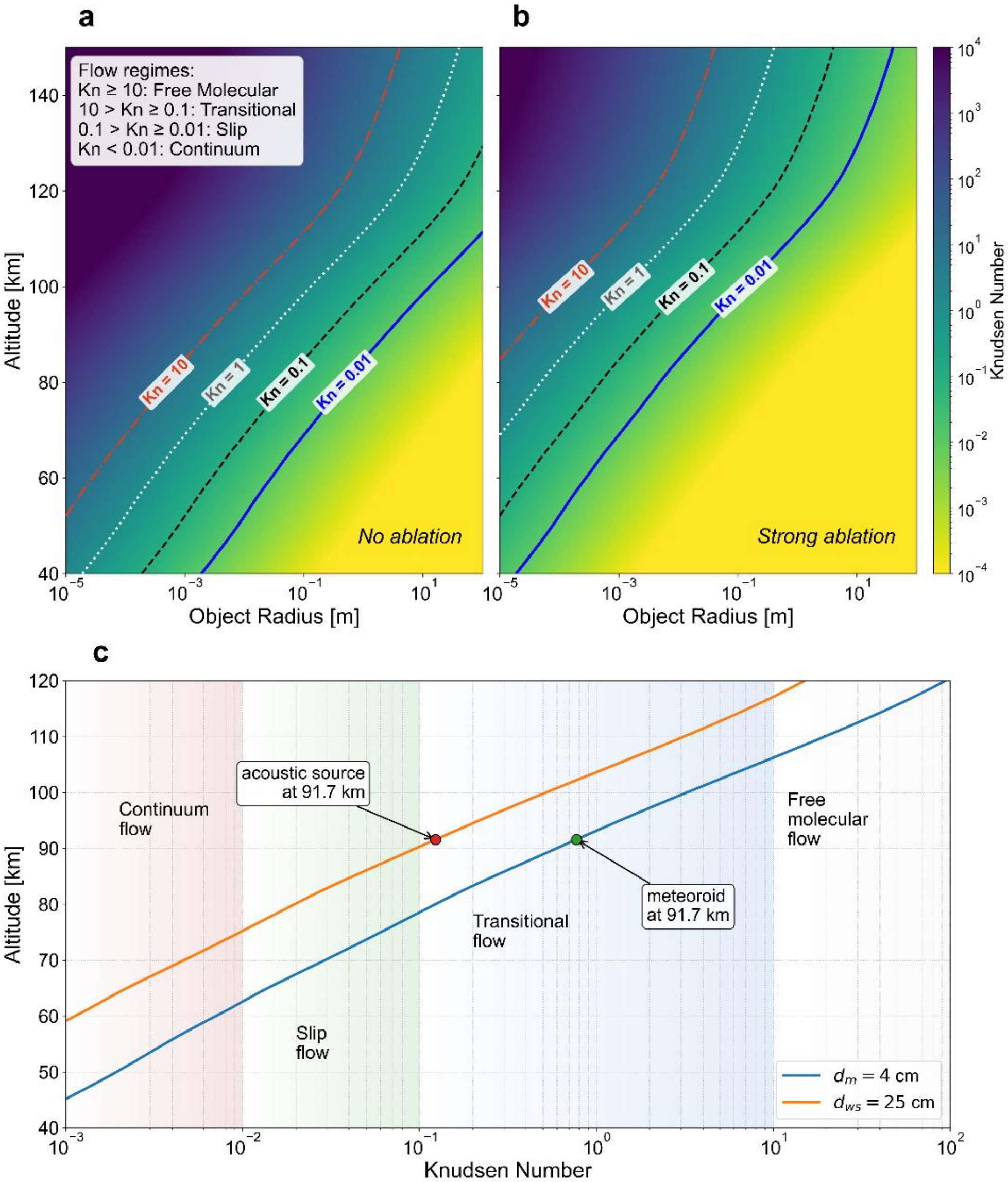

**Figure 11:** (a,b) Calculated Knudsen number (log scale) as a function of object size and altitude between 40 and 150 km, using mean free path values derived from G2S atmospheric specifications. Curves indicate conventional flow-regime boundaries: free-molecular ($Kn \geq 10$), transitional ($0.1 \leq Kn < 10$), slip-flow ($0.01 \leq Kn < 0.1$), and continuum ($Kn < 0.01$). (a) Under ambient conditions, centimeter-scale bodies at thermospheric altitudes (~90–110 km) reside in the transitional or free-molecular regime. (b) Strong ablation produces local density enhancements that reduce the effective Knudsen number, permitting shock-supporting flow. (c) Knudsen number evaluated for a 4 cm meteoroid (blue) and an acoustic-equivalent source diameter of ~25 cm (orange).

The optical observations and orbital constraints place the object in a dynamical regime populated by both cometary and carbonaceous meteoroids (**Section 4.1.1**), motivating the use of volatile-rich end members rather than a unique taxonomic assignment (Jenniskens and Devillepoix, 2025). Our calculation shows that cometary object, composed of ice, and serving as the upper bound approximation, would release a sufficient number of volatiles to satisfy Rankine-Hugoniot conditions for a strong shock wave formation.

To establish a conservative lower bound estimate on flow field mass loading, we consider a porous, CM-like meteoroid in which readily mobilized surface or near-surface volatiles are partially depleted during the earliest phase of the Earth-grazing trajectory and only silicate-dominated material contributes to vapor production. Given the tangential flight, the object traverses several hundred kilometers in highly rarefied flow and experiences sustained sputtering and low-level heating. Under this assumption, the perigee segment is treated as interaction with a devolatilized, phyllosilicate-rich surface layer, and we compute the maximum density enhancement achievable through impact-driven sputtering and thermal evaporation of silicate material alone. This treatment is consistent with the general thermochemical ordering reflected in differential ablation models in which volatile components are released earlier than silicates and are not assumed to contribute uniformly throughout the trajectory (e.g., Vondrak et al., 2008). The purpose of this simplified modeling exercise in **Appendix C** is not to represent the full physical evolution of the meteoroid, but to quantify the minimum particle supply that silicate evaporation can provide to the flow field.

For phyllosilicate-rich end members at thermospheric altitudes, a key limitation is the rate at which surface ablation and collision-driven sputtering can supply vapor and particles to the flow field surrounding the object. CM chondrites are dominated by phyllosilicates, with additional contributions from olivine, pyroxene, and minor Fe-bearing oxides (Suttle et al., 2021). Typical surface binding energies for silicates and oxides are on the order of 5–7 eV, implying that direct collisional yields of ejected atoms and molecules are limited for the atmospheric molecular energies relevant to this event. As shown in **Appendix C**, the sputtering-only yield is of order $Y_{\text{sput}} \sim 1$, under the favorable oblique-incidence assumptions used here. Popova et al. (2007) found lower near-threshold sputtering yields for fast meteoroids, so adopting their yield would further reduce the non-volatile supply and increase the inferred volatile mass-loading requirement. The total non-volatile ablation yield, including thermally enhanced surface evaporation at the computed surface temperature, is $Y_{\text{abl}} \approx 15\text{–}26$. We use $Y_{\text{abl}} = 20$ as a representative favorable value in the density-budget calculation. Even under these favorable assumptions, the non-volatile contribution falls short of the density required to sustain the shock-coupled flow field inferred from the infrasound observations. Direct atmospheric impingement, optimistic non-volatile ablation yield, and thermal evaporation jointly provide $\sim 4.70 \times 10^{18}\ \text{m}^{-1}$, or only $0.31 N_{\text{shock,req}}$ (**Appendix C**). This limitation is further exacerbated by the earthgrazing geometry of the event: the meteoroid remained confined to a narrow altitude band near the upper transitional flow regime and did not descend into denser atmospheric layers where ablation-driven shielding would naturally intensify through increased ambient mass flux.

To assess whether a density shortfall exists, and therefore whether volatile release is required, we evaluate the cumulative particle inventory available to the near-field flow. The total supplied line density is obtained by summing contributions from direct atmospheric impingement ($N_{imp}$), ablation-enhanced production ($N_{abl}$), and thermal evaporation ($N_{evap}$). Any remaining difference between this total and the shock-required line density,

$$N_{shock,req} - (N_{imp} + N_{abl} + N_{evap}) = N_{deficit},$$

quantifies the additional particle loading that must be supplied by volatile release. The remaining line-density deficit therefore defines the additional gas-phase mass loading required to sustain shock-bound flow. In **Appendix C**, this deficit is quantified as $N_{\text{deficit}} \approx 1.03 \times 10^{19}\ \text{m}^{-1}$. The result establishes that surface ablation of silicate-rich material alone cannot provide sufficient near-field mass loading at $\sim$ 92 km, and that an additional particle source is required to raise the local number density within the shielding envelope to shock-supporting levels.

#### *4.2.6 Volatile-Enhanced Mass Loading*

The required near-field density enhancement arises naturally if the meteoroid contains appreciable volatile-bearing phases. CM chondrites are known to host a substantial volatile inventory, dominated by structurally bound water in hydrated minerals along with additional volatile components (e.g., Cronin and Chang, 1993; Robert and Epstein, 1982; Suttle et al., 2021; Trigo-Rodríguez et al., 2019). A volatile fraction at the levels reported for CM material, up to ~14 wt%, represents a reservoir that can be mobilized as the surface heats, thermal penetration advances, and near-surface layers are mechanically eroded or exposed (e.g., Patzer et al., 2021; Suttle et al., 2021). Unlike silicate evaporation, which is governed by relatively high surface binding energies, volatile-bearing phases can release light molecular species through dehydration, dehydroxylation, decomposition, and related thermal processes (see **Appendix C**). The volatile-enhanced near-field flow structure and hydrodynamic shielding are schematically illustrated in **Figure 9**.

In the present context, volatile release is the most plausible mechanism for closing the remaining near-field density deficit through three coupled effects. First, it directly injects additional gas-phase material into the hydrodynamic shielding envelope, increasing the local number density and reducing the effective mean free path. Second, this release occurs within an already developing vapor-particle flow field where collisional interactions are enhanced, increasing the likelihood that released species contribute to sustaining a dense near-field envelope rather than being immediately dispersed into the ambient thermosphere. Third, light molecular species and volatile-bearing fragments can participate in dissociation, excitation, ionization, and thermochemical reactions within the hot flow, modifying the local pressure, density, and temperature structure across the interface between the shielding envelope and the surrounding atmosphere.

From the density-budget constraint developed in **Appendix C**, the remaining deficit corresponds to an equivalent volatile yield of $Y_{\mathrm{vol}} \approx 161$ released atoms or small molecules per incident atmospheric molecule. This value is not imposed as an assumed release rate; it is derived from the difference between the shock-supporting line-density requirement and the maximum non-volatile supply. With this yield, volatile release supplies the additional gas-phase mass loading needed to reconcile the acoustically inferred source scale with the rarefied ambient conditions of the thermosphere.

The optical and acoustic observations jointly support a distributed volatile-release interpretation rather than a single impulsive outgassing event. The erosion-controlled light curve indicates progressive exposure of mechanically weak, volatile-bearing material, while the multi-station infrasound detections show shock production over spatially separated source regions near perigee. These observations favor volatile release tied to thermal penetration, erosion, and exposure of fresh material along the near-perigee segment.

Classical meteor radar and photometric relations provide an additional consistency check on the required particle inventory. When the electron-line-density constraint inferred from the observed absolute magnitude is evaluated within the $d_{ws} \approx 0.25$ m shock-coupled flow-field scale, the implied particle loading exceeds what silicate-only ablation from the $\sim 4$ cm nucleus can provide. Depending on the adopted ionization relation, the inferred electron-derived line density is of the same order as the density required for shock-supporting near-field conditions. This independently supports the inference that an additional volatile contribution is needed. The full electron-line-density calculation, including the choice of $d_{ws}$ rather than the classical adiabatic trail radius as the early-time control volume, is given in **Appendix C**.

**Figure 12** summarizes the particle line-density budget within the acoustically inferred source volume ( $d_{ws} \approx 0.25$ m ). Under perigee conditions, direct atmospheric impingement, optimistic non-volatile ablation yield, and thermal evaporation jointly provide $\sim 4.70 \times 10^{18}$ m$^{-1}$, or only $0.31 N_{\mathrm{shock,req}}$. The shaded region marks the remaining deficit, $N_{\mathrm{deficit}} \approx 1.03 \times 10^{19}$ m$^{-1}$, that must be supplied by additional gas-phase mass loading. The derivation of each term, along with sensitivity tests and the grain-evaporation length-scale analysis, is provided in **Appendix C**.

The physical picture is internally consistent across the optical, acoustic, and density-budget constraints. A small, porous, mechanically weak meteoroid entered the lower thermosphere on a shallow trajectory. Direct collisions and surface heating initiated vapor production and hydrodynamic shielding, but non-volatile ablation and thermal evaporation alone could not supply enough near-field material to sustain shock-bound flow. Volatile release provided the additional gas-phase mass loading required to amplify hydrodynamic shielding, reduce the effective local Knudsen number, and allow the dense vapor-particle envelope to transition into a shock-producing flow field. This mechanism explains how a centimeter-scale earthgrazer could radiate detectable infrasound from altitudes near 92 km, where the ambient atmosphere alone would not support classical continuum shock formation.

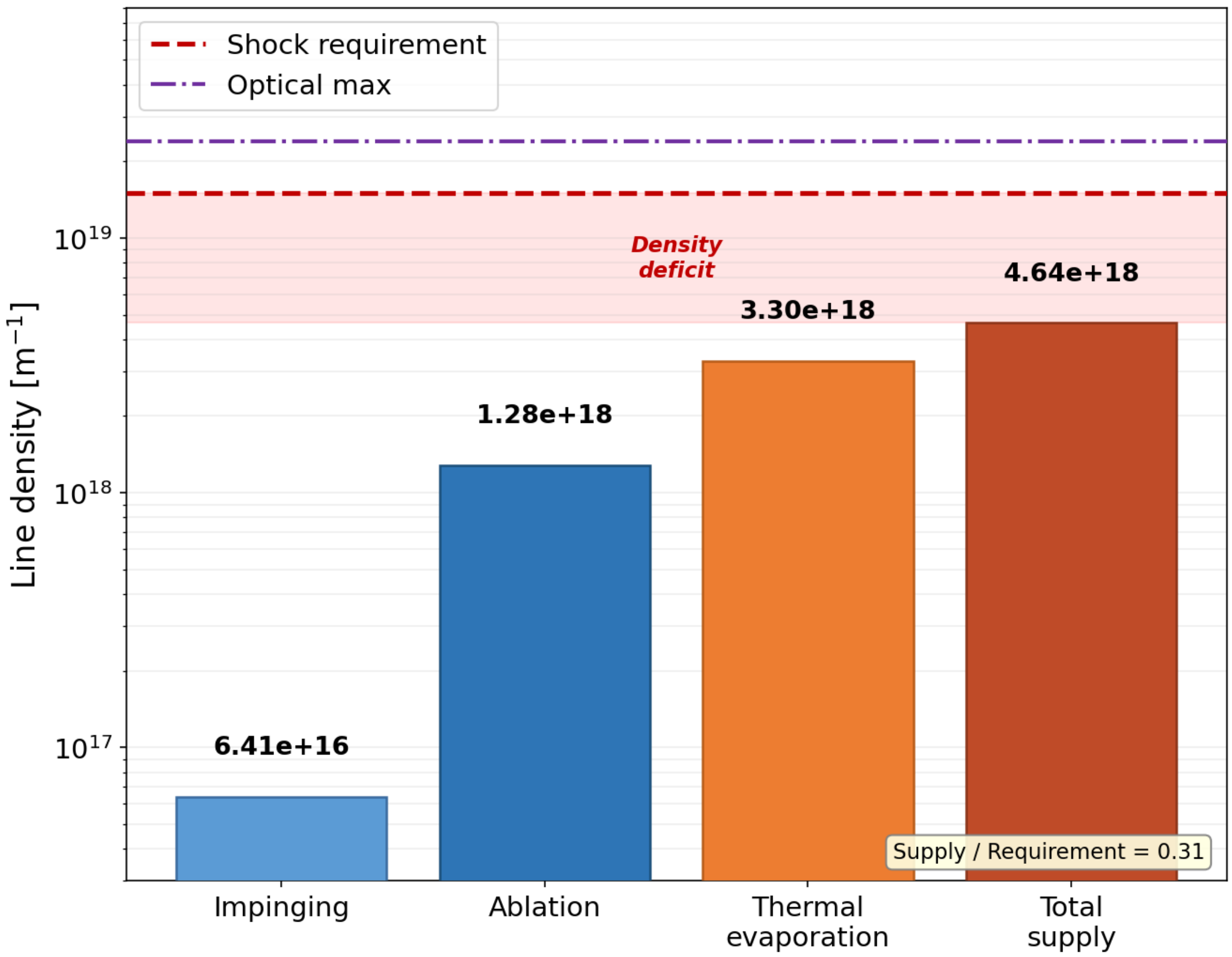

**Figure 12:** Comparison of particle line-density contributions within the acoustically inferred source volume. Bars show contributions from impinging atmospheric molecules, ablation-driven enhancement, and thermal evaporation, with their sum indicated as the total. The horizontal red line marks the line density required to sustain a shock within a 25 cm acoustic source diameter, while the dashed line indicates the maximum line density inferred from optical observations. The shaded region highlights the deficit between the total supplied particle density and the shock requirement. Note that the ordinate is in log scale.

### 4.3 Broader Implications

The results presented here have important implications for meteoroid aerodynamics, shock physics in rarefied flows, and interpretation of high-altitude acoustic observations. Through simultaneous optical and infrasound measurements, this study provides the first empirically constrained demonstration that a centimeter-scale earthgrazer can generate sustained, coherent shocks at thermospheric altitudes through ablation-driven hydrodynamic shielding amplified by volatile release.

These observations also show that centimeter-scale meteoroids can serve as natural analogs for much larger effective acoustic sources during early atmospheric entry. When hydrodynamic shielding and volatile release enlarge the effective flow field, small objects can exhibit aerodynamic and shock-generation behavior characteristic of substantially larger, structurally coherent sources at comparable altitudes. Earthgrazers therefore offer a rare opportunity to investigate continuum-like shock formation, energy coupling, and acoustic radiation under natural conditions that are otherwise accessible mainly through costly and infrequent artificial re-entry experiments. Leveraging such events can strengthen validation of atmospheric-entry and energy-deposition models relevant to high-altitude, non-terminal encounters. Further progress will benefit from coordinated infrasound, optical, radar, and satellite observations, which together can better constrain source evolution and flow-field properties.

The summary of implications is as follows:

i. Our results imply that shock formation in the thermosphere is controlled primarily by the properties of the ablation-generated vapor-particle flow field rather than by the solid meteoroid alone. This behavior is consistent with earlier theoretical and numerical studies showing that hydrodynamic shielding can enlarge the effective aerodynamic dimensions of a meteoroid and reduce the local Knudsen number (Anderson, 2006; Jenniskens et al., 2000; Popova et al., 2001; Popova et al., 1998; Silber et al., 2018). The present event provides quantitative observational support for this mechanism at thermospheric altitudes and demonstrates that it can operate even for small, mechanically weak bodies on shallow, non-terminal trajectories. It also shows that shock-related acoustic radiation in rarefied flow depends on the properties of the energy-coupled flow field rather than on the dimensions of the solid body alone.

ii. The nearly constant-altitude flight at $91.7 \pm 0.1$ km across a $\sim 164$ km trajectory segment offers an unusually favorable test bed for separating altitude-dependent effects from temporal evolution. The shock-producing region is confined to a narrow atmospheric layer while the source progresses laterally, allowing the onset and persistence of shock formation to be examined under conditions that are difficult to isolate in steeper entries.

iii. The multi-station, multi-point infrasound detections establish a benchmark for analyzing distributed acoustic sources along extended trajectories. The observations demonstrate that infrasound can resolve spatially separated shock-generation regions within a single event, enabling quantitative tracking of source evolution in the thermosphere.

iv. From a monitoring and planetary defense perspective, the results imply that even small, volatile-rich bodies can couple energy efficiently to the atmosphere at thermospheric altitudes in ways detectable by regional or global geophysical networks. Although high-altitude, non-terminal entries such as earthgrazers are not primary impact hazards, they provide important test cases for validating atmospheric-entry, fragmentation, and energy-coupling models used in early-warning, surveillance, and source-characterization applications. Incorporating thermospheric infrasound constraints can reduce uncertainties in how transient natural events are represented across a broader range of entry geometries and altitudes.

v. The processes documented here are also relevant beyond Earth. Understanding how vapor-cloud formation, mass loading, and local flow-regime transitions operate under different atmospheric compositions and densities is essential for interpreting meteoroid observations across the Solar System and for designing future entry probes and monitoring strategies. In lower-density atmospheres such as Mars, volatile-enhanced hydrodynamic shielding may help enable shock formation by small meteoroids at altitudes where continuum assumptions fail. In denser atmospheres such as Venus or Titan, where ambient collisionality is high but thermal and chemical environments differ substantially (e.g., Guinan et al., 2026), volatile release may modify the altitude, strength, and spatial extent of shock formation rather than determine its onset.

## 5. Conclusions

Hydrodynamic shielding has long been recognized theoretically as a critical but poorly constrained process in hypersonic rarefied flows, yet direct observational validation under natural conditions has remained limited. Rare earthgrazing meteoroids that generate infrasound detectable at the ground provide a powerful means of probing this phenomenon at thermospheric altitudes, where laboratory experiments and numerical models face significant limitations. In this study, we present the first coordinated optical and multi-station infrasound observations of a centimeter-scale earthgrazing meteoroid producing sustained shock-related infrasound near 92 km altitude. These observations provide direct constraints on the structure, density, and evolution of the shock-producing near-field flow.

Optical observations show that the meteoroid underwent early mechanical erosion at exceptionally low dynamic pressure, consistent with a weak, porous structure containing volatile-bearing material. The light curve is better explained by progressive erosion and exposure of fresh material than by sustained thermal ablation of a small body. Independent infrasound detections from three arrays localized acoustic radiation to distinct source regions along a narrow altitude band near perigee, confirming that shock production was distributed along multiple segments of the same earthgrazing trajectory rather than confined to a single impulsive fragmentation event.

Weak-shock modeling yields a consistent blast radius of ~30 m and an acoustic-equivalent source diameter of ~0.25 m, far exceeding the physical size of the ~45 g nucleus. These values should not be interpreted as the physical dimensions of the solid meteoroid, but as the effective scale of the energy-coupled vapor–particle flow field required to reproduce the observed infrasound periods. The results therefore show that the acoustic source was governed by the dimensions, density, and energy coupling of the near-field vapor-particle flow envelope rather than by the solid body alone. Classical ablation-driven hydrodynamic shielding is a necessary part of this interpretation, but by itself it is insufficient to explain the observed shock strength, source scale, and persistence under the ambient thermospheric conditions.

The density-budget analysis supports this conclusion. Regardless of whether the object is interpreted as an ice-rich cometary aggregate or as a volatile-bearing CM-like proxy, additional gas-phase mass loading is required for shock formation under thermospheric conditions at ~92 km. A conservative lower-bound calculation, in which the non-volatile supply is limited to direct atmospheric impingement, impact-driven sputtering, and thermal evaporation of silicate-dominated material, shows that these processes provide only a fraction of the particle line density required to sustain the shock-coupled structure inferred from the infrasound observations. This density shortfall persists even under favorable assumptions for silicate evaporation. The most plausible source of the remaining near-field mass loading is rapid release of volatile-bearing material from the interior of a porous body, including structurally bound water and hydroxyl-bearing phases in hydrated minerals. Such release amplifies hydrodynamic shielding, increases local collisionality, reduces the effective Knudsen number, and enables the transition from rarefied or transitional flow to a shock-bound flow field capable of radiating detectable infrasound.

This mechanism reconciles the observations with gas-dynamic theory and clarifies the apparent paradox of shock persistence at thermospheric altitudes for small bodies. The primary result is not simply that a centimeter-scale earthgrazer produced infrasound, but that optical and acoustic constraints together require an enlarged, mass-loaded, shock-supporting near-field flow. In this sense, the meteoroid acted as an effective cylindrical acoustic source much larger than the physical nucleus, with volatile-enhanced hydrodynamic shielding providing the additional density needed to support shock formation in an otherwise rarefied environment. This study demonstrates that optical observations, electron-line-density estimates, and infrasound measurements can be used together, while also providing independent diagnostic constraints on shock-wave formation.

Earthgrazers therefore provide rare high-altitude laboratories for characterizing the fundamental processes governing energy deposition, hydrodynamic shielding, and shock onset in the most rarefied reaches of an atmosphere. By providing the first empirically constrained demonstration of sustained shock-related infrasound from an earthgrazer at thermospheric altitude, this work establishes a basis for investigating how small bodies can transiently acquire effective source dimensions and collisional flow conditions far exceeding those expected from their physical size alone. Extending these efforts to comparative atmospheric environments, including Earth, Mars, Venus, Titan, and other planetary bodies with atmospheres, will help clarify how atmospheric density, molecular composition, volatile inventory, and thermal structure control shock formation and energy coupling for hypersonic objects. Future coordinated optical, infrasound, radar, and satellite observations of earthgrazers and other shallow-entry bolides will extend this multi-modal approach, providing essential empirical constraints for validating atmospheric-entry models, improving interpretation of high-altitude shock phenomena, and strengthening the physical basis of acoustic-monitoring approaches across planetary environments.

## Open Data

All infrasound data used in this study are freely available through the International Federation of Digital Seismograph Networks (FDSN) from the Royal Netherlands Meteorological Institute (KNMI) arrays (EXL, DBN, and CIA). The InfraPy signal analysis toolkit used for beamforming is open access and available at https://github.com/LANL-Seismoacoustics/InfraPy (Blom et al., 2016). The Cardinal array-processing software used for time–frequency–wavenumber analysis is freely available via https://github.com/sjarrowsmith/cardinal (Ronac Giannone et al., 2025). Atmospheric specifications for acoustic propagation modeling were obtained from the Ground-to-Space (G2S) model, hosted by the National Center for Physical Acoustics (NCPA), accessible at https://github.com/chetzer-ncpa/ncpag2s-clc (Hetzer, 2024). The weak shock modeling approach applied in this work follows the procedure fully presented and described in Silber et al. (2015), where the methodology, assumptions, and numerical implementation are documented in detail and can be replicated using the parameters provided therein. The camera network attributions come from the following sources: https://www.imonet.org/ (IMO VMN); https://www.dutch-meteor-society.nl/allsky/; http://www.nemetode.org/DataSet_2020.htm (NEMETODE); https://archive.ukmeteors.co.uk/ (UKMON); https://globalmeteornetwork.org/ (GMN); https://www.fripon.org/ (FRIPON).

## Acknowledgements

The authors sincerely thank all camera operators whose dedication and perseverance made optical observations possible. Their sustained commitment to monitoring the night sky and sharing high-quality data is essential to advances in meteoroid science. We also wish to explicitly recognize the camera operators whose data were directly used in this analysis. Any inadvertent omissions are unintentional, and all contributions are sincerely appreciated. Camera operators: Richard Bassom, Hans Betlem, Steve Bosley, Martin Breukers, Peter Carson, Jürgen Dörr, Richard Fleet, Erwin Harkink, Nick James, Klaas Jobse, Harry Kiewiet, Sirko Molau, Björn Poppe, Alex Pratt, Paul Roggemans, Jim Rowe, Hans Schremmer, Jörg Strunk, Arnold Tukkers, Marco Verstraaten. The authors thank Jelle Assink for assistance with KNMI infrasound station-related information.

Sandia National Laboratories is a multi-mission laboratory managed and operated by National Technology and Engineering Solutions of Sandia, LLC (NTESS), a wholly owned subsidiary of Honeywell International Inc., for the U.S. Department of Energy's National Nuclear Security Administration (DOE/NNSA) under contract DE-NA0003525. This written work is authored by an employee of NTESS. The employee, not NTESS, owns the right, title, and interest in and to the written work and is responsible for its contents. Any subjective views or opinions that might be expressed in the written work do not necessarily represent the views of the U.S. Government. The publisher acknowledges that the U.S. Government retains a non-exclusive, paid-up, irrevocable, world-wide license to publish or reproduce the published form of this written work or allow others to do so, for U.S. Government purposes. The DOE will provide public access to results of federally sponsored research in accordance with the DOE Public Access Plan.

## Funding

EAS, MRG, SA and DCB were supported by the Laboratory Directed Research and Development (LDRD) program at Sandia National Laboratories, project number 229346. DV and TD were partially funded by the NASA Meteoroid Environment Office under cooperative agreement 80NSSC24M0060.

## Conflict of Interest

The authors declare no conflict of interest.

# Appendix A: Gravity Correction Procedure

### *A.1. Refinement of the General Gravity Correction Method*

The gravity-correction procedure described here is applicable to long-duration, shallow (earthgrazing) trajectories where the strength of the gravitational field is nearly constant. For short-duration or steep-entry events, standard trajectory formulations remain sufficient and no additional correction is required. For fireballs of long duration, the trajectory curvature caused by Earth's gravity becomes significant relative to astrometric measurement errors. Vida et al. (2020) proposed a method to correct for this effect by calculating the vertical drop $\Delta h$ experienced over time $\Delta t$. However, we identified that the original implementation of this method omitted a step regarding the update of the trajectory reference point during the minimization procedure.

The standard solver projects a line of sight $U$ from a station $P$ to a radiant vector $R$ passing through a reference point $S$ (e.g., the fireball beginning point). This yields two closest points of approach: $Q$ (on the line of sight) and $T$ (on the radiant line). To account for gravity, the point $T$ is lowered by the calculated drop $\Delta h$ to a new position $T_G$ such that:

$$T_G = T - \Delta h \frac{T}{|T|}$$

The critical refinement implemented here is to update the reference point for the trajectory solution using $T_G$ instead of the initial reference point S. By shifting the reference point for each measurement, the modeled trajectory effectively bends to match the physical path. The recomputed closest points of approach $T'$ and $Q'$ derived from this gravity-corrected reference are then used for the final trajectory minimization. This correction is necessary for all fireballs lasting longer than ~3.2 seconds, where the gravitational drop exceeds typical astrometric uncertainties (~50 m).

### *A.2. Application to Earthgrazing Geometry*

While the geometric correction described above is general, the analytical formula typically used to calculate the magnitude of the drop $\Delta h$ assumes a steep entry where the vertical velocity component is maintained. The 22 September 2020 fireball presented a unique case: an earthgrazing trajectory confined to a narrow altitude band (91–114 km) that reached zero vertical velocity at perigee before ascending.

For this specific geometry, the standard variable-gravity integration is ill-suited. Instead, we calculated the gravitational acceleration g based on the mean geocentric radius of the trajectory. The average geodetic latitude of the track was determined to be 52.8°N, corresponding to a Mean Sea Level Earth radius $r_E$ of 6364.611 km. Adding the mean trajectory height of 95 km yielded a geocentric radius $r$ = 6459.611 km. Using the standard gravitational parameter GME ~ $3.986 \times 10^{14}$ $m^3/s^2$, the local acceleration of gravity was calculated as:

$$g = \frac{GM_E}{r^2} \approx 9.55 \text{ m/s}^2$$

To validate this constant-g approximation, we performed sensitivity analyses using g values derived for average heights of 90 km and 100 km. The resulting trajectory residuals differed by less than 2 meters. The difference in the calculated total drop $\Delta h$ between these extreme cases over the ~30-second duration was less than 15 meters. Given that this variation is well below the astrometric measurement error, the mean-gravity approach provides a sufficiently rigorous correction for the earthgrazing orbital solution.

### *A.3. Ensuring Accuracy of Camera Timings*

The gravity drop is time dependent, and a further challenge is to establish that apparent deceleration arises primarily from true aerodynamic drag, rather than from geometric and timing effects. This is because the solver makes timing adjustments to measurements to correct for camera clock errors. It was found that some combinations of data failed to show plausible deceleration behavior, which would also feed back into the gravity correction accuracy. The fireball only slowed by the equivalent of 0.2 seconds travel time over its 30 second duration, and this can be obscured if the solver treats it as clock error rather than drag. Stations broadside to the fireball recorded long arcs across the sky and gave better consistency. Measurements of the beginning of the visible path were hampered by the encroaching dawn, and at either end the fireball was too faint for the timing to be measured accurately. The data presented below were carefully chosen for timing consistency and gave a good measurement of deceleration.

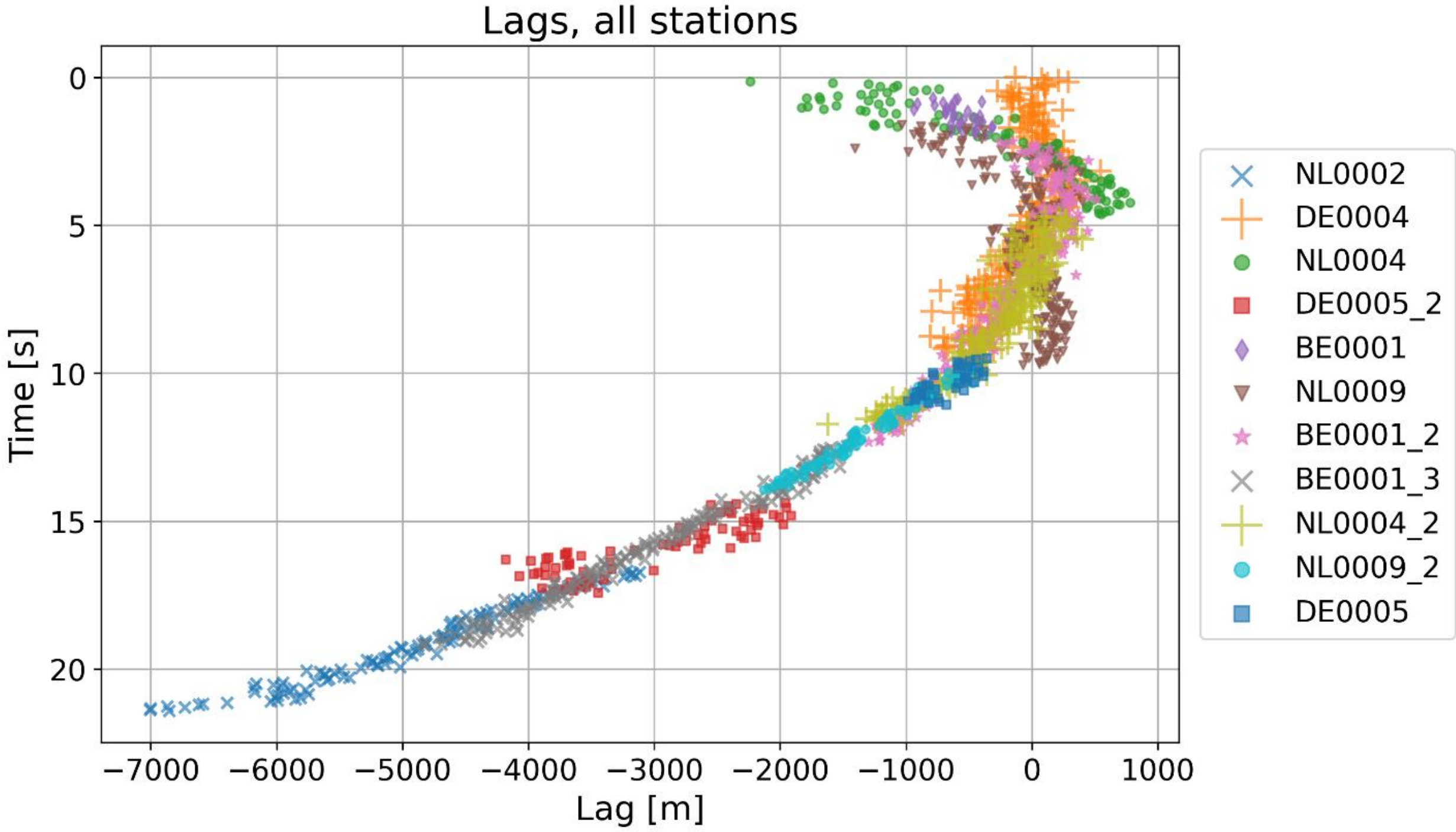

**Figure A1:** The 'lag' plot shows the distance the fireball drops behind over time due to atmospheric drag. Measurement accuracy is best when the fireball is brighter or cameras nearer.

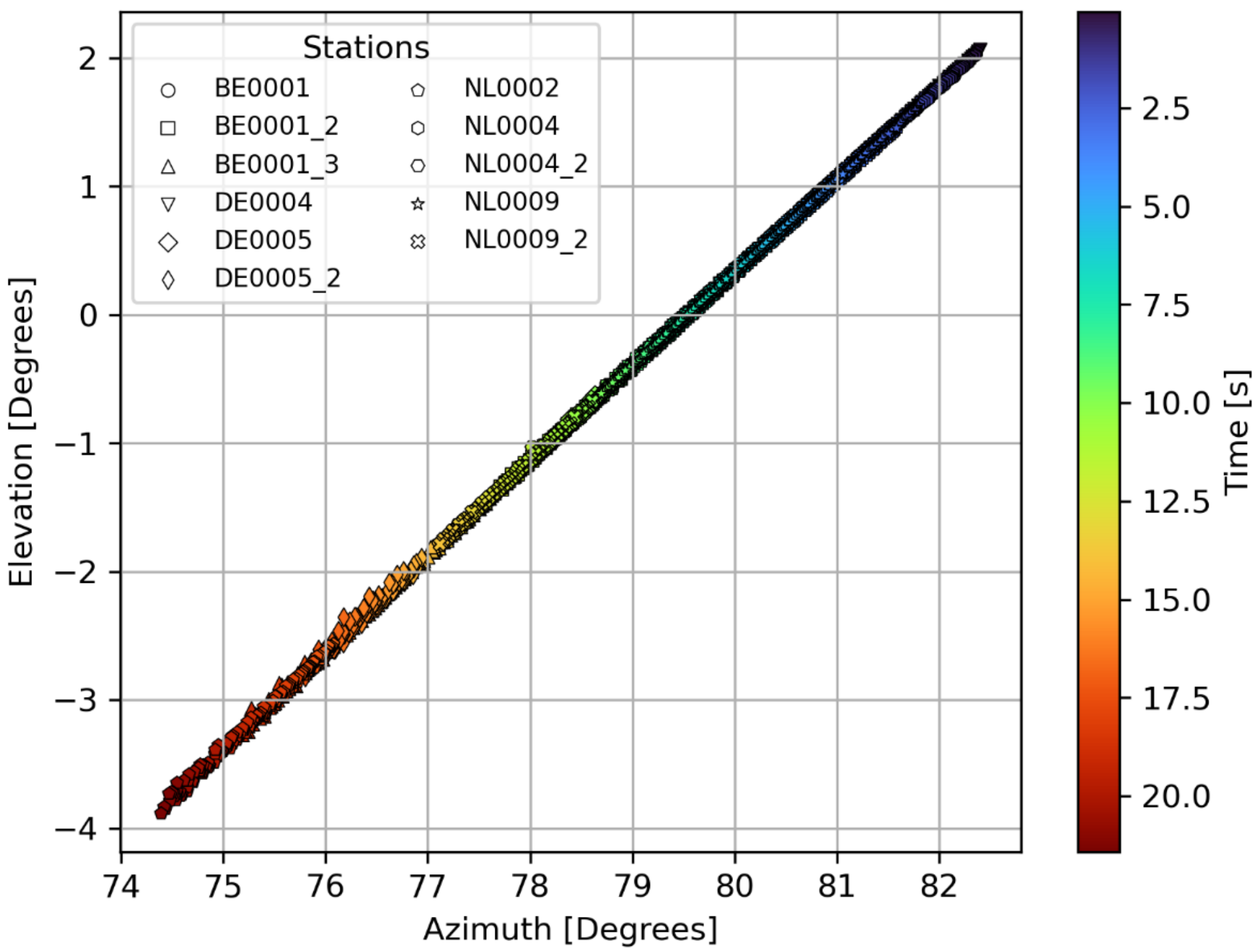

**Figure A2:** Evolution of the apparent radiant azimuth and elevation as the fireball traversed the curved Earth. When elevation passes from positive to negative, the object becomes a true earthgrazer.

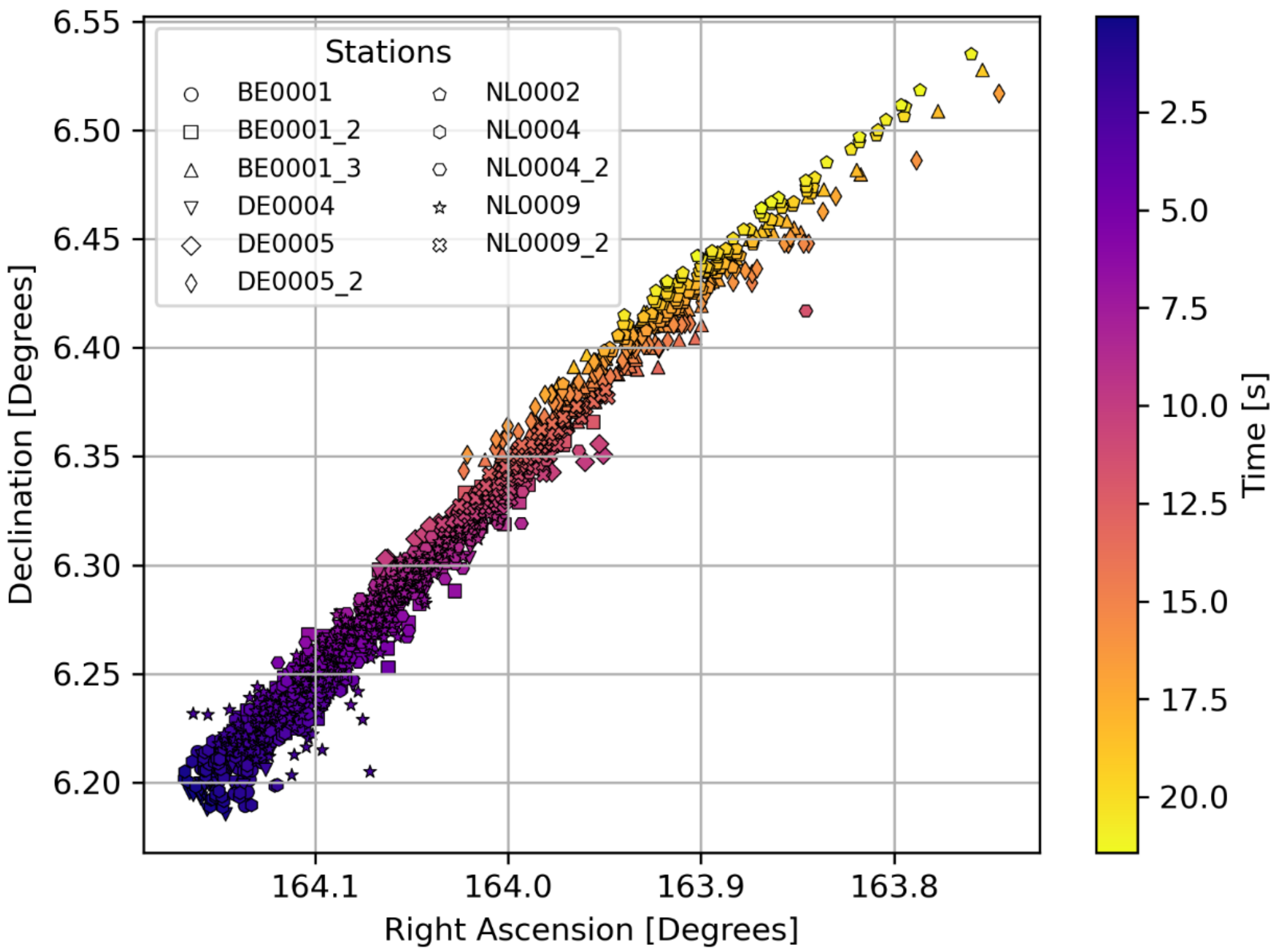

**Figure A3:** Point-to-point radiant positions for the central portion of observed flight. Measured relative to the fixed stars, the gravitational curvature of the trajectory amounts to more than 0.5° over just 20 s (apparent, ground-fixed radiant values, epoch of date).

# Appendix B

**Table B1:** Summary of infrasound results. Measured infrasound signal parameters (A), acoustic propagation modeling results (B), and weak shock modeling results (C). Listed velocities are local trajectory-interpolated/averaged values at the modeled acoustic source points. Small differences between stations reflect rounding and interpolation over the trajectory solution and are within the stated uncertainty; they do not imply physical acceleration.

| A | | | | |
|---|---|---|---|---|
| | **Signal measurements** | **EXL** | **DBN** | **CIA** |
| | Station latitude [deg] | 52.9068 | 52.0989 | 51.9688 |
| | Station longitude [deg] | 6.86555 | 5.172884 | 4.92793 |
| | Signal arrival time [UTC] | 3:58:45 | 4:00:20 | 4:00:46 |
| | Duration [s] | 2.2 | 2.0 | 2.5 |
| | Observed back azimuth [°] | 10.3 | 349.3 | 347.8 |
| | Amplitude, P2P [Pa] | 0.788 | 0.347 | 0.608 |
| | Amplitude, max [Pa] | 0.397 | 0.195 | 0.284 |
| | Period, zero crossings [s] | 1.23 ± 0.38 | 1.10 ± 0.07 | 1.38 ± 0.29 |
| | Period, PSD [s] | 0.93 | 1.00 | 1.43 |
| | Trace velocity [m/s] | 1702 | 537 | 445 |
| | Peak frequency [Hz] | 1.1 | 1.0 | 0.7 |
| | Frequency cutoffs [Hz] | 0.5 - 3.85 | 0.69 - 2.52 | 0.5 - 2.88 |
| | SNR | [68.37, 67.30, 65.89] | [37.81, 35.12, 39.83, 34.72] | [50.69, 49.22, 48.82, 55.30] |
| **B** | | | | |
| | **Propagation modeling** | **EXL** | **DBN** | **CIA** |
| | Shock source latitude [deg] | 52.979 ± 0.001 | 52.767 ± 0.004 | 52.711 ± 0.001 |
| | Shock source longitude [deg] | 6.853 ± 0.005 | 4.925 ± 0.031 | 4.455 ± 0.005 |
| | Shock source altitude [km] | 91.57 ± 0.01 | 91.78 ± 0.17 | 91.81 ± 0.03 |
| | Back azimuth [deg] | 10.6 ± 2.3 | 349.7 ± 1.5 | 341.7 ± 0.2 |
| | Ray inclination at source | -84.8 | -54.6 | -50.1 |
| | Fireball velocity [km/s] | 33.5 ± 0.3 | 33.5 ± 0.3 | 33.6 ± 0.4 |
| | Origin time [UTC] | 3:53:40 | 3:53:42 | 3:53:45 |
| **C** | | | | |
| | **Weak shock modeling results** | **EXL** | **DBN** | **CIA** |
| | Blast radius [m] | 30.8 | 30.6 | 30.6 |
| | Fundamental frequency [Hz] | 3.2 | 3.2 | 3.2 |
| | Fundamental period [s] | 0.31 | 0.31 | 0.31 |
| | Dominant frequency (weak shock) [s] | 0.8 | 0.7 | 0.7* |
| | Dominant frequency (linear) | 0.9 | 0.9 | 0.9 |
| | Dominant period (weak shock) [s] | 1.27 | 1.35 | 1.37 |
| | Dominant period (linear) [s] | 1.06 | 1.13 | 1.15 |
| | Slant range [km] | 92.0 | 119.3 | 127.5 |
| | Horizontal range [km] | 8.1 | 76.2 | 88.5 |
| | Arrival inclination [deg] | 84.9 | 50.3 | 46.0 |
| | Transition altitide [km] | 47.5 | 46.8 | 46.0 |
| | Zonal wind velocity at source [m/s] | 29.2 | 27.9 | 27.7 |
| | Meridional wind velocity at source [m/s] | 15.0 | 15.9 | 16.0 |

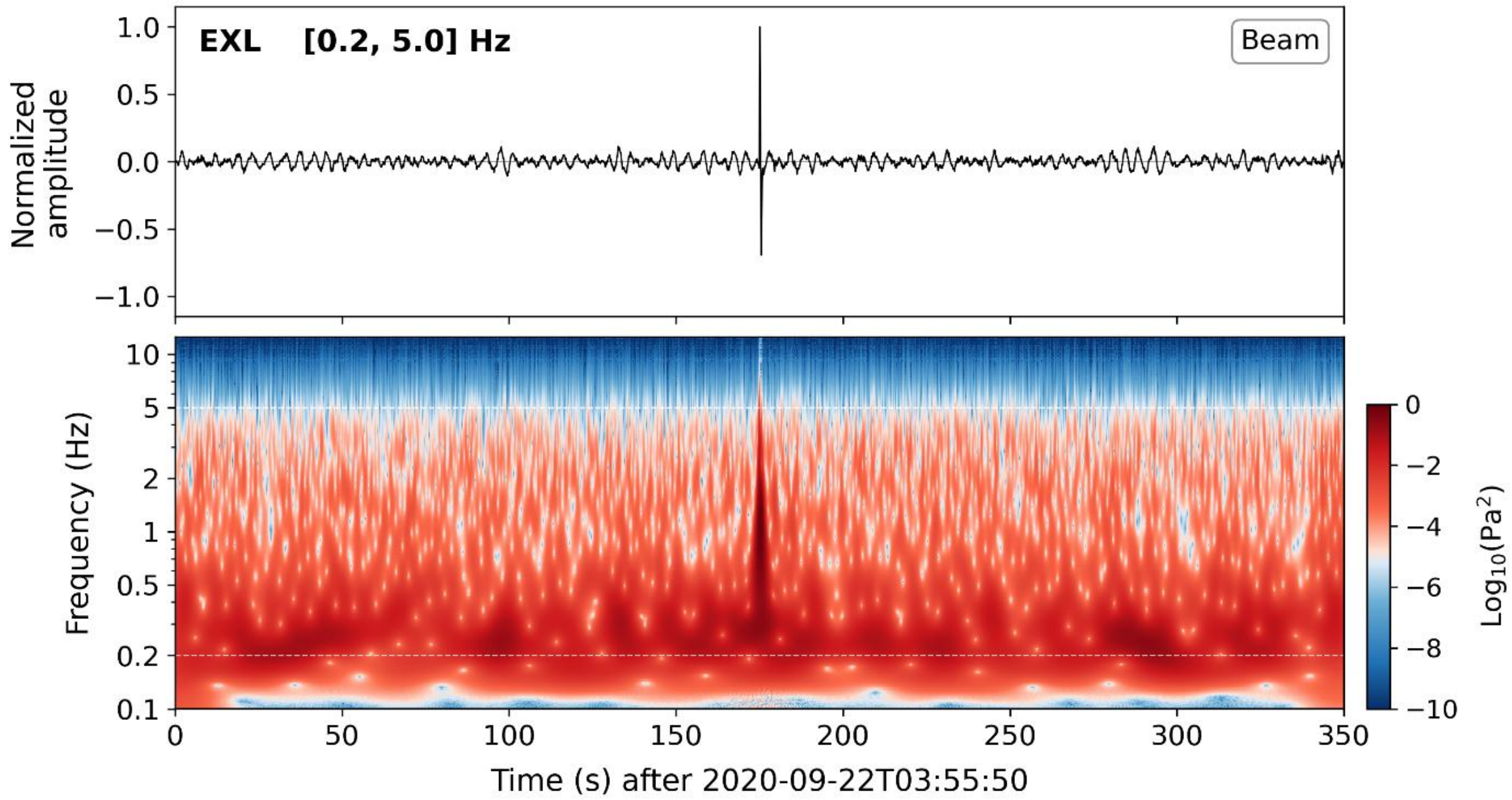

**Figure B1:** Infrasound signals detected at EXL. The top panel displays the beamformed time-series waveform (normalized amplitude vs. time), showing the characteristic N-wave morphology indicative of a ballistic shock arrival. The lower panel presents the corresponding scalogram (time-frequency analysis), illustrating that the acoustic energy is concentrated at low frequencies, primarily centered around a peak frequency of ~1 Hz. This dominant period (averaging 1.24±0.16 s) is consistent with a thermospheric source at ~92 km altitude, where the rarefied environment and subsequent propagation effects lead to the observed frequency dispersion.

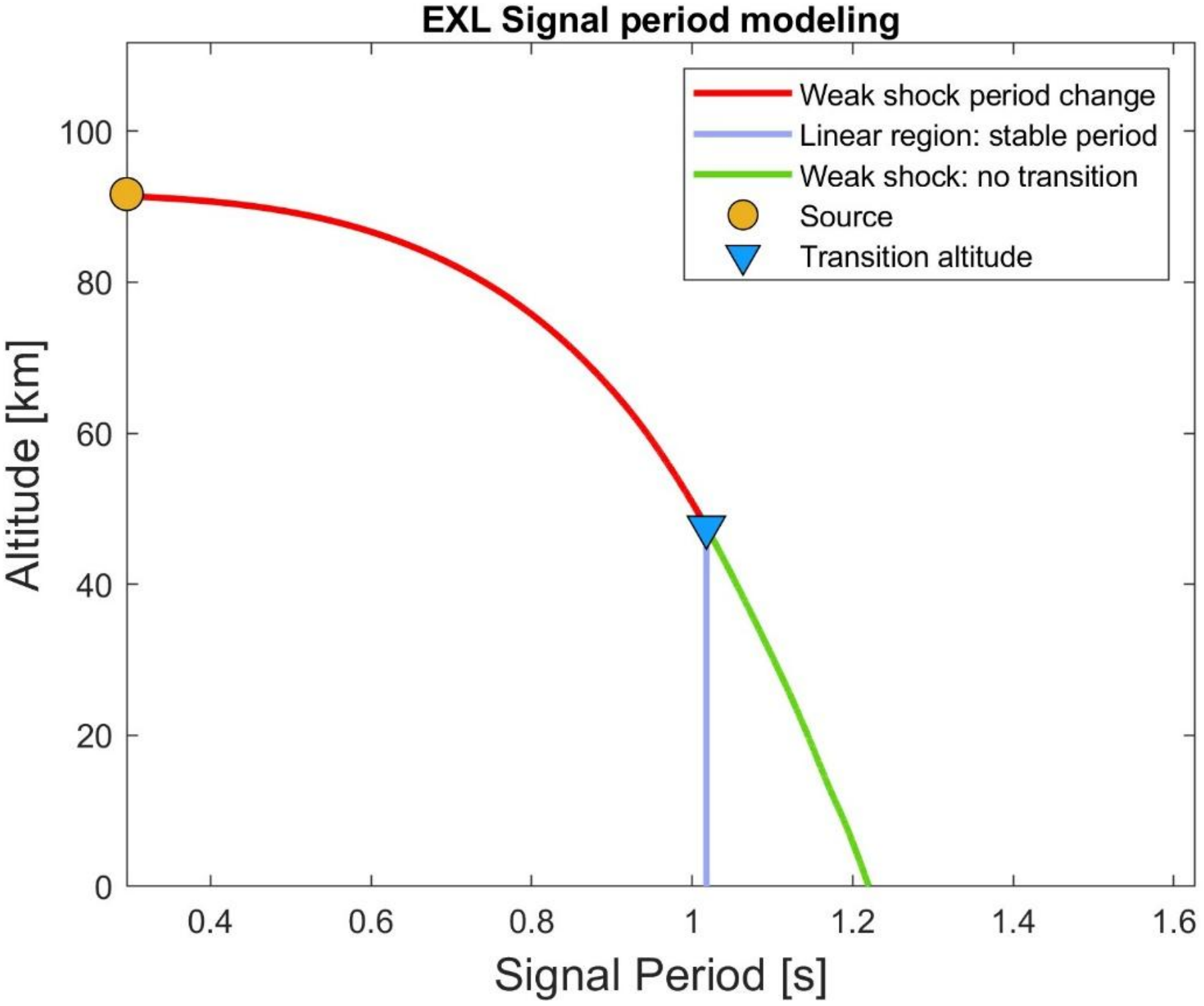

**Figure B2:** Example of the weak shock model output for the signal detected at station EXL. The plot shows the predicted change in the infrasound signal period as the wave propagates from the source altitude (~92 km) to the ground. The period evolves during the weak shock phase (red curve) until it reaches the transition altitude (blue triangle), after which it propagates linearly with a stable period (blue line). For comparison, the green line shows the predicted period evolution if the wave were to remain in the weak shock regime for its entire path without transitioning.

# Appendix C: Physical Constraints and the Role of Volatiles in Shock Formation

The calculations presented in this appendix are intentionally simplified and are intended as a bounding model for demonstration purposes. The objective is not to reproduce the full thermophysical evolution of a meteoroid or the detailed, time-dependent coupling between ablation, fragmentation, chemistry, and flow dynamics. Instead, the analysis isolates individual mass-loading pathways in order to establish conservative lower bounds on particle production and flow-field density enhancement under thermospheric conditions.

In particular, the non-volatile calculations developed in Sections C.2, C.4, and C.6 are designed to quantify the maximum contribution that impact-driven sputtering and thermal evaporation of CM-like, phyllosilicate-rich material can supply to the surrounding flow field in the absence of significant volatile release. This deliberately restrictive assumption allows us to assess whether classical ablation mechanisms alone are sufficient to provide the near-field density enhancement required for hydrodynamic shielding and subsequent shock formation under the observed thermospheric conditions. Any density shortfall identified under these conditions therefore represents a robust lower bound on the additional mass loading that must be provided by volatile release. While the absolute values derived here depend on simplified assumptions, the qualitative conclusion that silicate evaporation alone is insufficient, and that an additional gas-phase mass-loading source is required, is insensitive to reasonable variations in model parameters. Given the optical evidence for a porous, erosion-prone, volatile-bearing body and the inability of grain evaporation to supply vapor on the required near-field length scale, volatile release is identified as the physically most plausible source of this additional mass loading.

## C.1. Physical Constraints and Thresholds for Shock Formation in Rarefied Flow

Infrasound signatures of the event indicate that a strong shock wave, defined by a pressure ratio approaching or in the range $10^2 < p/p_0 < 10^4$, formed at an altitude of 91.7 km and subsequently produced an attenuated pressure wave detectable at the ground (Bronshten, 1983; Silber et al., 2018; Silber et al., 2017). Given the highly rarefied flow conditions in this region of the thermosphere, it is therefore necessary to establish a lower limit on the number of molecules, atoms, or ions that must be evaporated or ejected from the meteoroid surface per collision with an atmospheric molecule in order to support such a shock.

Rapid surface heating associated with high translational temperature, including rotational and vibrational excitation, is expected to contribute cascading release of meteoroid surface material. However, for the purpose of this first-order approximation, these effects are neglected because the goal here is to determine a conservative lower bound on the amount of evaporated material required to contribute to hydrodynamic shielding and subsequent shock formation. The intent of this analysis is to assess whether ablational evaporation of the earthgrazer alone is sufficient to reduce the local Knudsen number and enable the formation of strong shock waves under the observed thermospheric conditions (Silber et al., 2018; Silber et al., 2017).

From the optical methods employed here, reasonable estimates suggest that the earthgrazer had a characteristic diameter of 0.045 ± 0.002 m during the initial atmospheric entry and ~0.040 ± 0.002 m at the perigee, accounting for mass loss. At an altitude of 91.7 km, where the strongest shock waves were generated, the atmospheric number density is $n = 5.09 \times 10^{19}$ m$^{-3}$ and the mean free path is ~0.032 m. The corresponding Knudsen number is $Kn \approx 0.8$, placing the flow at the boundary of the upper and lower transitional regimes, where transitional flow is defined by $0.1 < Kn < 10$. At this altitude, sufficiently strong ablation, that is, meteoroid surface evaporation yielding at least several ejected surface molecules or atoms per collision with atmospheric $N_2$ or $O_2$, would be sufficient to reduce the local Knudsen number in the immediate vicinity of the meteoroid (Bronshten, 1983; Popova, 2004). Such a reduction would move the flow toward the boundary with the slip-flow regime and allow the formation of a sufficiently dense hydrodynamic shielding layer that may transition into a shock-bound flow field, provided that the Rankine–Hugoniot conditions are satisfied. Under such theoretically ideal circumstances, the resulting shock wave would be weak to moderate in strength, typically characterized by a flow-field to ambient pressure ratio $p/p_0$ (see Zel'dovich and Raizer (2002) and Bronshten (1983)). However, the number of ablated or evaporated molecules, atoms, or ions released from the meteoroid surface during collisions with the atmospheric column depends not only on the meteoroid velocity but, more importantly, on the material properties of the meteoroid, such as surface binding energy (Behrisch, 1981; Tielens et al., 1994).

For a non-ablating object to satisfy minimum requirements for the formation of a strong shock wave defined here by $p/p_0 \approx 10^2$, at a velocity of 33.5 km/s, its characteristic size would need to be ~3.2 m, in order to fall under continuum flow conditions. For an ablating object, theory indicates that the effective flow-field size may be up to two orders of magnitude larger than the characteristic dimensions of the meteoroid (function of material properties and velocity) (Boyd, 2000; Jenniskens et al., 2000; Silber et al., 2018). In the present case, the effective flow-field size is constrained by the acoustic volume inferred from weak-shock theory as the only experimental constraint: the acoustic-equivalent source diameter is $d_{ws} \approx 0.25$ m, which approximately corresponds to the size of the flow field enclosed by the initial shock envelope.

To place the acoustically inferred source scale in a gas-dynamic context, we treat the shock-coupled near-field region as an effective hypersonic source with characteristic diameter $d_{ws}$. This treatment is equivalent to asking what diameter a non-ablating hypersonic body would require, at the same Mach number and ambient atmospheric state, to generate the observed weak-shock source scale. In this analogy, $d_{ws}$ represents the acoustic-equivalent diameter of the energy-coupled vapor–particle flow field and provides the characteristic length scale for evaluating the local flow density, line-density budget, and transition toward collisional, shock-bound flow.

Using the Rankine–Hugoniot relations for a normal shock and a strong shock requirement, we establish the minimum required number density for strong shock formation:

$$n_{\text{shock,req}} \approx 3.05 \times 10^{20}\ \text{m}^{-3}. \qquad \text{(C1)}$$

For simplicity, we ignore the contribution of kinetic temperatures in evaluating this initial constraint. Converting this to a line density requirement over the acoustically inferred shock-coupled source cross-section ($d_{ws} \approx 0.25$ m, $A_{ws} = \pi(d_{ws}/2)^2 \approx 4.91 \times 10^{-2}$ m$^2$) yields:

$$N_{\text{shock,req}} \approx 1.50 \times 10^{19}\ \text{m}^{-1}. \qquad \text{(C2)}$$

This represents a mean (or line) number density, adopted for simplicity, as the actual number density within the flow field is expected to be non-uniform and approximately Gaussian in distribution rather than spatially constant. Thus, $N_{\text{shock,req}}$ is used as a physically motivated line-density scale for shock-supporting collisional flow rather than as a spatially uniform density prescription.

To identify the mechanism responsible for generating such a density discontinuity, and to assess whether the flow field contains sufficient material to satisfy the conditions for strong shock wave formation and a local Knudsen number approaching 0.01, three factors must be considered. These are: (a) the local atmospheric number density; (b) the minimum number of atoms, molecules, or ions evaporated or ejected from the meteoroid surface per collision with an atmospheric molecule; and (c) whether the combined number density within the flow-field volume, arising from both impinging atmospheric molecules per unit path length and evaporated meteoroid material, is sufficient to reduce the local Knudsen number, or whether an additional mechanism is required. By addressing these considerations in turn, Sections C.2–C.6 demonstrate that, in the case of this earthgrazer, volatile material is required to compensate for the density deficit in the flow field in order to generate strong shock waves under rarefied conditions.

### C.2. Evaluation of Thermal Evaporation and Surface Temperature Limits

Because the surface-temperature and thermal-evaporation estimates provide input values used throughout the density-budget exposition that follows, we present this calculation before evaluating the individual mass-loading pathways. The objective of this section is to estimate the surface temperature reached under the observed flight conditions and to quantify the corresponding silicate evaporation rate. This provides a favorable test of whether thermally driven silicate mass loss can close the near-field density deficit.

We begin with a comprehensive energy conservation equation for the meteoroid:

$$\frac{1}{2}\Lambda\rho_a v^3 A_c = mC_p\frac{dT}{dt} - L\frac{dm}{dt} + \epsilon\sigma A_s(T^4 - T_0^4) + kA_c\frac{dT}{dx}. \quad \text{(C3)}$$

$$\textbf{Heating input} = \textbf{Heating} + \textbf{Ablation} + \textbf{Radiative Cooling} + \textbf{Conduction}$$

Here the left-hand side is the total energy flux absorbed by the meteoroid cross-section $A_c$, where $\Lambda \approx 0.5$ is the heat transfer coefficient (Briani et al., 2013; Rogers et al., 2005). On the right: $mC_p\, dT/dt$ is internal heating $\left(C_p \approx 1200\ \mathrm{J/(kg\ K)}\right)$; $L\, dm/dt$ is ablation cooling $(L \approx 6.05\ \mathrm{MJ/kg})$ for stony material; $\epsilon\sigma A_s(T^4 - T_0^4)$ is radiative cooling from the full surface area $A_s = 4\pi r^2 (\epsilon \approx 0.9)$ (Rogers et al., 2005); and $kA_c\, dT/dx$ is conductive loss into the interior.

We note an important geometric factor: heating is absorbed over the cross-section $A_c = \pi r^2$, while evaporation and radiation occur over the full surface $A_s = 4\pi r^2$. For a sphere, $A_s/A_c = 4$.

The radiative term is retained as a first-order surface-energy sink in the bounding energy balance. In the developing vapor/shock layer, radiative exchange with hot gas may further modify the surface energy budget. The sensitivity analysis therefore treats heat transfer as a controlling parameter and evaluates whether enhanced silicate evaporation can close the density deficit under favorable thermal assumptions (**Figure C1**).

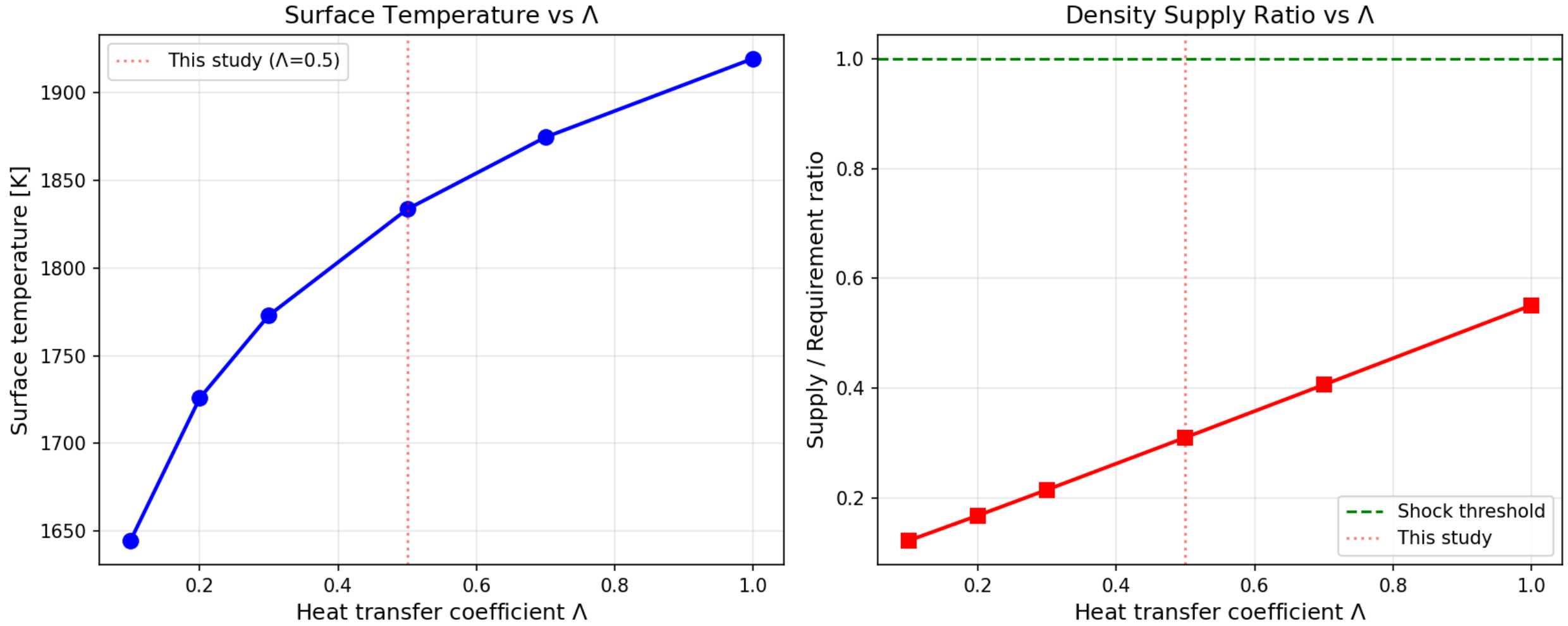

**Figure C1:** Sensitivity of the density-budget result to the heat transfer coefficient Λ. Left axis: equilibrium surface temperature; right axis: ratio of total silicate supply to the shock requirement $\left(\sum N/N_{\text{shock,req}}\right)$. Even at the upper bound $\Lambda = 1.0$, which maximises energy deposition and silicate evaporation, the supply ratio reaches only ~0.55, demonstrating that the density deficit persists across the full physically reasonable parameter range.

We adopt: $\bar{\mu} \approx 23$ amu $= 3.82 \times 10^{-26}$ kg, evaporation coefficient $\alpha_e \approx 1$, $\rho_m = 1000$ kg/m$^3$, consistent with the porous, volatile-bearing structure inferred from the optical analysis, $\bar{\mu}_{air} \approx 28.97$ amu, $v = 3.35 \times 10^4$ m/s, and $h = 91.7$ km. The atmospheric density from G2S provides $n_a = 5.09 \times 10^{19}$ m$^{-3}$, giving $\rho_a = 2.44 \times 10^{-6}$ kg/m$^3$. The assumed mass at perigee is $m = 3.35 \times 10^{-2}$ kg. The input energy flux per unit cross-sectional area is:

$$\frac{1}{2}\Lambda\rho_a v^3 = 2.30 \times 10^7 \text{ W/m}^2. \qquad \text{(C4)}$$

The total heating power absorbed by the meteoroid is $\frac{1}{2}\Lambda\rho_a v^3 A_c \approx 28.9$ kW. To determine the surface temperature equilibrium, we set $dT/dt = 0$ (quasi-steady state) and neglect the conduction term, justified by the small skin depth. Dividing both sides by $A_c$ and noting that cooling occurs over $A_s = 4A_c$:

$$\frac{1}{2}\Lambda\rho_a v^3 = \frac{A_s}{A_c}\left[L\alpha_e P_v(T)\sqrt{\frac{\mu}{2\pi k_B T}} + \epsilon\sigma(T^4 - T_0^4)\right]. \qquad \text{(C5)}$$

We employ the Clausius–Clapeyron relation for the saturated vapor pressure:

$$P_v(T) = A_C \exp\left(-\frac{L_{mol}}{RT}\right), \qquad \text{(C6)}$$

commonly written in log-linear form:

$$\log_{10}(P_v) = A - \frac{B}{T}, \qquad \text{(C7)}$$

where $A \approx 10.6$ and $B \approx 13{,}500$ K for chondritic materials, with $P_v$ in Pa (Rogers et al., 2005).

Following Romig (1965), the kinetic energy lost by the body is partitioned such that the energy used for heating, light, and ionization scales approximately as 100:10:1. As a consistency check, solving for the surface temperature based solely on radiative cooling, ignoring ablation, yields $T \approx 3800$–$4600$ K, depending on the specific choice of $\Lambda$; the inclusion of ablation cooling brings the equilibrium temperature down to $\sim 1835$ K as derived below. Following the parameters established by Sekanina and Chodas (2012), the pre-exponential constant $A_C \approx 1.07 \times 10^{13}$ Pa is adopted. Note that the value of the molar latent heat $L_{mol}$ depends on the choice of the $A$ and $B$ coefficients; the latent heat used in the Clausius–Clapeyron derivation ($L_{mol}/M_{mol}$) is therefore not identical to the bulk effective heat of ablation $L$ that appears in Eq. (C3), where $M_{mol}$ is the molar mass.

Solving Eq. (C5) iteratively with the Hertz–Knudsen–Langmuir evaporation flux:

$$J \equiv \left(\frac{dm}{dt}\right)_v = \alpha_e P_v(T)\sqrt{\frac{\mu}{2\pi k_B T}}, \qquad \text{(C8)}$$

where $J$ denotes the per-area evaporative mass flux ($\text{kg m}^{-2}\ \text{s}^{-1}$), we obtain an equilibrium surface temperature of $T \approx 1835 \pm 25$ K, with $P_v \approx 1.75$ kPa. At this temperature, $\log_{10}(P_v) = 10.6 - 13500/1835 \approx 3.24$, giving $P_v = 10^{3.24} \approx 1750$ Pa.

The energy dissipation is heavily partitioned: approximately 90% through the latent heat flux of ablation and ~10% via radiative cooling. For the total meteoroid body, ablation cooling accounts for $Q_{lat} \approx 26.2$ kW and radiative cooling for $Q_{rad} \approx 2.9$ kW, summing to ~29.1 kW, consistent with the total heating input. For the purpose of maintaining a conservative estimate, this calculation ignores heat loss due to thermal conduction into the meteoroid's interior. Incorporating internal conduction would likely result in an even lower maximum surface temperature, further reinforcing the established thermal limits of the silicate nucleus. We also note that, besides the direct heating of the meteoroid surface and the surrounding atmosphere, a significant portion of the total energy is partitioned into auxiliary processes including molecular dissociation, electronic excitation, and mechanical work such as shock wave generation.

The total mass loss rate from evaporation is obtained by integrating the per-area flux $J$ over the meteoroid surface area $A_s = 4\pi r^2$:

$$\left(\frac{dm}{dt}\right)_{\text{total}} = J \cdot A_s \approx 4.28 \times 10^{-3}\ \text{kg/s}. \qquad \text{(C9)}$$

Using a mean atomic mass of 23 u ≈ $3.82 \times 10^{-26}$ kg per evaporated CM chondrite atom, the number loss rate is $dN/dt \approx 1.12 \times 10^{23}$ atoms/s. Converting to a line density:

$$\frac{dN_{\text{evap}}}{ds} = \frac{dN_{\text{evap}}/dt}{v} = 3.36 \times 10^{18}\ \text{m}^{-1}. \qquad \text{(C10)}$$

For comparison, applying the same approach with Clausius–Clapeyron coefficients more appropriate to cometary/meteoroid material (Rogers et al., 2005) yields a slightly higher equilibrium surface temperature and, for "dirty" cometary material containing a substantial fraction of low-temperature volatile-bearing phases, an effective evaporation rate that can be up to an order of magnitude higher than the chondritic value adopted here. This comparison is consistent with the assumed role of volatile-bearing constituents in providing the additional gas-phase mass loading required by the density-budget analysis of **Section C.5**. These theoretical maxima demonstrate that thermally driven silicate evaporation remains below the particle inventory required by the observed infrasonic and optical constraints within the acoustically inferred near-field volume. The remaining density deficit is closed by localized gas-phase mass loading from volatile release, as quantified in **Section C.5**.

### C.3. Sputtering Yields from Silicate-Rich Surfaces

We include sputtering in this analysis to account for the full range of particle contributions from the non-volatile surface. Sputtering is driven by direct impacts of atmospheric molecules onto the meteoroid surface and is most effective during the earliest phase of atmospheric entry, before sufficient ablated and vaporized material has accumulated in the near-field flow to deflect incoming molecules away from the first order surface collisions. As hydrodynamic shielding develops, that is, as the near-surface vapor density, ahead of the object, increases toward levels that modify the local flow, the rate of direct atmospheric impacts on the solid surface progressively decreases. However, the meteoroid surface temperature increases simultaneously, meaning that each remaining collision deposits energy into an increasingly hot surface from which atoms are more easily ejected. The per-collision sputtering yield therefore increases with surface temperature even as the collision rate decreases. By adopting an upper-bound yield estimate, we capture this tradeoff and maximize the silicate-only mass supply. Even under these favorable assumptions, the resulting non-volatile supply remains well below the shock formation requirement (**Section C.5**), demonstrating that additional mass-loading enhancement is needed.

To bound the upper range of direct impact-driven ejection from the non-volatile mineral component of the CM-like lower-bound end member, it is first necessary to consider the composition of CM chondrites. The CM-like proxy is phyllosilicate-rich and volatile-bearing, but this subsection isolates the silicate-rich surface contribution. Their matrix, comprising ~70 vol%, consists primarily of phyllosilicates belonging to the serpentine mineral group. CM chondrites also contain olivine- and pyroxene-rich chondrules, along with approximately 3–12 wt% water and ~2 wt% organic carbon (e.g., Rubin et al., 2007; Sanchez et al., 2024; Suttle et al., 2021; Trigo-Rodríguez et al., 2019). For the purposes of this first-order analysis, the CM chondrite matrix is treated as a pseudo-element represented by serpentine ($\mathrm{Mg_3Si_2O_5(OH)_4}$) as a commonly used proxy.

We adopt representative properties based on the well-studied Murchison CM chondrite (Cronin and Chang, 1993): mean atomic mass $M_t \approx 23.3$ u, mean atomic number $Z_t \approx 11.5$, surface binding energy $U_0 \sim 5.7$ eV, and bulk number density $n \approx 9.0 \times 10^{22}$ atoms/cm$^3$. The velocity of impacting atmospheric molecules is taken to be 33.5 km/s, consistent with the meteoroid velocity at perigee. We first calculate the kinetic energy of atmospheric $N_2$ and $O_2$ molecules impinging on the meteoroid surface:

$$E = \frac{1}{2}mv^2. \qquad \text{(C11)}$$

At 33.5 km/s, this yields $E_{N_2} \approx 163$ eV per $N_2$ molecule and $E_{O_2} \approx 186$ eV per $\mathrm{O_2}$ molecule. The dissociation energies of $\mathrm{N_2}$ and $\mathrm{O_2}$ are 9.8 eV and 5.2 eV, respectively. These impact energies exceed both dissociation thresholds, making dissociation energetically accessible and an important contributor to the near-surface and non-equilibrium collision cascade. The available kinetic energy per atom of the dissociated projectiles is then approximately 81.5 eV for N and 93.1 eV for O.

At these impact energies, the interaction of the dissociated atomic projectiles with the meteoroid surface can be treated as a collision cascade process. Following common practice for first-order sputtering estimates, we assume that nuclear stopping contributes significantly to energy deposition in the near-surface region, while electronic stopping is neglected in order to obtain a conservative lower-bound estimate of the sputtering yield (Hoang and Lee, 2020). This assumption is appropriate for establishing minimum material removal rates and does not preclude additional contributions from electronic excitation or thermal processes. The Sigmund sputtering formalism describes the ejection of surface atoms resulting from energetic particle impacts through the development of atomic collision cascades. In this formulation, energy deposition by the incident particle is treated statistically in an effectively infinite target medium, allowing the sputtering yield to be expressed in terms of projectile energy, target material properties, and surface binding energy. The model distinguishes between high-energy interactions, where linear collision cascades dominate, and lower-energy regimes characterized by single knock-on events or thermal-spike processes (Hobler et al., 2016; Sigmund, 1969). Although originally developed for ion–solid interactions, the formalism is commonly used as a first-order approximation for energetic neutral projectiles following dissociation, as assumed here. While a more rigorous treatment would require computationally intensive molecular dynamics simulations, this semi-empirical approach is sufficient to provide the order-of-magnitude estimates needed to establish the basis of the present argument. We employ the linear collision cascade formulation of Sigmund's sputtering theory. The sputtering yield $Y_{\mathrm{sput}}$ is:

$$Y_{\mathrm{sput}}(E,0) = \frac{3}{4\pi^2}\frac{\alpha_S(\mu)S_n(E)}{C_0 U_0}, \qquad \text{(C12)}$$

where $U_0$ is the surface binding energy, $\alpha_S(\mu)$ is a dimensionless function of the mass ratio $\mu = M_t/M_p$ (subscript $S$ for Sigmund, to distinguish from the electron line density $\alpha$ used in **Section C.6**), $S_n(E)$ is the nuclear stopping cross section in eV Å$^2$, and $C_0 \approx 1.81$ Å$^2$ arises from the Born–Mayer interatomic potential approximation. For the present case, $\alpha_S(\mu) \approx 0.2$ over the relevant mass-ratio range (Vondrak et al., 2008).

The nuclear stopping cross section is computed using the Lindhard–Scharff–Schiøtt (LSS) formalism. The reduced energy $\epsilon$ is given by:

$$\epsilon = \frac{0.8853 a_0 M_t E}{Z_p Z_t (M_p + M_t)(Z_p^{2/3} + Z_t^{2/3})^{0.5}}, \qquad \text{(C13)}$$

where $E$ is the projectile kinetic energy per atom in eV, $Z_p$ and $Z_t$ are the atomic numbers, $M_p$ and $M_t$ are the atomic masses, and $a_0 = 0.529$ Å is the Bohr radius. For N and O atoms at 81.5 eV and 93.1 eV respectively, we obtain $\epsilon = 6.95 \times 10^{-3}$ and $\epsilon = 6.47 \times 10^{-3}$.

The reduced nuclear stopping function is evaluated using the Ziegler–Biersack–Littmark (ZBL) universal function:

$$s_n(\epsilon) = \frac{0.5 \ln(1 + 1.1383\epsilon)}{\epsilon + 0.01321\epsilon^{0.21226} + 0.19593\epsilon^{0.5}}. \qquad \text{(C14)}$$

Substituting the values of $\epsilon$ yields $s_n(\epsilon) \approx 0.139$. The nuclear stopping cross section in eV Å$^2$ is then:

$$S_n(E) = 4\pi Z_p Z_t a \left(\frac{e^2}{4\pi\epsilon_0}\right) \frac{M_p}{M_p + M_t} s_n(\epsilon), \qquad \text{(C15)}$$

yielding $S_n(N) \approx 122.3$ eV Å$^2$ and $S_n(O) \approx 144.3$ eV Å$^2$. This provides a first-order estimate of the direct sputtering yield under the adopted collision-cascade assumptions. Additional enhancement from thermal spikes, electronic excitation, or chemically assisted desorption would raise the yield, while the progressive development of a near-surface vapor layer would reduce the rate of direct atmospheric impacts on the solid surface.

For oblique incidence at angle $\theta$ measured from the surface normal, a larger fraction of the projectile energy is deposited closer to the surface, increasing the probability that atoms are ejected. The angular dependence of the sputtering yield is described by the Yamamura formulation, which for projectile energies below ~200 eV reduces to the commonly used simplified form:

$$Y_{\text{sput}}(\theta) = Y_{\text{sput}}(0)(\cos\,\theta)^{-f}, \qquad \text{(C16)}$$

where $Y_{\text{sput}}(0)$ is the sputtering yield at normal incidence. In this energy regime, the parameter $f$ typically lies between 1.0 and 2.0; for light projectiles impacting rocky or silicate-rich surfaces, representative values are near $f \approx 1.5$. Experimental and theoretical studies show that sputtering yields typically peak at oblique incidence angles between approximately 60° and 80° from the normal, where the impacting atoms deposit a greater fraction of their energy closer to the surface (Tolmachev, 2011; Wasa, 2012). The optimal angle depends on target material properties. We adopt an incidence angle $\theta = 75°$ from the normal as a reasonable first-order approximation for a rotating, obliquely oriented meteoroid surface. The resulting yields per atmospheric molecule are approximately $Y_{\text{sput}} \approx 1.1$ for $N_2$ and $Y_{\text{sput}} \approx 1.2$ for $O_2$.

The primary species ejected from the surface of a CM chondrite under such collisions are expected to be oxygen, silicon, magnesium, and hydrogen, reflecting the dominant mineralogical constituents of the target material. If volatile-bearing phases are present within the immediate surface of an object, a small fraction of the sputtered material may be released in molecular form, such as OH derived from phyllosilicates. For the purposes of this lower-bound estimate, we assume that such molecular species constitute no more than 1–2% of this initial total sputtering yield.

Thus, if we neglect the partitioning of translational energy into rotational and vibrational excitation of surface molecules and atoms and consider only direct impact-induced ejection under sustained atmospheric bombardment, we find that the minimum number of atoms or molecules ejected from a CM chondrite surface is of order unity per incident atmospheric molecule. For the earthgrazer considered here, this result corresponds to a sputtering-only mass-loss rate of approximately $1.04 \times 10^{-3}$ kg/s, which remains below the mass-loss rate inferred from optical observations. Using the Popova et al. (2007) near-threshold yield would further reduce the non-volatile supply and increase the inferred volatile mass-loading requirement.

Both the impacting atmospheric gas and ejected meteoroid material are expected to exist in a non-equilibrium state within the developing vapor cap, with translational temperatures exceeding rotational and vibrational temperatures ($T_{\mathrm{tran}} \gg T_{\mathrm{rot}} \gg T_{\mathrm{vib}}$). Under these conditions, surface heating, impact-driven ejection, and thermally enhanced evaporation act together to increase the near-field vapor population. Radiative losses from the meteoroid surface, estimated using the Stefan–Boltzmann law, imply that the surface temperature stabilizes and remains substantially lower than the translational temperature of the impacting atmospheric molecules. Nevertheless, surface temperatures are still expected to substantially exceed 1800 K, corresponding to the boiling or evaporation temperatures of silicate melts. Previous observational and computational studies suggest that actual meteoroid surface temperatures will not exceed values approaching the vaporization temperature of their primary constituents, i.e., for chondrites $T_{\mathrm{vap}} \sim 2100$–$2500$ K (e.g., Borovička, 1994; Bronshten, 1983; Loehle et al., 2017; Sears, 2004). Under these conditions, thermally enhanced evaporation increases relative to purely impact-induced ejection despite the relatively high surface binding energy of approximately 5.7 eV. This enhancement arises from the high-energy tail of the Maxwell–Boltzmann distribution. The quantitative surface-temperature and evaporation calculation is given in **Section C.2**.

Despite these combined effects, the resulting number density of evaporated material remains insufficient to close the density gap required to reduce the local Knudsen number to values characteristic of strong shock formation. Using a more rigorous treatment of sputtering and evaporation initially described by Tielens et al. (1994) and subsequently applied by Hoang and Lee (2020), we obtain somewhat smaller lower-bound yields for evaporated atoms. When surface-temperature corrections to the binding energy, $U_0(T)$, are included, the maximum yield increases to approximately 15–26 released atoms per first order collision with an atmospheric molecule. This of course will depend on physical properties of the impacted surface as well as the temperature. Even under this favorable upper-limit assumption and accounting for the meteoroid cross-sectional area of approximately $9.6 \times 10^{-4}$ m$^2$, the total number of evaporated surface constituents from the non-volatile component of a CM chondrite remains insufficient to establish a continuous-flow regime or to reduce the local Knudsen number to values near 0.01 within the acoustic flow field defined earlier. The preceding estimates motivate the full quantitative density budget in **Section C.5**, where we test whether the non-volatile supply can close the required near-field line-density deficit. We next examine whether an additional mass source is available to close this deficit.

### C.4. Volatile Release and Grain-Evaporation Length Scales

CM chondrites contain a substantial inventory of volatile species, dominated by water and structurally bound hydroxyl groups in hydrous minerals (Lee et al., 2023). The total water content can reach up to approximately 12 wt%, with the majority present as structural water incorporated within the lattice of phyllosilicate minerals. Carbon is the next most abundant volatile constituent, reaching up to about 2.5 wt% (roughly 70% in insoluble organic matter). Nitrogen and sulfur-bearing phases may contribute up to approximately 3 wt% of the total mass. Sodium and other alkali species may also be released earlier than major silicate components and can contribute disproportionately to high-altitude visible emission, although their low abundance makes them a minor contributor to the total gas-phase mass budget (e.g., Vondrak et al., 2008).

At the computed equilibrium surface temperature of $T \sim 1835$ K (**Section C.2**), the principal thermally labile phases considered here, including adsorbed and weakly bound water, organic matter, structural hydroxyl in phyllosilicates, and carbonates, are above their relevant decomposition or release thresholds. Here, 'water-bearing' refers primarily to structurally bound water and hydroxyl in hydrated minerals, released during dehydration and dehydroxylation, rather than preserved free water or ice (e.g., Suttle et al., 2021; Trigo-Rodríguez et al., 2019). These volatile species reside predominantly within the meteoroid interior and are progressively exposed as thermal penetration and surface erosion advance during the near-perigee segment.

**Figure C2** illustrates the relationship between the surface temperature evolution and the qualitative release intensity of each volatile phase along the trajectory. More specifically, free and loosely bound water is released at temperatures between 373 K and 450 K. This is followed by the release of soluble organic compounds, including $CO_2$, CO, and in some cases $CH_4$ and other small hydrocarbons, depending on oxygen fugacity, catalytic surface effects, and local pressure conditions. Structural water associated with phyllosilicates is released during dehydration at higher temperatures, typically between 700 K and 900 K. In particular, dehydroxylation of serpentine near 750 K is expected to represent a major source of volatile release. This process is endothermic and can liberate significant quantities of steam, potentially inducing internal overpressure and contributing to mechanical weakening of the meteoroid.

At still higher temperatures, ~900–1100 K, carbonates trapped within the matrix decompose and release additional volatiles. At temperatures above this range, sulfides and sulfates undergo thermal decomposition, further contributing to volatile liberation. Owing to the low thermal conductivity and thermal diffusivity of CM chondritic material, with a representative thermal diffusivity of $k \approx 5 \times 10^{-7}$ m$^2$/s, the interior of a meteoroid with a diameter of ~4 cm is expected to remain at substantially lower temperatures than the surface over much of the luminous trajectory. Nevertheless, the combination of internal heating, high porosity, and progressive exposure of fresh material promotes sequential and relatively rapid volatile release throughout atmospheric passage.

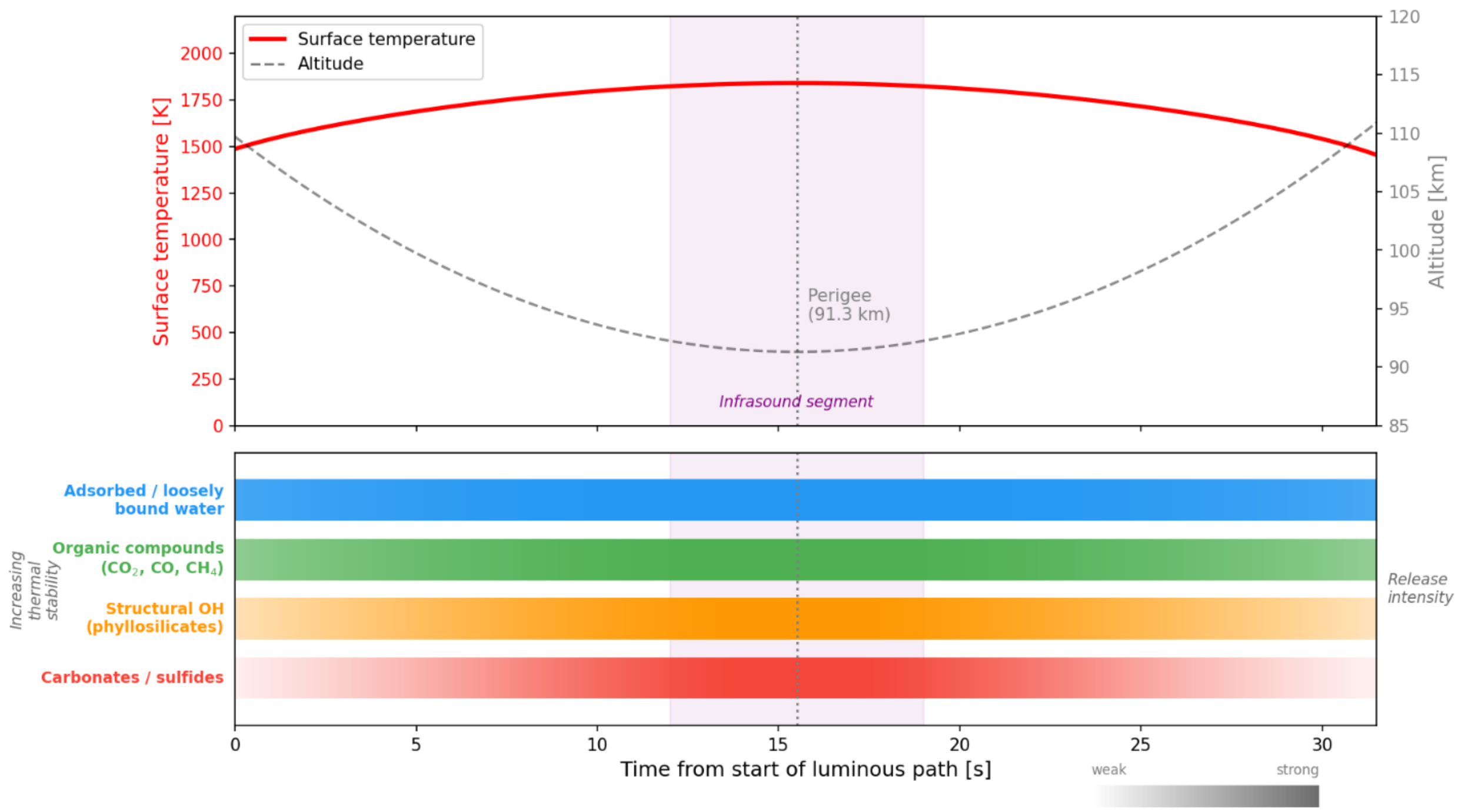

**Figure C2:** Schematic overview of volatile release along the earthgrazer trajectory. Top panel: Equilibrium surface temperature (red) and altitude (dashed grey) as a function of time from the start of the luminous path. The purple-shaded region marks the infrasound-generating segment. Bottom panel: Qualitative release intensity for four volatile-bearing phases, ordered by increasing thermal stability from top to bottom. Color saturation indicates release intensity, which scales with surface temperature. The least stable phases (adsorbed water) release over the broadest portion of the trajectory, while the most stable phases (carbonates, sulfides) contribute predominantly near perigee where surface temperatures are highest. The ordering reflects the relative thermal stabilities of these phases in carbonaceous chondrite material rather than phase-specific kinetic thresholds for this event.

The thermal skin depth, defined as the depth over which the temperature of a solid body is raised significantly, is estimated to be about 0.003 m, with temperatures reaching ~1000 K at that depth. This level of heat penetration is sufficient to mobilize volatile-rich interior material that is mechanically weaker and more susceptible to devolatilization than the surface layers.

Volatile-bearing phases provide a highly efficient source of gas-phase material on the near-field length scale. Dehydration, dehydroxylation, and decomposition release small molecules directly into the surrounding flow, while eroded silicate grains contribute vapor only after heating and surface evaporation over much longer downstream distances. This length-scale contrast makes volatile liberation the dominant candidate for localized mass loading within the $d_{ws}$-scale shock-coupled source region. **Section C.5** quantifies this requirement as an equivalent volatile yield needed to close the density deficit.

Typically, in the region immediately behind the shock front that delineates the flow field and in the near vicinity of a hypersonic object, gas temperatures may reach values on the order of $4 \times 10^3$ K and, for small and strongly ablating meteoroids, can exceed $10^4$ K (Hocking et al., 2016; Silber et al., 2018). These temperatures depend on several factors, including entry velocity, material properties, radiative cooling efficiency, and the degree of ablation and mass loading in the flow field. This thermal environment plays an important role in governing surface evaporation, volatile release, and the subsequent evolution of the near-field density. At $T \sim 1835$ K and $v = 33.5$ km/s, released grains in the erosion-model size range vaporize over distances far exceeding the acoustically inferred source scale. A 10 μm grain evaporates over $\sim 10^2$ m, and a 100 μm grain evaporates over $\sim 10^3$ m, whereas the relevant near-field scale is $d_{ws}/2 = 0.125$ m. Erosion therefore increases the total surface area available to the broader luminous wake, while volatile liberation supplies gas-phase material on the length scale required for shock-coupled near-field mass loading. As shown in **Figure C3**, this length-scale mismatch prevents erosion-released grains from closing the immediate near-field density deficit. Volatile release therefore provides localized gas-phase mass loading on the $d_{ws}$-scale, whereas grain evaporation contributes primarily to the broader luminous wake and meteor trail.

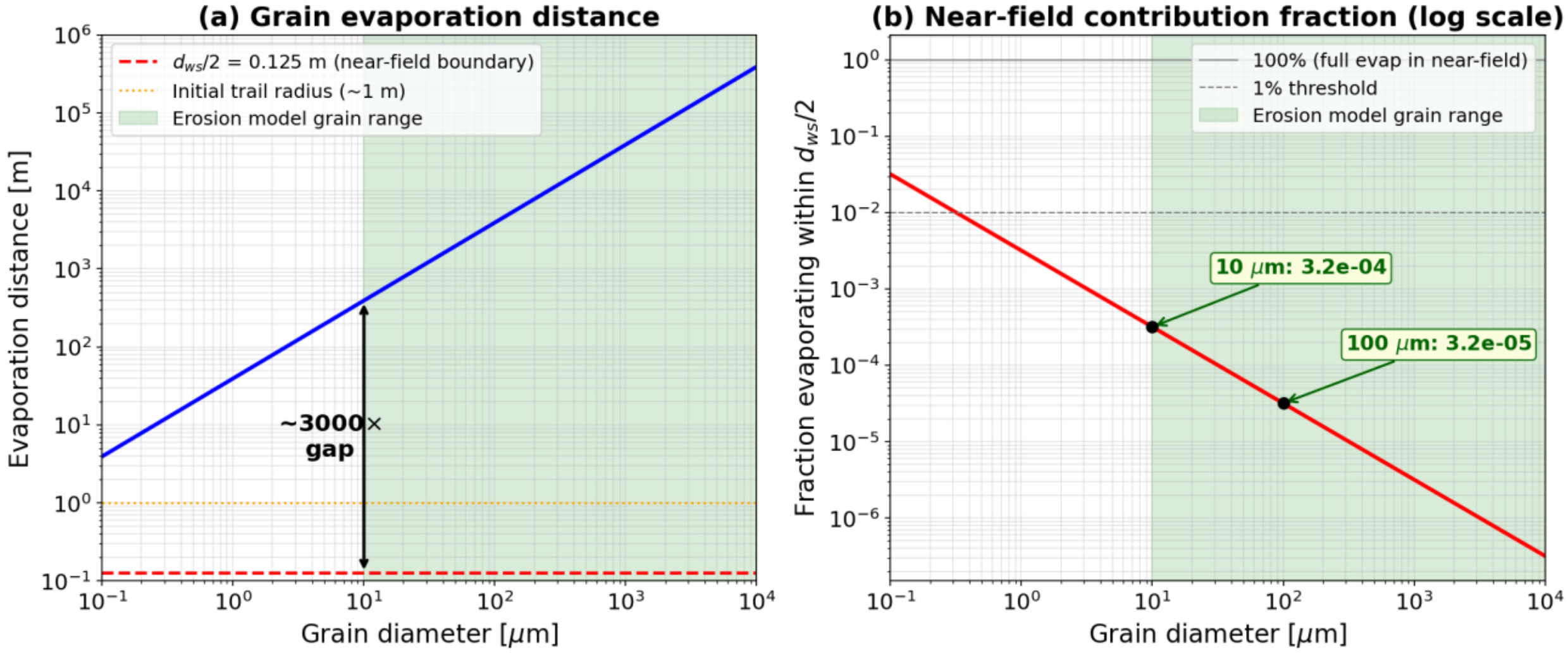

**Figure C3:** Evaporation distance for released silicate grains as a function of initial grain diameter at $v = 33.5$ km/s and $T_{\text{surf}} \approx 1835$ K (log scale). The horizontal dashed line marks the acoustic source half-diameter $d_{ws}/2 = 0.125$ m, which defines the relevant near-field length scale for shock-coupled mass loading. This length-scale mismatch of more than three orders of magnitude for even the smallest erosion-model grains shows that grain release contributes to the broader luminous wake but cannot supply near-field shock-coupled density.

## C.5. Density Budget and the Mass-Loading Deficit

Using the event parameters $v = 33.5$ km/s, meteoroid diameter $d_m = 0.04$ m, and altitude $h = 91.7$ km (with the associated G2S atmospheric specifications), we evaluate the cumulative particle line density available to the near-field flow. In mass calculations, we use a representative mean mass per released particle $m_p \approx 3.82 \times 10^{-26}$ kg (~23 amu).

The meteoroid cross-sectional area is:

$$A_c = \pi r^2 = \pi(0.02)^2 = 1.26 \times 10^{-3} \text{ m}^2. \qquad \text{(C17)}$$

The shielding vapor envelope is initiated as atmospheric impacts deposit tens to hundreds of eV into a thin surface layer of the meteoroid, releasing surface constituents through a combination of thermal evaporation and impact-driven sputtering. It should be noted that following the formation of the hydrodynamic shielding, the number of first order collisions drops exponentially, while the primary collisions with the surface of the object are of higher order. The effective yield, defined as the number of particles ejected per incident atmospheric molecule, depends strongly on entry velocity, surface temperature, and material properties. For example, extremely fast Leonid meteoroids have been inferred to eject several hundred atoms per collision under favorable conditions (e.g., Jenniskens et al., 2000; Zinn et al., 2004). In volatile-bearing objects, the number of released atoms and molecules in the near-field flow can be substantially larger than that produced by impact-induced evaporation alone, thereby enhancing hydrodynamic shielding and facilitating shock formation. This behavior is consistent with the concept of "screening by ablation vapor" described by Popova et al. (2000) and Popova et al. (2001).

**Impinging atmospheric molecules.** The number of atmospheric molecules intercepted per meter of travel by the meteoroid cross-section is:

$$N_{\mathrm{imp}} = nA_c(1\ \mathrm{m}) = (5.09 \times 10^{19})(1.26 \times 10^{-3}) = 6.41 \times 10^{16}\ \mathrm{m}^{-1}. \qquad \text{(C18)}$$

The corresponding mass of swept ambient air per meter is $M_{\mathrm{imp}} \approx 3.08 \times 10^{-9}$ kg/m.

**Ablation yield ($Y_{\mathrm{abl}}$).** At the computed surface temperature $T \sim 1835$ K, the combined yield from impact-driven atomic and molecular ejections and thermally enhanced surface evaporation is estimated to be $Y_{\mathrm{abl}} \approx 15$–$26$ released atoms or molecules per atmospheric collision, where we distinguish this total ablation yield $Y_{\mathrm{abl}}$ from the sputtering-only yield $Y_{\mathrm{sput}} \approx 1$ computed in **Section C.3**. Using a representative average of $Y_{\mathrm{abl}} = 20$:

$$N_{\mathrm{abl}} = Y_{\mathrm{abl}} \cdot N_{\mathrm{imp}} = 20 \times 6.41 \times 10^{16} \approx 1.28 \times 10^{18}\ \mathrm{m}^{-1}. \qquad \text{(C19)}$$

**Thermal evaporation ($N_{\mathrm{evap}}$).** From the energy balance analysis in **Section C.2**:

$$N_{\mathrm{evap}} \approx 3.36 \times 10^{18}\ \mathrm{m}^{-1}. \qquad \text{(C20)}$$

**Total supply and deficit.** The total supplied particle line density from impinging, ablation, and evaporation sources is:

$$\sum N = N_{\mathrm{imp}} + N_{\mathrm{abl}} + N_{\mathrm{evap}} \approx 4.70 \times 10^{18}\ \mathrm{m}^{-1}. \qquad \text{(C21)}$$

The density deficit is:

$$N_{\mathrm{deficit}} = N_{\mathrm{shock,req}} - \sum N = 1.50 \times 10^{19} - 4.70 \times 10^{18} \approx 1.03 \times 10^{19}\ \mathrm{m}^{-1}. \qquad \text{(C22)}$$

The non-volatile supply isolated from the CM-like lower-bound end member reaches $0.31\,N_{\mathrm{shock,req}}$, leaving a factor-of-3.2 line-density deficit. Under the optical and compositional constraints considered here, this deficit defines the localized gas-phase mass loading supplied most plausibly by volatile release. As shown in **Figure C4**, this deficit persists along the near-perigee segment associated with the infrasound source projections, where the non-volatile supply remains below the Rankine–Hugoniot shock requirement despite the increase in atmospheric density near minimum altitude.

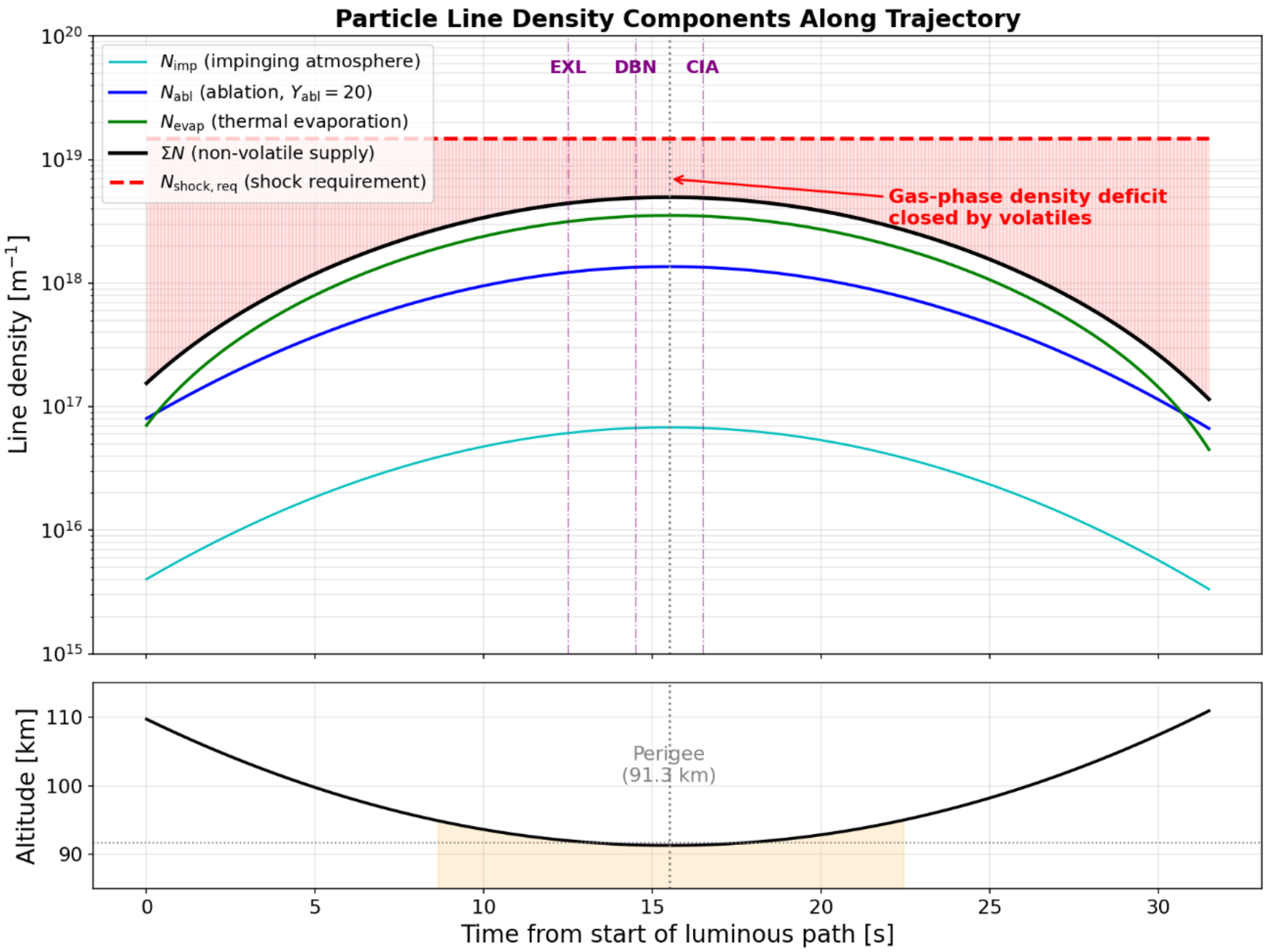

**Figure C4:** Particle line density components along the earthgrazer trajectory. Top panel: Individual contributions from impinging atmosphere $(N_{\text{imp}})$, ablation $(N_{\text{abl}}, Y_{\text{abl}} = 20)$, and thermal evaporation $(N_{\text{evap}})$, together with the total non-volatile supply $N_{\text{nonvol}}$(black) and the Rankine–Hugoniot shock requirement $N_{\text{shock,req}}$(dashed red). The pink-shaded region marks the gas-phase density deficit remaining after the non-volatile contribution is maximised. Purple dash-dotted lines indicate the trajectory segments projected onto the three infrasound stations. Bottom panel: Altitude profile, with the perigee region shaded.

**Required equivalent volatile yield.** Closing this deficit corresponds to an equivalent volatile yield of

$$Y_{\text{vol}} = \frac{N_{\text{deficit}}}{N_{\text{imp}}} \approx 161 \qquad \text{(C23)}$$

released volatile atoms or small molecules per atmospheric collision.

With this equivalent yield, $N_{\text{vol}} = Y_{\text{vol}} N_{\text{imp}} = 1.03 \times 10^{19}\ \text{m}^{-1}$, which closes the line-density deficit. Adopting a representative mean mass of ~23 amu per released particle, the volatile mass per meter of travel is $M_{\text{vol}} \approx 3.81 \times 10^{-7}$ kg/m, nearly two orders of magnitude larger than the ambient atmospheric mass swept per meter. Such mass loading provides a physically plausible explanation for how a volatile-enhanced near-field flow could satisfy the density conditions needed for shock generation at unexpectedly high altitudes despite the small size of the meteoroid. The qualitative physical picture and the detailed processes governing shock formation under these conditions are discussed in Silber et al. (2018) and in the classic treatment by Bronshten (1983).

### C.6. Independent Constraint from Electron Line Density

This subsection provides an independent constraint on the near-field density enhancement required for thermospheric shock formation by using the electron line density from classical meteor radar/photometric physics. Empirical relationships linking absolute visual magnitude to electron line density have been developed and validated over many decades, and therefore offer a well-established observational pathway to infer the electron (and, via ionization efficiency, the associated ablated/entrained particle inventory) in the early wake.

We use the observed luminosity to infer the electron line density and thereby estimate the electron/ablated-particle inventory implied by observations ('requirement') and then compare that requirement to the maximum particle supply expected from non-volatile (silicate/oxide) ablation of the physical nucleus ('supply'). Consistent with Sections C.1–C.4, the supply-side calculation is controlled by the physical meteoroid diameter ( $d_m \approx 0.04$ m ), whereas the requirement is evaluated within a control volume defined by the acoustically inferred characteristic flow-field diameter ( $d_{ws} \approx 0.25$ m ).

In meteor radar physics, the electron line density (often denoted $q$ in the radar literature) is the number of free electrons per unit trail length ($s$) measured along the trajectory (in $\mathrm{m}^{-1}$). Here we denote it by $\alpha$ (to avoid confusion with the aerodynamic dynamic pressure $q$) and define it as:

$$\alpha = \frac{dN_e}{ds}. \qquad \text{(C24)}$$

For a trail segment of length $\Delta s$, the total number of free electrons is $N_e(\Delta s) = \alpha \Delta s$. If we associate those electrons with a cylindrical control volume of cross-sectional area $A_{ws}$ and length $\Delta s$, the mean electron number density is:

$$\bar{n}_e = \frac{N_e}{V} = \frac{\alpha \Delta s}{A_{ws} \Delta s} = \frac{\alpha}{A_{ws}}. \qquad \text{(C25)}$$

This construction is self-consistent: integrating the (uniform, volume-averaged) density over the cross section recovers the line density, $N_e = \bar{n}_e A_{ws} \Delta s = \alpha \Delta s$.

**Choice of control volume: $r_0$ vs. $d_{ws}$.** In classical meteor-trail theory, an initial trail radius $r_0$ is often introduced to parameterize the Gaussian radial distribution of electrons in an adiabatically formed meteor trail (after rapid initial expansion). We do not use $r_0$ to define the control volume in this subsection. Instead, because our objective is to evaluate whether the early-time, shock-coupled near-field can achieve shock-supporting continuum-like conditions (i.e., a sufficiently low effective Knudsen number within the shock-bound flow field), we define the control volume using the acoustically inferred characteristic flow-field diameter $d_{ws} \approx 0.25$ m. Thus, the area in Eq. (C17) is $A_{ws} = \pi(d_{ws}/2)^2$. We mention $r_0$ only to emphasize that the classical adiabatic trail radius is a different (and typically much larger) scale than the shock-coupled control volume relevant here. For the object considered in this study, the initial radius of the adiabatically formed trail would yield a cross-sectional area at least a factor of four larger than that of the shock-bound flow field. The initial radius forms on timescales of order $10^{-4}$ s, whereas the shock-coupled near-field controlling shock onset evolves over the earliest $10^{-5}$–$10^{-4}$ s, depending on velocity and ablation rate. For the present event, with $v = 33.5$ km/s, a 1 m trail segment corresponds to $\sim 3 \times 10^{-5}$ s, which lies inside the $10^{-5}$–$10^{-4}$ s window. We therefore use $\Delta s = 1$ m as a convenient bookkeeping length for "how many electrons are present at early time", while noting that $\bar{n}_e$ is independent of the particular choice of $\Delta s$ because $\Delta s$ cancels exactly.

We relate the electron inventory to the total ablated-particle inventory $\left(N_{\text{part}}\right)$ using the ionization coefficient $\beta$, defined as the average number of free electrons produced per ablated meteoric atom (Bronshten, 1983; Jones, 1997):

$$\beta \equiv \frac{N_e}{N_{\text{part}}}. \qquad \text{(C26)}$$

The electron line density can be related to the mass loss rate as:

$$\alpha = -\frac{\beta}{\mu_0 v}\frac{dm}{dt}, \qquad \text{(C27)}$$

where $\mu_0$ is the mean mass per ablated atom (kg), $v$ is the meteoroid speed (m/s), and $dm/dt$ is the mass loss rate (negative during ablation). We note that the kinetic energy loss from deceleration ($mv\, dv/dt$) is neglected in this formulation. For the present earthgrazer, the total drag-induced fractional velocity change was less than 2%. This term is therefore negligible.

The mass loss rate from single-body ablation can be written as:

$$\frac{dm}{dt} = -\frac{\mathcal{A}\Lambda}{2L}\left(\frac{m}{\rho_m}\right)^{2/3}\rho_a v^3, \qquad \text{(C28)}$$

where $\mathcal{A}$ is the shape factor, $\Lambda$ is the heat transfer coefficient, $L$ is the effective heat of ablation, $\rho_a$ is atmospheric density, and $\rho_m$ is meteoroid bulk density. This can equivalently be expressed as:

$$-\frac{dm}{dt} = \frac{2I_v}{\tau_v v^2}, \qquad \text{(C29)}$$

where $I_v$ is the optical power in a defined band and $\tau_v$ is the luminous efficiency (~5–6%).

Combining Eqs. (C27) and (C28):

$$\alpha = \frac{\beta \mathcal{A} \Lambda}{2\mu_0 L} \rho_a v^2 \left(\frac{m}{\rho_m}\right)^{2/3}. \qquad \text{(C30)}$$

Alternatively, combining Eqs. (C27) and (C29) (e.g., Verniani and Hawkins, 1964):

$$I_v = \tau_v \frac{1}{2} v^2 \left(-\frac{dm}{dt}\right), \qquad \text{(C31)}$$

we can remove the ablation per unit time and write:

$$\alpha = \frac{2\beta}{\mu_0 \tau_v} \frac{I_v}{v^3}. \qquad \text{(C32)}$$

Therefore, $\alpha$ depends on $I_v$, $\beta/\tau_v$, composition (inferred via $\mu_0$), and $v^{-3}$.

We infer $\alpha$ from the observed absolute visual magnitude $M_p$ using the long-established empirical relations summarized in **Table C1** (Hawkins et al., 1964; Verniani and Hawkins, 1964; Weryk and Brown, 2013). We then convert $\alpha$ to $\bar{n}_a$ using $\beta$, and compare this observation-implied requirement to the maximum particle supply expected from non-volatile ablation of a $d_m \approx 0.04$ m nucleus. Using the absolute magnitude derived from the light curve $\left(M_p = -3.7\right)$, we obtain the electron line density $\alpha$between $0.91 \times 10^{17}$ m$^{-1}$ and $3.02 \times 10^{17}$ m$^{-1}$.

**Table C1:** Summary of calculated $\alpha$, $\bar{n}_e$, and $\bar{n}_a$. The effective area of the flow field is calculated using the diameter of 25 cm ($A_{ws} = 0.0491$ m$^2$). The relations are from Weryk and Brown (2013) (WB2013), Verniani and Hawkins (1964) (VH1964), and Hawkins et al. (1964) (HLS1964). Ionization coefficients: (a) $\beta = 0.0126$, (b) $\beta = 0.0133$, (c) $\beta = 0.0862$, (d) $\beta = 0.0899$.

| **Relation** | $\alpha$ **[10$^{17}$]** | $\bar{n}_e$ **[10$^{18}$]** | $\bar{n}_a^{(a)}$ **[10$^{20}$]** | $\bar{n}_a^{(b)}$ **[10$^{20}$]** | $\bar{n}_a^{(c)}$ **[10$^{19}$]** | $\bar{n}_a^{(d)}$ **[10$^{19}$]** |
|---|---|---|---|---|---|---|
| WB2013 $M_p = 38.7 - 2.5\log\,\alpha$ | 0.91 | 1.86 | 1.48 | 1.40 | 2.16 | 2.07 |
| VH1964 $M_p = 39.4 - 2.5\log\,\alpha$ | 1.74 | 3.55 | 2.81 | 2.66 | 4.11 | 3.94 |
| HLS1964 $M_p = 40.0 - 2.5\log\,\alpha$ | 3.02 | 6.15 | 4.88 | 4.63 | 7.14 | 6.84 |

The electron-line-density estimate provides an independent consistency check on the density-budget result. The observed optical luminosity implies a particle inventory of the same order as that required for shock-supporting near-field conditions. The comparison depends on the adopted ionization coefficient and the $d_{ws}$-scale control volume, but it independently supports the inference that the particle inventory exceeds what the silicate-only budget can provide under the adopted control-volume assumptions.

This agreement between the optical electron-line-density constraint and the acoustic shock-density requirement strengthens the inference that additional gas-phase mass loading was present in the near-field source region. The inferred particle inventory is consistent with the level of near-field density enhancement required to reduce the effective local Knudsen number toward shock-supporting collisional conditions.

**Table C2** summarizes the main result of the bounding calculation. Direct atmospheric impingement, optimistic non-volatile ablation yield ($Y_{\text{abl}} = 20$), and thermal evaporation together provide $4.70 \times 10^{18}\ \text{m}^{-1}$, or $0.31 N_{\text{shock,req}}$. Erosion-released grains increase the surface area contributing to the broader wake, but their evaporation length scales exceed the $d_{ws}$-scale source region by orders of magnitude. The remaining line-density deficit requires localized gas-phase mass loading.

This lower-bound analysis demonstrates that surface ablation and sputtering of the non-volatile component of the CM-like lower-bound end member cannot account for the near-field density required for strong shock formation. Consequently, non-volatile ablation and evaporation alone cannot sustain hydrodynamic shielding at the level needed to reduce the local Knudsen number to shock-supporting values under the observed thermospheric conditions. The sequential release of interior volatiles provides the necessary additional mass loading, amplifying hydrodynamic shielding and enabling the transition from rarefied, transitional flow to a shock-bound flow field capable of radiating the infrasound detected at the ground.

**Table C2:** Density-budget summary at $h = 91.7$ km for the CM-like conservative lower-bound volatile-bearing end member.

| Quantity | Value | Unit |
|---|---|---|
| Shock requirement, $N_{\text{shock,req}}$ | $1.50 \times 10^{19}$ | $\text{m}^{-1}$ |
| Impinging atmosphere, $N_{\text{imp}}$ | $6.41 \times 10^{16}$ | $\text{m}^{-1}$ |
| Ablation ($Y_{\text{abl}} = 20$), $N_{\text{abl}}$ | $1.28 \times 10^{18}$ | $\text{m}^{-1}$ |
| Thermal evaporation, $N_{\text{evap}}$ | $3.36 \times 10^{18}$ | $\text{m}^{-1}$ |
| Total non-volatile/silicate supply, $\sum N$ | $4.70 \times 10^{18}$ | $\text{m}^{-1}$ |
| Remaining gas-phase density deficit, $N_{\text{deficit}}$ | $1.03 \times 10^{19}$ | $\text{m}^{-1}$ |
| Supply / Requirement | 0.31 | – |
| Required equivalent volatile yield, $Y_{\text{vol}}$ | 161 | atoms/collision |